\documentclass[12pt,a4paper,final]{article}
\usepackage[utf8]{inputenc}
\usepackage[english]{babel}
\usepackage{amsmath}
\usepackage{amsfonts}
\usepackage{amssymb}
\usepackage{appendix}
\usepackage{faktor}
\usepackage[hidelinks]{hyperref}
\hypersetup{
    colorlinks=true,
    urlcolor=blue,
    linkcolor=blue,
    citecolor=red,
    }
\usepackage{amsthm}
\usepackage{bbold}
\usepackage{mathrsfs}
\usepackage{wasysym}
\usepackage{mathtools}
\usepackage{bm}
\usepackage{cancel}
\usepackage{color}
\usepackage{tikz-cd}
\usepackage[Glenn]{fncychap}
\usepackage{subfig}
\usepackage{accents}
\usepackage{slashed}
\usepackage{psfrag}
\usepackage{stackengine}
	
\usetikzlibrary{babel}
\makeatletter
\newcommand*\owedge{\mathpalette\@owedge\relax}
\newcommand*\@owedge[1]{%
	\mathbin{%
		\ooalign{%
			$#1\m@th\bigcirc$\cr
			\hidewidth$#1\m@th\wedge$\hidewidth\cr
		}%
	}%
}
\makeatother
\newenvironment{diagram}
{
	
	\centering
	\begin{tikzcd}
}
{ \end{tikzcd}\par }

\newtheorem{teo}{Theorem}[section]
\newtheorem{cor}[teo]{Corollary}
\newtheorem{prop}[teo]{Proposition}
\newtheorem{lema}[teo]{Lemma}
\newtheorem{defi}[teo]{Definition}

\newtheorem{obs}[teo]{Remark}

\newtheorem{rmk}[teo]{Remark}

\usepackage{makeidx}
\usepackage{graphicx}
\usepackage[left=2.60cm, right=2.60cm, top=2.40cm, bottom=2.60cm]{geometry}

\usepackage{scalerel}
\usepackage{stackengine,wasysym}

\usepackage{color}
    \makeatletter
\renewcommand\part{%
	\if@openright
	\cleardoublepage
	\else
	\clearpage
	\fi
	\thispagestyle{empty}%
	\if@twocolumn
	\onecolumn
	\@tempswatrue
	\else
	\@tempswafalse
	\fi
	\null\vfil
	\secdef\@part\@spart}
\makeatother

\newcommand{\wt}{\widetilde}
\newcommand{\wh}{\widehat}

\newcommand{\ul}{\underline}
\newcommand{\ol}{\overline}

\newcommand{\rad}{\operatorname{Rad}}

\newcommand{\bY}{\textup{\textbf Y}}

\newcommand{\bQ}{\textup{\textbf Q}}
\newcommand{\bF}{\textup{\textbf F}}
\newcommand{\bK}{\textup{\textbf K}}
\newcommand{\bPi}{\bm\Pi}
\newcommand{\bg}{\bm{\gamma}}

\newcommand{\real}{\mathbb R}
\newcommand{\tr}{\operatorname{tr}}

\renewcommand{\div}{\operatorname{div}}
\newcommand{\grad}{\text{grad}}

\newcommand{\st}{\stackrel}

\newcommand{\para}{\parallel}

\renewcommand{\d}{\coloneqq}

\newcommand{\dd}{\text{d}}

\newcommand{\ric}{\operatorname{Ric}}

\newcommand{\lie}{\mathcal{L}}
\newcommand{\mc}{\mathcal}

\renewcommand{\to}{\longrightarrow}

\title{\vspace*{-1.35cm}\textbf{Double Null Data and the Characteristic Problem in General Relativity}}
\author{Marc Mars\footnote{\href{mailto:marc@usal.es}{marc@usal.es}}\,\, and Gabriel Sánchez-Pérez\footnote{\href{mailto:gasape21@usal.es}{gasape21@usal.es}} \\
Departamento de Física Fundamental, Universidad de Salamanca\\
Plaza de la Merced s/n, 37008 Salamanca, Spain }
\date{\today}

\begin{document}
	\maketitle
\vspace{-2mm}
\begin{abstract}
	General hypersurfaces of any causal character can be studied abstractly using the hypersurface data formalism. In the null case, we write down all tangential components of the ambient Ricci tensor in terms of the abstract data. Using this formalism, we formulate and solve in a completely abstract way the characteristic Cauchy problem of the Einstein vacuum field equations. The initial data is detached from any spacetime notion, and it is fully diffeomorphism and gauge covariant. The results of this paper put the characteristic problem on a similar footing as the standard Cauchy problem in General Relativity.
	\end{abstract}

\section{Introduction}

The study of the Einstein field equations (EFE) is of unquestionable interest both in mathematics and in physics. From a physical point of view, the solutions of the EFE describe the gravitational field outside the matter sources and model a large amount of physical phenomena including black holes or propagation of gravitational waves. From a mathematical perspective they constitute a system of geometric PDE for a Lorentzian metric which can be studied in many different ways. One particularly relevant is the initial value (or Cauchy) problem. The fundamental well-posedness result of Choquet-Bruhat \cite{Choquet} established that prescribing the (full) ambient metric and its first normal derivative on a spacelike hypersurface subject to certain constraint equations gives rise to an ambient metric solving the vacuum EFE realizing this initial data. Since the Einstein equations are of geometric nature, the strategy was to solve a reduced system of quasi-linear hyperbolic PDE, called the reduced Einstein equations, obtained from the original EFE under the assumption of harmonic coordinates, and then proving that the solution of the reduced equations is indeed Ricci flat by showing that the coordinates in which the solution is obtained are actually harmonic. A detailed account of this beautiful argument, and extensions thereof, can be found in the excellent work \cite{ringstrom}.\\

This original statement for the initial value problem has evolved over time into a much more abstract notion where, instead of prescribing a full metric and its normal derivative, one prescribes an abstract $m$-dimensional manifold $\Sigma$, together with a Riemannian metric $h$ and a symmetric two covariant tensor field $K$ satisfying the so-called constraint equations. In this approach, the well-posedness of the initial value problem becomes a completely geometric (i.e. independent of any choice of coordinates) statement, namely that there exists an $(m+1)$-dimensional ambient manifold $(\mc M,g)$ and an embedding $f : \Sigma \rightarrow \mc M$ such that $h$ is the induced (pull-back) metric and $K$ the second fundamental form. Thus, the initial data has become completely detached from the spacetime one wishes to construct. This is very satisfactory because it puts the initial data at the same geometric level as the field equations themselves.\\

One of the main objectives of this paper is to achieve a similar geometrization for the characteristic initial value problem, where the data is posed, instead of on a time slice, on a pair of null hypersurfaces that intersect transversely. The essential result of Rendall \cite{Rendall} is that by prescribing the full spacetime metric on two transversely intersecting null hypersurfaces there exists a unique spacetime solution of the reduced Einstein equations in a future neighbourhood of the intersection compatible with the initial data. Moreover, working in the so-called \textit{standard coordinates}, Rendall gives a procedure to reconstruct all the components of the spacetime metric on the hypersurfaces from a minimal amount of initial data in such a way that the solution of the reduced equations is indeed a solution of the vacuum Einstein equations. Other approaches to this problem can be found in \cite{cabet,It,Hilditch_Kroon_Zhao,Book,Luk,Yau}. In \cite{Luk} Luk extends Rendall's result provided the intersection surface is a topological 2-sphere by showing that the spacetime exists in a future neighbourhood of the full two initial hypersurfaces (and not only of their intersection). In \cite{rodnianski} the authors state that Luk's result is valid in arbitrary dimension, and in \cite{cabet} this result is extended for a large class of symmetric hyperbolic systems including the Einstein equations in four spacetime dimensions without any assumption on the topology of the intersection spacelike surface. The characteristic problem is also known to be well-posed when the data is given on the future null cone of a point \cite{cone}.\\

Of special interest for the present work is the paper \cite{CandP} in which Chruściel and Paetz present new approaches to the characteristic problem and review the existing ones. The main difference with the approach followed by Rendall is that, instead of prescribing the minimal amount of data, they provide redundant initial data (and therefore this data must fulfill several constraint equations). In fact, one could add more initial data on the hypersurfaces by providing additional constraint equations (which in the resulting spacetime will become the null structure equations, i.e., the pullback of the Einstein equations on the hypersurfaces (cf. \cite{Book})). One of the main purposes of our paper is to follow this philosophy by providing enough abstract initial data (with the corresponding equations constraining it) to reconstruct the whole spacetime metric and the transverse derivative of its tangent components in the embedded case. One of the advantages of our approach is that it encompasses all the possible initial data constructed from the metric and its first derivatives for the characteristic initial value problem of the Einstein field equations. In order to write the initial data and its constraints from a geometrically satisfactory point of view, it is necessary to employ some abstract formalism capable of detaching the initial data from the spacetime one wishes to construct.\\

In \cite{Marc1,Marc2} a formalism to study hypersurfaces at the abstract level (i.e. by making no reference to any ambient space) has been developed. This data is called hypersurface data and it consists of an abstract manifold $\mc H$ together with a set of tensors on it. From such tensors one is able to reconstruct the full spacetime metric at the hypersurface as well as the transverse derivatives of its tangent components in the embedded case. This data is endowed with an internal gauge structure related to the freedom present in the choice of an everywhere transverse vector field in the embedded case. In this paper, we employ this formalism in its null version to formulate and solve the characteristic problem from an abstract point of view. In particular we show that all tangential components of the ambient Ricci tensor in the embedded case can be written in terms of the abstract data on (null) the hypersurface. This allows us to encode the tangential components of the $\Lambda$-vacuum field equations as a set of constraint equations written fully in terms of the hypersurface data. Moreover, these constraints turn out to be gauge-covariant, which means that once they are satisfied in a particular gauge, they are necessarily satisfied in any other gauge. In order to define the abstract notion of the two null and transverse hypersurfaces we need to extract the essential properties that two null hypersurface data must have so that they can be simultaneously embedded as two null hypersurfaces with a common spacelike boundary. It turns out that one requires a number of compatibility conditions at the boundary. These conditions define the abstract notion of double null data (Definition \ref{def_DND}).\\

The next step is to find a solution of the EFE from an initial double null data satisfying the abstract constraint equations. Before solving the reduced equations and proving that this solution is also a solution of the EFE we need to capture at the abstract level the notion of the harmonic condition on the coordinates. Given a set of $m$ independent functions on the abstract hypersurface we construct a vector field depending on these functions and on the hypersurface data. This vector field has the crucial property that it vanishes in the embedded case if and only if the given functions are harmonic at the hypersurfaces. The gauge behaviour of this vector field allows us to prove that there always exists an essentially unique gauge where it vanishes. This gauge, called ``harmonic gauge'', is still fully covariant and plays a key role in solving the characteristic problem in the present ~framework.\\

Given double null data written in the harmonic gauge and satisfying the constraint equations we solve the reduced Einstein equations with the metric data as initial condition. In order to prove that the solution of the reduced equations is indeed a solution of the EFE we follow a different approach to that of Rendall. While Rendall reconstructs the data integrating 2nd order ODE for the unknown components of the metric in a specific coordinate system, we have the full metric as given data, so we can solve the reduced Einstein equations straight away. From this solution we can build new embedded data which is a priori different from the original one. To conclude the proof we must show two things. Firstly, that the two data are actually the same, so that we have been able to embed the full given data (and not only the metric part of it). Secondly, that the spacetime is actually a solution of the Einstein field equations. We achieve both things in one go by combining the constraint equations and the gauge conditions. An informal version of the main Theorem that we prove is:\\

\textbf{Theorem 1.} Given double null data satisfying the abstract constraint equations, then there exists a unique spacetime $(\mc M,g)$ solution of the $\Lambda$-vacuum Einstein equations such that the the double null data is embedded on $(\mc M,g)$.\\

This result achieves our two main goals in this paper. Firstly, the initial data and the constraint equations are fully detached from the spacetime one wishes to construct, so we obtain a satisfactory geometrization of the characteristic initial value problem similar to the one of the standard Cauchy problem. Secondly, since in the resulting spacetime the initial data corresponds to the full metric and the transverse derivatives of its tangent components, our result encompasses all possible initial data constructed from them. There are many possible ways of trading prescribed data by constraints. Each consistent choice would lead to a different version of the characteristic initial value problem, but all of them can be viewed as a particularization of Theorem 1. This recovers (and greatly extends) ``The many ways of the characteristic Cauchy problem'' in \cite{CandP}. Moreover, we discuss how the Theorem can be accommodated to cover matter fields. We believe that the abstract geometrization of the characteristic initial value problem of General Relativity provides interesting new insights into the problem and that it will open up a new world of possibilities.\\

This paper is structured as follows. In Section \ref{preliminaries} we recall the definitions and general results of hypersurface data. Sect. \ref{section_CHD} is devoted to studying the notion of characteristic hypersurface data and its fundamental properties. In Section \ref{section_gauge} we compute the components of the ambient Ricci tensor in terms of the abstract (null) data and then promote these expressions into the abstract definition of the so-called constraint tensor. Sect. \ref{sec_fol} is devoted to computing the constraint tensor adapted to the foliation. The paper includes two Appendices. In Appendix \ref{appendix} we prove the gauge covariance of two tensors ($A$ and $B$) defined in terms of the curvature tensor of the hypersurface data connection. This result is used in Sect. \ref{section_gauge}. In Appendix \ref{appendixB} we derive a number of contractions of the tensors $A$ and $B$ which are needed in Sect. \ref{sec_fol}. In Section \ref{sec_HG} we introduce the harmonic gauge which plays a crucial role in solving the characteristic problem from an abstract point of view. Finally in Sect. \ref{sec_CP} we define the notion of double null data, we study some of its fundamental properties and we finish the section with the statement and proof of the main Theorem of this paper. 
	
\section{Preliminaries}

\label{preliminaries}

In this section we summarize the hypersurface data formalism, which is the general framework of this paper. This formalism was developed in \cite{Marc1,Marc2} with precursor \cite{Marc3}. Let us start by fixing some notation and conventions. In a differentiable manifold $\mc H$ we let $\mc F(\mc H)=\mc C^{\infty}(\mc H,\real)$ and $\mc{F}^{\star}(\mc H)$ be the subset of nowhere-vanishing functions. $T\mc H$ is the tangent bundle of $\mc H$ and $\Gamma(T\mc H)$ the sections of $T\mc H$. Similarly, $T^{\star}\mc H$ is the cotangent bundle of $\mc H$ and $\Gamma(T^{\star}\mc H)$ the corresponding space of one-forms. The interior contraction of a covariant tensor $T$ with a vector $X$ is the tensor $i_X T \d T(X,\cdot,\cdots,\cdot)$. Abstract indices on $\mc H$ are denoted with small Latin letters from the beginning of the alphabet ($a,b,c,...$). As usual, parenthesis (resp. brackets) denote symmetrization (resp. antisymmetrization) of indices, and the symbol $\otimes_s$ denotes the symmetrized tensor product $T_1\otimes_s T_2 = \frac{1}{2}(T_1\otimes T_2 + T_2\otimes T_1)$. In this paper spacetime means a smooth, orientable manifold endowed with a time-oriented Lorentzian metric. All manifolds are connected unless otherwise indicated.

\begin{defi}
	\label{defi_hypersurfacedata}
	Let $\mc H$ be a smooth $m$-dimensional manifold, $\bg$ a symmetric two-covariant tensor field, $\bm\ell$ a one-form and $\ell^{(2)}$ a scalar on $\mc H$. A four-tuple $\mc D^{met}=\{\mc H,\bg,\bm\ell,\ell^{(2)}\}$ defines metric hypersurface data (of dimension $m$) provided that the symmetric two-covariant tensor $\mc A|_p$ on $T_p\mc H\times\real$ defined by $$\mc A|_p\left((W,a),(Z,b)\right) \d \bg|_p (W,Z) + a\bm\ell|_p(Z)+b\bm\ell|_p(W)+ab\ell^{(2)}|_p$$ is non-degenerate at every $p\in\mc H$. A five-tuple $\mc D=\mc D^{met}\cup \{\bY\}$, where $\bY$ is a symmetric two-covariant tensor field on $\mc H$, is called hypersurface data.
\end{defi}

Since $\mc A|_p$ is symmetric and non-degenerate for every $p\in \mc H$, there exists a unique symmetric, two-contravariant tensor $\mc A^{\sharp}|_p$ defined by $\mc A^{\sharp}|_p(\mc A|_p(V,\cdot),\cdot) = V$ for every $V\in T_p\mc H\times \real$. Let $a,b\in\real$ and $\bm\alpha,\bm\beta\in T^{\star}_p\mc H$. Then we can define a symmetric, two-contravariant tensor $P|_p$, a vector $n|_p$ and a scalar $n^{(2)}|_p$ on $T_p\mc H$ by 
\begin{equation}
	\label{Asharp}
	\mc A^{\sharp}\left((\bm\alpha,a),(\bm\beta,b)\right) = P|_p (\bm\alpha,\bm\beta)+a n|_p(\bm\beta)+bn|_p(\bm\alpha)+ab n^{(2)}|_p.
\end{equation} 
The definition of $\mc A^{\sharp}$ is equivalent to
\begin{align}
	\bg_{ab}n^b+n^{(2)}\bm\ell_a&=0,\label{gamman}\\
	\bm\ell_a n^a+n^{(2)}\ell^{(2)}&=1,\label{ell(n)}\\
	P^{ab}\bm\ell_b+\ell^{(2)} n^a&=0,\label{Pell}\\
	P^{ab}\bg_{bc} +n^a\bm\ell_c &= \delta^a_c.\label{Pgamma}
\end{align}

As usual, the radical of $\bg$ is defined by $$\rad(\bg)_p \d \left\{X\in T_p\mc H\ : \ \bg(X,\cdot)=0\right\}\subset T_p\mc H.$$ An important property is that $\rad(\bg)_p$ is either empty or one-dimensional \cite{Marc2}. The points $p\in\mc H$ such that $\rad(\bg)_p\neq \langle 0 \rangle$ are called null points. The condition for $p$ being a null point is equivalent to $\rad(\bg)_p = \langle n|_p\rangle$ and also equivalent to $n^{(2)}|_p=0$. Despite its name, the notion of (metric) hypersurface data does not view $\mc H$ as a hypersurface of another manifold. The connection between the abstract data and the standard notion of hypersurface is as follows.
\begin{defi}
	\label{defi_embedded}
	A metric hypersurface data $\{\mc H,\bg,\bm\ell,\ell^{(2)}\}$ is embedded in a pseudo-Riemannian manifold $(\mc M,g)$ if there exists an embedding $f:\mc H\hookrightarrow\mc M$ and a vector field $\xi$ along $f(\mc H)$ everywhere transversal to $f(\mc H)$, called rigging, such that
	\begin{equation}
		f^{\star}(g)=\bg, \hspace{0.5cm} f^{\star}\left(g(\xi,\cdot)\right) = \bm\ell,\hspace{0.5cm} f^{\star}\left(g(\xi,\xi)\right) = \ell^{(2)}.
	\end{equation}
The hypersurface data $\{\mc H,\bg,\bm\ell,\ell^{(2)},\bY\}$ is embedded provided that, in addition, 
\begin{equation}
	\label{Yembedded}
	\dfrac{1}{2}f^{\star}\left(\lie_{\xi} g\right) = \bY.
\end{equation}
\end{defi}
Motivated from this geometric picture, hypersurface data satisfying $n^{(2)}=0$ everywhere on $\mc H$ will be called \textbf{null hypersurface data}. The necessary and sufficient condition for $f(\mc H)$ to admit a rigging is that $f(\mc H)$ is orientable (see \cite{Marc1}). Observe that the signature of the ambient metric $g$ and of the tensor $\mc A$ are necessarily the same.\\

Given hypersurface data one defines the tensor 
\begin{equation}
	\label{defK}
	\bK\d n^{(2)} \bY +\dfrac{1}{2}\lie_n\bg + \bm{\ell}\otimes_s \dd n^{(2)},
\end{equation} 
which when the data is embedded coincides \cite{Marc1} with the second fundamental form of $f(\mc H)$ with respect to the unique normal one-form $\bm{\nu}$ satisfying $\bm{\nu}(\xi)=1$. For embedded hypersurfaces the set of transversal vector fields is given by $z(\xi+f_{\star}\zeta)$, where $z\in\mc{F}^{\star}(\mc H)$ and $\zeta\in\Gamma(T\mc H)$. In terms of the abstract data, this translates into a gauge freedom.
\begin{defi}
	\label{defi_gauge}
	Let $\{\mc H,\bg,\bm\ell,\ell^{(2)},\bY\}$ be hypersurface data. Let $z\in\mc{F}^{\star}(\mc H)$ and ${\zeta\in\Gamma(T\mc H)}$. The gauge transformed hypersurface data with gauge parameters $(z,\zeta)$ are 
	\begin{align}
		\mc{G}_{(z,\zeta)}\left(\bg \right)&\d \bg,\label{transgamma}\\
		\mc{G}_{(z,\zeta)}\left( \bm{\ell}\right)&\d z\left(\bm{\ell}+\bg(\zeta,\cdot)\right),\label{tranfell}\\
		\mc{G}_{(z,\zeta)}\left( \ell^{(2)} \right)&\d z^2\left(\ell^{(2)}+2\bm\ell(\zeta)+\bg(\zeta,\zeta)\right),\label{transell2}\\
		\mc{G}_{(z,\zeta)}\left( \bY\right)&\d z \bY + \bm\ell\otimes_s \dd z +\dfrac{1}{2}\lie_{z\zeta}\bg.\label{transY}
	\end{align}
\end{defi}


The set of gauge transformations defines a group with composition law and inverse \cite{Marc1}
\begin{align}
	\mc{G}_{(z_1,\zeta_1)}\circ \mc{G}_{(z_2,\zeta_2)} &= \mc{G}_{(z_1 z_2,\zeta_2+z_2^{-1}\zeta_1)}\label{group}\\
 	\mc{G}_{(z,\zeta)}^{-1}& = \mc{G}_{(z^{-1},-z\zeta)}\label{gaugelaw}
 \end{align} 
and the identity element is $\mc G_{(1,0)}$. A gauge transformation on the data induces another on the contravariant data given by 
\begin{align}
\mc{G}_{(z,\zeta)}\left(P \right) &= P + n^{(2)}\zeta\otimes\zeta-2\zeta\otimes_s n,\label{gaugeP}\\
\mc{G}_{(z,\zeta)}\left( n \right)&= z^{-1}(n-n^{(2)}\zeta),\label{transn}\\
\mc{G}_{(z,\zeta)}\left( n^{(2)} \right)&= z^{-2}n^{(2)}.
\end{align}
In agreement with its geometric interpretation in the embedded case,
\begin{equation}
	\label{Ktrans}
	\mc{G}_{(z,\zeta)}\left(\bK \right)= z^{-1}\bK.
\end{equation} 
Hypersurface data admits a torsion-free connection $\ol\nabla$ defined by the conditions \cite{Marc1}
\begin{align}
	\left(\ol\nabla_X\bg\right)(Z,W) &= - \bm\ell(Z) \bK(X,W)- \bm\ell(W) \bK(X,Z),\label{olnablagamma}\\
	\left(\ol\nabla_X\bm{\ell}\right)(Z)& = \bY(X,Z) + \bF(X,Z)-\ell^{(2)} \bK(X,Z).\label{olnablaell}
\end{align}
where
\begin{equation}
	\label{def_F}
	\bF\d \dfrac{1}{2}\dd\bm\ell.
\end{equation} 
Equations \eqref{olnablagamma} and \eqref{olnablaell} can be thought as a generalization of the Koszul formula. Under a gauge transformation $\ol\nabla$ transforms as \cite{Marc1}
\begin{equation}
	\label{gaugeconnection}
\mc G_{(z,\zeta)}\left(\ol\nabla\right) = \ol\nabla + \zeta\otimes \bK.
\end{equation}
In the context of embedded data (with ambient $(\mc M,g)$, rigging $\xi$ and corresponding Levi-Civita connection $\nabla$), it satisfies \cite{Marc1,Marc3}
\begin{equation}
	\label{nablambient}
	\nabla_X Z = \ol\nabla_X Z - \bK(X,Z)\xi,
\end{equation}
 where $X,Z$ in the LHS of the previous equation are the push forward of $X,Z\in\Gamma(T\mc H)$. Here and in the rest of this paper we will make the (standard) abuse of notation of identifying a vector field and its image under $f_{\star}$ and let the context determine the meaning. Let $\{e_a\}$ be a (local) basis on $\Gamma(T\mc H)$. The derivatives $\nabla_{e_a}e_b$ and $\nabla_{e_a}\xi$ can be decomposed as \cite{Marc1}
 \begin{align}
 	\nabla_{e_a} e_b &= \ol\Gamma_{ba}^c e_c-\bK_{ab}\xi,\label{nablatt}\\
 	\nabla_{e_a}\xi &= \left(\dfrac{1}{2}n^{(2)}\ol\nabla_a\ell^{(2)}+(\bY_{ab}+\bF_{ab})n^b\right)\xi + \left(P^{bc}(\bY_{ac}+\bF_{ac})+\dfrac{1}{2}n^b\ol\nabla_a\ell^{(2)}\right)e_b,\label{nablatxi}
 \end{align}
 where $\ol\Gamma_{ba}^c$ are the connection coefficients of $\ol\nabla$. Equations \eqref{olnablagamma}-\eqref{olnablaell} together with \eqref{gamman}-\eqref{Pgamma} yields \cite{Marc1}
\begin{align}
	\ol\nabla_a n^b & = P^{bc}\left(\bK_{ac}-n^{(2)}(\bF_{ac}+\bY_{ac})\right)-(\bY_{ac}+\bF_{ac})n^b n^c -n^{(2)} n^b \ol\nabla_a \ell^{(2)},\label{olnablan}\\
	\ol\nabla_a P^{bc} & = -P^{cd}(\bY_{ad}+\bF_{ad}) n^b -P^{bd}(\bY_{ad}+\bF_{ad})n^c -n^c n^b \ol\nabla_a\ell^{(2)}.\label{olnablaP}
\end{align}

\section{Characteristic Hypersurface Data}
\label{section_CHD}
In this Section we particularize the general results of Section \ref{preliminaries} to the specific setup of characteristic data, which is the central object of this paper.

\begin{defi}
	\label{def_CHD}
	Let $\mc D=\{\mc H,\bg,\bm\ell,\ell^{(2)},\bY\}$ be hypersurface data of dimension $m$. We say that the set $\mc D$ is ``characteristic hypersurface data'' (CHD) provided that 
	\begin{enumerate}
		\item $\rad(\bg)\neq\{0\}$ and $\bg$ is semi-positive definite.
		\item There exists $\ul u\in\mc{F}(\mc H)$ satisfying $\lambda\d n(\ul u)\neq 0$. Such functions are called ``foliation functions'' (FF).
		\item The leaves $\mc S_{\ul u} \d \left\{p\in\mc H: \ \ul u(p)=\ul u\right\}$ are all diffeomorphic.
	\end{enumerate}
\end{defi} 

\begin{rmk}
	All the results of Sections \ref{section_CHD}-\ref{sec_HG} are insensitive to the signature of $\bg$ provided $\rad(\bg)\neq\{0\}$. The condition that $\bg$ is semi-positive definite is only needed in Sec. \ref{sec_CP}.
\end{rmk}
 
Let $\mc S$ be the underlying topological space of each $\mc S_{\ul u}$. Then the topology of $\mc H$ is fixed to be a product of the form $\mc H\simeq \mc I\times\mc S$, where $\mc I\subset \real$ is an interval, and $\{\mc S_{\ul u}\}$ defines a foliation on $\mc H$. In this paper we will always assume that $\mc S$ is orientable, and thus $\mc H$. With a foliation on $\mc H$, we can define the following decomposition of the tangent space. Let $p\in\mc H$ and $X\in T_p\mc H$. We say that $X$ is tangent to the leaf $\mc S_{\ul u(p)}$ at the point $p$ provided that $X(\ul u)|_p = 0$. The set of all these vectors defines $T_p\mc S_{\ul u(p)}$. Hence, at every point $p\in\mc H$, the tangent space $T_p\mc H$ decomposes as 
\begin{equation}
	\label{decomposition}
	T_p\mc H = T_p\mc S_{\ul u(p)}\oplus \langle n|_p\rangle.
\end{equation} 
This decomposition induces another on the cotangent space, $T_p^{\star}\mc H = T_p^{\star}\mc S_{\ul u(p)}\oplus \langle \dd\ul u|_p\rangle$. Now let $T\mc S_{\ul u}=\bigcup_{q\in\mc S_{\ul u}} T_q\mc S_{\ul u}$ and $T\mc S = \bigcup_{\ul u\in\mc I} T\mc S_{\ul u}$. Then, the tangent bundle $T\mc H$ is the direct sum $T\mc H= T\mc S\oplus\langle n\rangle$, and therefore every vector field $X\in\Gamma(T\mc H)$ can be written in a unique way as $X = X_{\para}+ X_{n} n$, with $X_{\para}\in \Gamma(T\mc S)$ and $X_{n}\in\mc F(\mc H)$. A vector field $X\in\Gamma(T\mc H)$ is said to be tangent to the foliation $\{\mc S_{\ul u}\}$ provided that $X_n$ vanishes on $\mc H$. Analogously we have the cotangent bundle decomposition 
\begin{equation}
	\label{decocotang}
	T^{\star}\mc H=T^{\star}\mc S\oplus \langle\dd \ul u\rangle
\end{equation} 
and we can talk about 1-forms tangent to the foliation when they belong to $T^{\star}\mc S$. Finally one can generalize all this to any tensor field. In this paper we will abuse the notation and denote with $T_p\mc S_{\ul u}$ (resp. $T_p^{\star}\mc S_{\ul u}$) both the tangent (resp. cotangent) space of the manifold $\mc S_{\ul u}$ and the subset of $T_p\mc H$ (resp. $T_p^{\star}\mc H$) as defined above. The precise meaning will be clear from the context. Let $\mc D$ be CHD endowed with a foliation function $\ul u$ and consider the unique vector field $N\in\rad(\bg)$ such that $N(\ul u)=1$\footnote{$N$ and $n$ are proportional to each other, and related by $n=n(\ul u)N=\lambda N$.}. Given $v\in \Gamma(T\mc S_{\ul u_0})$ for some $\ul u_0\in\mc I$, we can define a unique vector field $X$ on $\mc H$ by integrating the equation $\lie_N X=0$ with the initial conditions $X_{\para}|_{\mc S_{\ul u_0}} = v$ and $X_n|_{\mc S_{\ul u_0}} =0$. This field is tangent to the foliation because $X(\ul u)=0$ on $\mc S_{\ul u_0}$ and $\lie_N X = 0$ implies $N\left(X(\ul u)\right) = X\left(N(\ul u)\right)=0$, so $X(\ul u)$ is constant along $N$. By virtue of decomposition \eqref{decomposition}, a torsion-free connection $\ol\nabla^{\mc S}$ on the leaves $\{\mc S_{\ul u}\}$ can be defined by 
\begin{equation}
	\label{decompnabla}
	\ol\nabla_X Z = \ol\nabla_X^{\mc S} Z - \bQ(X,Z)n, \hspace{1cm} X,Z\in \Gamma(T\mc S_{\ul u}),
\end{equation} 
where $\ol\nabla_X^{\mc S} Z\in\Gamma(T\mc S_{\ul u})$ and $\bQ$ is a symmetric\footnote{If $X,Z\in \Gamma(T\mc S_{\ul u})$, then $X(\ul u)=Z(\ul u)=0$ and therefore $[X,Z](\ul u)=0$, so $[X,Z]\in \Gamma(T\mc S_{\ul u})$. Since $\ol\nabla$ is torsion-free, $\ol\nabla_X Z -\ol\nabla_Z X =[X,Z]$ is tangent to the foliation. This proves both that $\ol\nabla^{\mc S}$ is torsion-free and that $\bQ$ is symmetric.}, two-covariant tensor field on $\mc S_{\ul u}$. In order to identify this tensor, we recall that $\lambda= n(\ul u)\neq 0$ and $\dd\ul u(\ol\nabla_X^{\mc S} Z)=0$, so
\begin{equation}
	\label{Q1}
	\lambda \bQ(X,Z) = -\dd\ul u \left(\ol\nabla_X Z\right) = -\ol\nabla_X\left(\dd\ul u(Z)\right)+\left(\ol\nabla_X\dd\ul u\right)(Z) = \left(\ol\nabla_X\dd\ul u\right)(Z),
\end{equation} 
where in the last equality we used that $\dd\ul u(Z)=Z(\ul u)=0$. Let $\phi_{\ul u}:\mc S_{\ul u}\hookrightarrow\mc H$ be an embedding. Defining the one-form $\bm\ell_{\para} \d \phi_{\ul u}^{\star}\bm\ell\in T^{\star}\mc S_{\ul u}$ we can decompose 
\begin{equation}
	\label{elldu}
	\dd\ul u = \lambda \left(\bm\ell-\bm\ell_{\para}\right),
\end{equation} 
which follows at once by applying both sides to tangential vectors and to $n$. Here we are abusing the notation by identifying $\bm\ell_{\para}$ with the one-form on $\mc H$ that coincides with $\bm\ell_{\para}$ acting over tangent vectors and vanishes when it acts on $n$. Then, 
\begin{align*}
	\left(\ol\nabla_X\dd\ul u\right)(Z)& = \lambda \left(\ol\nabla_X\bm\ell\right)(Z) - \lambda\left(\ol\nabla_X\bm\ell_{\para}\right)(Z)\\
	& = \lambda \left(\ol\nabla_X\bm\ell\right)(Z) - \lambda X\left(\bm\ell_{\para}(Z)\right) + \lambda \bm\ell_{\para}\big(\ol\nabla_X^{\mc S} Z - \bQ(X,Z)n\big)\\
	&=\lambda \left(\ol\nabla_X\bm\ell\right)(Z) - \lambda\big(\ol\nabla^{\mc S}_X\bm\ell_{\para}\big)(Z),
\end{align*} 
since $\bm\ell_{\para}(n)=0$. Inserting this together with \eqref{olnablaell} into \eqref{Q1}, it follows 
\begin{equation}
	\label{tensorQ}
	\bQ(X,Z) = \bY(X,Z) + \bF(X,Z) - \ell^{(2)} \bK(X,Z) - \big(\ol\nabla^{\mc S}_X\bm\ell_{\para}\big)(Z).
\end{equation}

Let $\mc D$ be CHD and $p\in \mc H$. Since $\rad(\bg)|_p=\langle n|_p\rangle$ the tensor $h\d\phi_{\ul u}^{\star}\bg$ is a metric on $\mc S_{\ul u}$. It is convenient to have an explicit relation between its Levi--Civita connection, $\nabla^h$, and the induced connection $\ol\nabla^{\mc S}$. First we prove some intermediate results.

\begin{prop}
	\label{nablah}
	Let $h=\phi_{\ul u}^{\star}\bg$ be the induced metric on $\mc S_{\ul u}$ and ${\bm\chi}\d \phi_{\ul u}^{\star} \bK$. Then, 
	\begin{equation}
		\label{eqnablah}
		\ol\nabla^{\mc S}_X h = -2 \bm\ell_{\para}\otimes_s {\bm\chi}(X,\cdot) \hspace{0.5cm} \forall X\in\Gamma(T\mc S).
	\end{equation}
	\begin{proof}
		Let $V,X,Z\in \Gamma(TS_{\ul u})$. Then, 
		\begin{align*}
			\big(\ol\nabla^{\mc S}_X h\big) (V,Z)&= \ol\nabla^{\mc S}_X \left(h(V,Z)\right) - h (\ol\nabla^{\mc S}_X V,Z)-h(V, \ol\nabla^{\mc S}_X Z)\\
			&=\ol\nabla_X \left(\bg(V,Z)\right) -  \bg(\ol\nabla_X V,Z)-\bg(V, \ol\nabla_X Z)\\
			&=\phi_{\ul u}^{\star}\left(\ol\nabla_X\bg\right)(V,Z).
		\end{align*}
which, upon inserting \eqref{olnablagamma}, gives \eqref{eqnablah}.
	\end{proof}
\end{prop}
In order to obtain a general relation between $\ol\nabla^{\mc S}$ and $\nabla^h$ we first determine how $\ol\nabla^{\mc S}$ transforms under a gauge transformation. 

\begin{prop}
	\label{gaugeconection}
	Let $\mc D=\{\mc H,\bg,\bm\ell,\ell^{(2)},\bY\}$ be CHD and $(z,\zeta)\in\mc F^{\star}(\mc H)\times\Gamma(T\mc H)$. Then,
\begin{equation*}
	\mc{G}_{(z,\zeta)}(\ol\nabla^{\mc S}) = \ol\nabla^{\mc S} + \zeta_{\para}\otimes {\bm\chi}.
\end{equation*}
	\begin{proof}
 Let $\ol\nabla'=\mc G_{(z,\zeta)}\ol\nabla$. Since by \eqref{gaugeconnection} $$\ol\nabla'_X Z = \ol\nabla_X Z +\bK(X,Z)\zeta = \ol\nabla_X^{\mc S} Z + \bK (X,Z)\zeta_{\para} + \left(\bK(X,Z)\zeta_n-\bQ(X,Z)\right)n$$ for every $X,Z\in\Gamma(T\mc S)$, the result follows.
	\end{proof}
\end{prop}

Now we can find the relation between $\nabla^{h}$ and the induced connection $\ol\nabla^{\mc S}$ by making use of this gauge transformation.

\begin{prop}
	\label{nablaSandnablah}
	Let $h=\phi^{\star}_{\ul u}\bg$ be the induced metric on $\mc S_{\ul u}$, $\nabla^{h}$ its Levi--Civita connection and $\ol\nabla^{\mc S}$ the induced one. Then, $$\ol\nabla^{\mc S} = \nabla^{h}+\ell^{\sharp}\otimes {\bm\chi},$$ where $\ell^{\sharp}\d h^{\sharp}(\bm\ell_{\para},\cdot)$ and $h^{\sharp}$ is the inverse metric of $h$.
	\begin{proof}
Since $\ol\nabla^{\mc S}$ is torsion-free, by Proposition \ref{nablah} the connection $\ol\nabla^{\mc S}$ coincides with $\nabla^{h}$ when $\bm\ell_{\para}=0$. Given CHD $\mc D=\{\mc H,\bg,\bm\ell,\ell^{(2)},\bY\}$, the transformed data $\mc D'=\mc G_{(1,-\ell^{\sharp})}\mc D$ satisfies $\bm\ell_{\para}'=0$, so $$\ol\nabla^{\mc S} = \mc G_{(1,\ell^{\sharp})}\mc G_{(1,-\ell^{\sharp})}(\ol\nabla^{\mc S}) = \mc G_{(1,\ell^{\sharp})}(\nabla^{h}) = \nabla^{h}+ \ell^{\sharp}\otimes {\bm\chi},$$ where we have used that $(1,-\ell^{\sharp})=(1,\ell^{\sharp})^{-1}$ (see \eqref{gaugelaw}) and Proposition \ref{gaugeconection}.
	\end{proof}
\end{prop}

In the following sections we will need the transformation law of the curvature tensor $\ol R$ of the connection $\ol\nabla$ under a gauge transformation. This can be computed from the transformation law of Proposition \ref{gaugeconection}. 

\begin{prop}
	\label{curvatura}
	Let $\mc D=\{\mc H,\bg,\bm\ell,\ell^{(2)},\bY\}$ be hypersurface data and $(z,\zeta)\in\mc F^{\star}(\mc H)\times\Gamma(T\mc H)$. Let $\ol R$ be the curvature tensor of $\ol\nabla$. Then,
	$$\mc G_{(z,\zeta)}\left(\ol R^{f}{}_{bcd}\right) = \ol R^f{}_{bcd}+2\ol\nabla_{[c}\left(\zeta^f \bK_{d]b}\right)+2\zeta^f\zeta^g \bK_{g[c}\bK_{d]b}.$$ 
	\begin{proof}
Let ${}^{(1)}\nabla$ and ${}^{(2)}\nabla$ be two connections and let $S\d {}^{(2)}\nabla-{}^{(1)}\nabla$. Then the curvatures of both connections are related by (see e.g. \cite{Wald}) $${}^{(2)}R^f{}_{bcd} = {}^{(1)}R^f{}_{bcd} + 2{}^{(1)}\nabla_{[c}S^f{}_{d]b}+2S^f{}_{d[c}S^d{}_{d]b}.$$ Particularizing to ${}^{(2)}\nabla=\mc G_{(z,\zeta)}\left(\ol\nabla\right)$, ${}^{(1)}\nabla  = \ol\nabla$ and $S=\zeta\otimes \bK$ (see Proposition \ref{gaugeconection}), the result follows at once.	
	\end{proof}
\end{prop}
From equation \eqref{olnablan} with $n^{(2)}=0$ it follows
\begin{equation}
	\label{normalndeco}
	\ol\nabla_X n = \chi^{\sharp}(X) - \left(\bPi(X,n)+{\bm\chi}(X,\ell^{\sharp})\right)n, \quad X\in\Gamma(T\mc S),
\end{equation}
 where $\chi^{\sharp}$ is the endomorphism defined by $h\big(\chi^{\sharp}(X),\cdot) = {\bm\chi}(X,\cdot)$ and we introduce the tensor $\bPi\d \bY+\bF$ because it will arise frequently below. An interesting property of $\bPi$ is 
\begin{equation}
	\label{Pipi}
	\bPi(n,X) -\bPi(X,n) = \left(\lie_n\bm\ell\right)(X)=\left(\lie_n\bm\ell_{\para}\right)(X)+\ol\nabla_X^{\mc S}\log\left|\lambda\right|.
\end{equation}

The first equality follows from the Cartan formula $\lie_n \bm\ell = \dd(\bm\ell(n)) + \dd\bm\ell (n,\cdot)=\dd\bm\ell (n,\cdot)$,
\begin{equation}
\label{PiLienell}
\bPi(n,X) -\bPi(X,n) = 2 \bF(n,X) = \dd\bm\ell(n,X) =\left(\lie_n\bm\ell\right)(X),
\end{equation} 
and the second follows from $$\dd\bm\ell(n,X) = \lambda^{-1}\dd\lambda(X) + \dd\bm\ell_{\para}(n,X) = X\left(\log\left|\lambda\right|\right) + \left(\lie_n\bm\ell_{\para}\right)(X),$$ where we have used the differential of $\bm\ell = \lambda^{-1}\dd\ul u + \bm\ell_{\para}$ and that $\left(\dd\lambda\wedge\dd\ul u\right)(n,X)= -\lambda\dd\lambda(X),$ as well as the Cartan formula applied to $\bm{\ell}_{\para}$.\\

Next we compute the gauge transformation of $\bPi(\cdot,n)$. From \eqref{olnablaell} and the property $\bK(\cdot,n)=0$ we have the equality $\bPi(\cdot,n)=\left(\ol\nabla\bm\ell\right)(n)$. Using the transformation law of $\ol\nabla$ in \eqref{gaugeconnection},
\begin{align*}
	\bPi'(\cdot,n')  = \left(\ol\nabla'\bm\ell'\right)(n')= \left(\ol\nabla\bm\ell'\right)(n')=\left(\ol\nabla\bm\ell\right)(n) + z^{-1} dz + \left(\ol\nabla\bg(\zeta,\cdot)\right)(n).
\end{align*}
The last term is $\left(\ol\nabla\bg(\zeta,\cdot)\right)(n) = -\bK(\cdot,\zeta)$ because of \eqref{olnablagamma} and recalling $\bg(n,\cdot)=0$ and $\bm\ell(n)=1$. So finally 
\begin{equation}
	\label{transPiXn}
	\bPi'(\cdot,n')  = \bPi(\cdot,n) - \bK(\cdot,\zeta) + d\log|z|.
\end{equation}

We now can relate the curvature tensor of $\ol\nabla^{\mc S}$ with that of $\ol\nabla$.

\begin{prop}[Gauss identity]
	\label{gauss}
	Let $X,Z,V,W\in\Gamma(T\mc S_{\ul u})$. Then,
	\begin{equation}
		\label{gausseq}
		\bg( V, \ol R(X,W)Z) = h(V,R^{\mc S}(X,W)Z)-\bQ(W,Z){\bm\chi}(X,V)+\bQ(X,Z){\bm\chi}(W,V),
	\end{equation} 
where $R^{\mc S}$ is the curvature of $\ol\nabla^{\mc S}$ and $\bQ$ is given by \eqref{tensorQ}.
	\begin{proof}
By definition of the curvature tensor and decompositions \eqref{decompnabla} and \eqref{normalndeco},
\begin{align*}
	\ol R(X,W)Z & = \ol\nabla_X\ol\nabla_W Z -\ol\nabla_W\ol\nabla_X Z-\ol\nabla_{[X,W]} Z\\
	&= \ol\nabla_X\big(\ol\nabla^{\mc S}_W Z-\bQ(W,Z)n\big)-\ol\nabla_W\big(\ol\nabla^{\mc S}_X Z-\bQ(X,Z)n\big) - \ol\nabla^{\mc S}_{[X,W]}Z+\bQ\left([X,W],Z\right)n\\
	&=\ol\nabla_X^{\mc S}\ol\nabla_W^{\mc S} Z - \bQ\big(X,\ol\nabla_W^{\mc S} Z\big)n-\ol\nabla^{\mc S}_X\left(\bQ(W,Z)\right) n - \bQ(W,Z)\big(\ol\nabla^{\bot}_X n+\chi^{\sharp}(X)\big) \\
	&\quad - \left( X\leftrightarrow W\right) - \ol\nabla^{\mc S}_{[X,W]}Z+\bQ\left([X,W],Z\right)n,
\end{align*}
which, after identifying $R^{\mc S}(X,W)Z =\ol\nabla_X^{\mc S}\ol\nabla_W^{\mc S} Z -\ol\nabla_W^{\mc S}\ol\nabla_X^{\mc S} Z- \ol\nabla^{\mc S}_{[X,W]}Z$, and using that $\ol\nabla^{\mc S}$ is torsion-free, $[X,W]=\ol\nabla^{\mc S}_X W - \ol\nabla^{\mc S}_W X$, simplifies to
\begin{align*}
\ol R(X,W)Z & = R^{\mc S}(X,W)Z -\chi^{\sharp}(X)\bQ(W,Z) +\chi^{\sharp}(W)\bQ(X,Z) \\
&\quad +\big(\big(\ol\nabla^{\mc S}_W \bQ\big)(X,Z)-\big(\ol\nabla^{\mc S}_X \bQ\big)(W,Z)\big)n +\bQ(X,Z)\ol\nabla^{\bot}_W n-\bQ(W,Z)\ol\nabla^{\bot}_X n.
\end{align*}
The $\bg$-product of this expression with $V$ gives \eqref{gausseq} after taking into account $\bg(n,\cdot)=0$ and the definition of $\chi^{\sharp}$.
	\end{proof}
\end{prop}

To conclude this section we define a connection induced from $\ol\nabla$. Let $X$ be a vector field tangent to the foliation. Then one can decompose $\ol\nabla_n X$ as 
\begin{equation}
	\label{nablan}
	\ol\nabla_n X = \wt\nabla_n X +{\bm{\eta}}(X)n,
\end{equation} 
where $\wt\nabla_n X$ is tangent to the foliation and, as we show next, ${\bm{\eta}}\in\Gamma(T^{\star}\mc S_{\ul u})$ given by
\begin{equation}
	\label{Psin}
	\begin{aligned}
		{\bm{\eta}}(X)&=-\bPi(X,n)-{\bm\chi}(X,\ell^{\sharp})-X\left(\log\left|\lambda\right|\right) \\
		&= \left(\lie_n\bm\ell_{\para}\right)(X)-\bPi(n,X)-{\bm\chi}(X,\ell^{\sharp}).
	\end{aligned}
\end{equation}
Indeed, $\lambda{\bm{\eta}}(X)  = \dd\ul u\left(\ol\nabla_n X\right)=\ol\nabla_n\left(\dd\ul u(X)\right) - \left(\ol\nabla_n\dd\ul u\right)(X)$, and since $X(\ul u)=0$ and the Hessian of a function is symmetric, it follows $\lambda{\bm{\eta}}(X) = -\left(\ol\nabla_X \dd\ul u\right)(n) = \dd\ul u\left(\ol\nabla_X n\right)-\ol\nabla_X\left(\dd\ul u(n)\right)$. Then, making use of \eqref{normalndeco} and $\dd\ul u(n)=\lambda\neq 0$, the first equality in \eqref{Psin} follows. The second is a direct consequence of \eqref{Pipi}. The connection $\wt\nabla_n$ extends to a covariant derivative acting on general tensors tangent to the foliation. For example, if $\bm{\alpha}\in\Gamma(T^{\star}\mc S)$, we can define
\begin{equation}
	\label{nablatildeeta}
	\big(\wt{\nabla}_n\bm{\alpha}\big)(X) \d n\left(\bm{\alpha}(X)\right)-\bm{\alpha}\big(\wt\nabla_n X\big),
\end{equation}
and analogously for any other tensor field tangent to the foliation. A key property of the connection $\wt\nabla_n$ is that it is independent of the tensor $\bY$. Indeed, taking into account equations \eqref{normalndeco}, \eqref{nablan} and \eqref{Pipi} and recalling that $\ol\nabla$ is torsion-free, $$[X,n] = \ol\nabla_X n-\ol\nabla_n X = \chi^{\sharp}(X)-\wt\nabla_n X +X(\log|\lambda|)n,$$ from where it yields 
\begin{equation}
	\label{indpendentY}
	\wt\nabla_n X = \lie_n X +\chi^{\sharp}(X)+X(\log|\lambda|)n,
\end{equation} 
which does not involve the tensor $\bY$.

\section{Covariance of the constraint tensors}
\label{section_gauge}
Let $\mc D$ be embedded null hypersurface data on an ambient manifold $(\mc M,g)$ with rigging $\xi$. It is possible to write all the tangent components of the Ricci tensor of $g$ on the hypersurface $\mc H$ in terms of the data. In this section we obtain those expressions and promote them to abstract tensors without mention of any ambient space. By construction these tensors will coincide with the tangent components of the Ricci when the data is embedded.\\

Assume that the data is embedded on an ambient manifold whose Riemann tensor is $R$. By Corollary 4 of \cite{Marc1},
\begin{align}
	R_{\alpha\beta\mu\nu}\xi^{\alpha}e^{\beta}_be^{\mu}_ce^{\nu}_d&\st{\mc H}{=} \bm\ell_a\ol{R}^a{}_{bcd}+2\ell^{(2)}\ol{\nabla}_{[d} \bK_{c]b} +\bK_{b[c}\ol\nabla_{d]} \ell^{(2)}, \label{proj1}\\
	R_{\alpha\beta\mu\nu} e^{\alpha}_ae^{\beta}_be^{\mu}_ce^{\nu}_d&\st{\mc H}{=} \bg_{af}\ol{R}^f{}_{bcd}+2\ol\nabla_{[d}\left(\bK_{c]b}\bm\ell_a\right)+2\ell^{(2)}\bK_{b[c}\bK_{d]a},\label{proj2}
\end{align}
 where $\{e_a\}$ is a (local) frame on $\mc H$ and the Greek letters denote ambient indices. We promote the right-hand side into new tensors on $\mc H$ and define $A$ and $B$ on any hypersurface data by means of
\begin{align}
	A_{bcd}&\d\bm\ell_a\ol{R}^a{}_{bcd}+2\ell^{(2)}\ol{\nabla}_{[d} \bK_{c]b} +\bK_{b[c}\ol\nabla_{d]} \ell^{(2)},\label{A} \\
	B_{abcd}&\d\bg_{af}\ol{R}^f{}_{bcd}+2\ol\nabla_{[d}\left(\bK_{c]b}\bm\ell_a\right)+2\ell^{(2)}\bK_{b[c}\bK_{d]a},\label{B}
\end{align}
so that for embedded data it holds $R_{\alpha\beta\mu\nu}\xi^{\alpha}e^{\beta}_be^{\mu}_ce^{\nu}_d \st{\mc H}{=} A_{bcd}$ and $R_{\alpha\beta\mu\nu} e^{\alpha}_ae^{\beta}_be^{\mu}_ce^{\nu}_d \st{\mc H}{=} B_{abcd}$. In Appendix \ref{appendix} the gauge behaviour and the symmetries of the tensors $A$ and $B$ are studied. From \eqref{Asharp} and $n^{(2)}=0$, the inverse metric $g^{\alpha\beta}$ on $\mc H$ is given by
\begin{equation}
	\label{inverse}
	g^{\alpha\beta}\st{\mc H}{=}P^{cd} e_c^{\alpha} e_d^{\beta} + n^c (e_c^{\alpha}\xi^{\beta}+\xi^{\alpha} e_c^{\beta}),
\end{equation}
and thus the pullback to $\mc H$ of the ambient Ricci tensor can be written as
\begin{align*}
	g^{\alpha\beta} R_{\alpha \mu \beta \nu} e_a^{\mu} e_b^{\nu}&\st{\mc H}{=}\left[P^{cd} e_c^{\alpha} e_d^{\beta} + n^c (e_c^{\alpha}\xi^{\beta}+\xi^{\alpha} e_c^{\beta})\right]  R_{\alpha \mu \beta \nu} e_a^{\mu} e_b^{\nu}\\
	&\st{\mc H}{=} B_{cadb}P^{cd}+A_{bca}n^c + A_{acb}n^c\\
	&\st{\mc H}{=} B_{acbd}P^{cd}- (A_{bac}+A_{abc})n^c,
\end{align*}
where in the last equality we used the symmetries of $B$ in Prop. \ref{symmetries}. This leads naturally to the introduction of the abstract tensor
\begin{equation}
	\label{ricci}
{\bm{\mc R}}_{ab}\d B_{acbd}P^{cd}- (A_{bac}+A_{abc})n^c
\end{equation}
in the context of null hypersurface data. Inserting \eqref{A} and \eqref{B}, 
\begin{align}
{\bm{\mc R}}_{ab}&= \left(\bg_{af}\ol{R}^f{}_{cbd} + 2\ol\nabla_{[d}\left(\bK_{b]c}\bm\ell_a\right)+2\ell^{(2)} \bK_{c[b}\bK_{d]a}\right)P^{cd}\label{ricci2}\\
		&\quad\, - \left(\bm\ell_d\ol{R}^d{}_{bac} + 2\ell^{(2)}\ol\nabla_{[c}\bK_{a]b} + \bK_{b[a}\ol\nabla_{c]}\ell^{(2)} + \bm\ell_d\ol{R}^d{}_{abc} + 2\ell^{(2)}\ol\nabla_{[c}\bK_{b]a} + \bK_{a[b}\ol\nabla_{c]}\ell^{(2)}\right)n^c.\nonumber
\end{align}
By construction, when the data is embedded ${\bm{\mc R}}_{ab}$ coincides with the pull-back of the ambient Ricci tensor into the hypersurface. As expected from the embedded case, ${\bm{\mc R}}_{ab}$ is symmetric. Indeed, using that $P$ is symmetric and item 5. of Prop. \ref{symmetries} one gets
\begin{align*}
{\bm{\mc R}}_{ab} & = \dfrac{1}{2}\left(B_{acbd}+ B_{adbc}\right)P^{cd} -\left(A_{bac}+A_{abc}\right)n^c\\
	&= \dfrac{1}{2}\left(B_{bdac}+ B_{bcad}\right)P^{cd} - \left(A_{bac}+A_{abc}\right)n^c\\
	&={\bm{\mc R}}_{ba}.
\end{align*}
For future convenience we define the following two contractions of ${\bm{\mc R}}_{ab}$
\begin{align}
	H&\d -\dfrac{1}{2}P^{ab}{\bm{\mc R}}_{ab}=-\dfrac{1}{2} B_{cadb} P^{cd} P^{ab}+A_{bac}P^{ab}n^c,\label{hamil2}\\
{\bm J}_b &\d {\bm{\mc R}}_{ab}n^a=B_{acbd} n^a P^{cd}-A_{abc} n^c n^a .\label{momentum2}
\end{align}

For obvious reasons we will refer to Definitions \eqref{ricci}, \eqref{hamil2} and \eqref{momentum2} as \textbf{constraint tensors}. In the following Proposition we introduce a special class of gauges that will be used frequently below.

\begin{prop}
	\label{propcaracteristico}
	Let $\mc D$ be CHD and $\ul u$ a FF. Then there exists $(z,\zeta)\in\mc F^{\star}(\mc H)\times \Gamma(T\mc H)$ such that $\mc D'=\mc G_{(z,\zeta)}\mc D$ satisfies the following properties
	\begin{enumerate}
		\item $\bm\ell'(X)=0$ for all $X\in\Gamma(T\mc S)$,
		\item $\ell'{}^{(2)}=0$.
	\end{enumerate}
	The gauge transformations respecting 1. and 2. are $\mc{G}_{(z,0)}$ for arbitrary $z\in\mc{F}^{\star}(\mc H)$. Moreover, there exists a unique pair $(z,\zeta)\in\mc F^{\star}(\mc H)\times \Gamma(T\mc H)$ such that, in addition, 
	\begin{enumerate}
		\item[3.] $\lambda'\d n'(\ul u)=1$.
	\end{enumerate}
	\begin{proof}
		Directly from the transformations in Def. \ref{defi_gauge} and the decomposition $\zeta = \zeta_{\para} + \zeta_n n$,
		\begin{align}	
			\bm\ell(X)+h(\zeta_{\para},X)&=0,\label{xinormal2}\\
			\ell^{(2)}+2\bm\ell(\zeta_{\para})+2\zeta_n+h(\zeta_{\para},\zeta_{\para})&=0.\label{xinull2}
		\end{align}
		Equation \eqref{xinormal2} admits a unique solution for $\zeta_{\para}$, which substituted in the equation \eqref{xinull2} fixes completely $\zeta_n$. Therefore, there always exist a gauge satisfying the conditions (1) and (2). Any transformation of the form $(z,0)$ keeps these two conditions invariant and by the uniqueness of $\zeta$ it is clear that no other transformation does. To fulfill (3) simply apply an additional gauge transformation with $(z=\lambda,\zeta=0)$, and use \eqref{transn}. Uniqueness of the gauge satisfying (1), (2), (3) is immediate from the argument.
	\end{proof}
\end{prop}

\begin{defi}
	\label{def_ch_gauge}
	A gauge in which (1) and (2) hold is called \textbf{characteristic gauge} (CG). The gauge satisfying (1), (2), (3) is called \textbf{adapted characteristic gauge} (ACG). We emphasize that the ACG is \textit{unique} once the FF $\ul u$ is chosen. A change in $\ul u$ affects the corresponding ACG gauge.
\end{defi}

\begin{rmk}
	\label{rmkCG}
	When the data is embedded on an ambient manifold $(\mc M,g)$ with rigging $\xi$, the abstract conditions (1) and (2) are equivalent to $\xi$ being orthogonal to the leaves and null, respectively.
\end{rmk}

The proof of Proposition \ref{propcaracteristico} has the following immediate Corollary.
\begin{cor}
	\label{CGcorollary}
	Let $\mc D$ be CHD and $\mc S\subset\mc H$ any section. Let $f\in\mc F(\mc S)$ and $\bm\alpha\in \Gamma(T^{\star}\mc S)$. Then there always exist a gauge in which $$\bm\ell_{\para}|_{\mc S}=\bm\alpha, \hspace{1cm} \ell^{(2)}|_{\mc S}=f.$$ Moreover, the freedom of this gauge is parametrized by the pair $(z,\zeta)$ satisfying $\zeta|_{\mc S}=0$.
\end{cor}

\begin{obs}
	\label{observación}
	By Proposition \ref{nablah}, in a CG the induced connection on the foliation, $\ol\nabla^{\mc S}$, coincides with the Levi--Civita connection, $\nabla^h$. Moreover $\bm\ell_{\para}=0$ and hence $\dd \ul u=\lambda\bm\ell$ by \eqref{elldu}. Then, $\dd\lambda\wedge\bm\ell+\lambda\dd\bm\ell=0$, so $\bF=-\frac{1}{2\lambda}\dd\lambda\wedge\bm\ell$ and hence $\bPi_{AB}=\bY_{AB}$, where here capital Latin letters denote abstract indices on the foliation. Moreover, from equation \eqref{tensorQ}, $\bQ_{AB}=\bY_{AB}$ in a CG and therefore the Gauss identity \eqref{gausseq} takes the form
	\begin{equation}
		\label{gaussfacil}
		\bg( V, \ol R(X,W)Z) = h(V,R^{h}(X,W)Z)-\bY(W,Z){\bm\chi}(X,V)+\bY(X,Z){\bm\chi}(W,V).
	\end{equation}
\end{obs} 

Next we prove that the constraint tensors are gauge-covariant. This is the expected behaviour if one thinks the data as embedded. However, we prove this statement in full generality without assuming the existence of any ambient spacetime.

\begin{teo}
	\label{teo_gauge}
	Let $\mc D=\{\mc H,\bg,\bm\ell,\ell^{(2)},\bY\}$ be CHD and $(z,\zeta)\in\mc F^{\star}(\mc H)\times\Gamma(T\mc H)$. Then given a gauge transformation $\mc D' = \mc G_{(z,\zeta)}\mc D$, the tensors $H$, ${\bm J}$ and ${\bm{\mc R}}_{ab}$ transform as:
	\begin{enumerate}
		\item $H' = H + \zeta^a{\bm J}_a$,
		\item ${\bm J}'_a = z^{-1} {\bm J}_a$,
		\item ${\bm{\mc R}}_{ab}'={\bm{\mc R}}_{ab}$.
	\end{enumerate}
	\begin{proof}
		Suppose that (3) is true. Then, \eqref{transn} and \eqref{gaugeP} imply 
		\begin{align*}
{\bm J}'_a &= {\bm{\mc R}}_{ab}' n'{}^b = z^{-1} {\bm{\mc R}}_{ab} n^b = z^{-1} {\bm J}_a,\\
			H' &= -\dfrac{1}{2}P'{}^{ab} {\bm{\mc R}}'_{ab} = -\dfrac{1}{2} P^{ab} {\bm{\mc R}}_{ab} + {\bm{\mc R}}_{ab}\zeta^an^b = H + {\bm J}_a\zeta^a,
		\end{align*}
		where in the equation for $H'$ we use the symmetry of ${\bm{\mc R}}_{ab}$. So it suffices to show (3). Using again \eqref{transn}-\eqref{gaugeP} as well as the transformation laws of $A$ and $B$ of Proposition \ref{symmetries},
		\begin{align*}
{\bm{\mc R}}'_{ab} & = B_{cadb}' P'{}^{cd}-n'{}^c (A_{bac}'+A_{abc}')\\
			&={\bm{\mc R}}_{ab}-\zeta^c n^d B_{cadb}-\zeta^d n^c B_{cadb}-\zeta^d n^c B_{dbac}-n^c\zeta^d B_{dabc}\\
			&={\bm{\mc R}}_{ab},
		\end{align*}
		where the last equality follows from the symmetries of Proposition \ref{symmetries}.
	\end{proof}
\end{teo}

Once the constraint tensors have been defined, the next step is to write them in a frame adapted to the foliation. Let $\{e_A\}$ be a (local) basis of $T\mc S_{\ul u}$ with dual basis $\{\theta^A\}$, i.e., $\theta^A(e_B)=\delta^A_B$, where recall that capital Latin letters denote abstract indices on the foliation. Then $\{n,e_A\}$ can be considered as a (local) basis of $T\mc S_{\ul u}\oplus \langle n\rangle$ with dual basis $\{\bm q\d \lambda^{-1}\dd\ul u,\theta^A\}$, where as before $\lambda=n(\ul u)$ and $\bm{q}(n)=1$, $\bm{q}(e_A)=0$, $\theta^A(n)=0$. This decomposition motivates the following definitions 
\begin{equation}
	\label{constrainttensors}
	J(n)\d {\bm J}_a n^a,\hspace{1cm} {\bm J}_A\d {\bm J}_a e^a_A\hspace{1cm} \text{and} \hspace{1cm} {\bm{\mc R}}_{AB}\d {\bm{\mc R}}_{ab}e^a_A e^b_B.
\end{equation} 
These objects are not all independent as shown next.

\begin{prop}
	\label{constraint}
	Let $\mc D$ be CHD and let $H$, $J(n)$, ${\bm J}_A$ and ${\bm{\mc R}}_{AB}$ be the constraint tensors as before. Let $\ell^{\sharp}\d h^{\sharp}(\bm\ell_{\para},\cdot)$, $\ell^A \d (\ell^{\sharp})^A$ and $\ell_{\sharp}^{(2)}\d h_{AB}\ell^A\ell^B$. Then, the following identity holds 
	\begin{equation}
		\label{RJH}
		h^{AB}{\bm{\mc R}}_{AB}-2{\bm J}(\ell^{\sharp})+\big(\ell_{\sharp}^{(2)}-\ell^{(2)}\big)J(n)+2H=0.
	\end{equation}
	\begin{proof}
In the frame $\{n,e_A\}$ we can decompose
	\begin{align}
		\bm{\ell}&=\bm q+\bm\ell_A\theta^A,\\
		P&=P^{AB}e_A\otimes e_B + 2P^{n\, A} n\otimes_s e_A + P^{n\, n}n\otimes n.
	\end{align}
Since in this basis $n^A=0$ and $\bg_{n \, b}=0$ it follows from \eqref{Pgamma} that $P^{AB}\bg_{BC}=\delta^A_C$, and hence we can identify the components $P^{AB}$ with the inverse metric $h^{\sharp}$, that is, $P^{AB}=h^{AB}$. To compute $P^{n\,A}$ and $P^{n\, n}$ we first note that \eqref{Pell}, namely $P(\cdot,\bm\ell)=-\ell^{(2)}n$, gives $P(\theta^A,\bm\ell)=0$ and $P(\bm\ell,\bm\ell)=-\ell^{(2)}$ (by \eqref{ell(n)}). Hence,
\begin{align*}
P^{n\,A}&=P(\bm q,\theta^A) = P(\theta^A,\bm\ell-\bm\ell_B\theta^B) = -\bm\ell_BP^{AB}=-h^{AB}\bm\ell_B=- \ell^A,\\
P^{n\,n}&=P(\bm q,\bm q)=P(\bm\ell,\bm\ell)-2P(\bm\ell,\bm\ell_A\theta^A)+P(\bm\ell_A\theta^A,\bm\ell_B\theta^B) = \ell_{\sharp}^{(2)}-\ell^{(2)}.
\end{align*}  
Inserting this decomposition into the definition of $H$ in \eqref{hamil2},
\begin{align*}
-2H &=	P^{AB}{\bm{\mc R}}_{AB} + 2P^{n\,A} {\bm{\mc R}}_{A\,n}+P^{n\,n}{\bm{\mc R}}_{n\,n}\\
	&=h^{AB}{\bm{\mc R}}_{AB}-2{\bm J}(\ell^{\sharp})+(\ell_{\sharp}^{(2)}-\ell^{(2)})J(n),
\end{align*}
so \eqref{RJH} is established.
	\end{proof}
\end{prop}
A by-product of the proof is that in a general gauge the tensor $P$ decomposes as 
\begin{equation}
	\label{Pdecomposition}
P = h^{AB} e_A\otimes e_B -2\ell^A n\otimes_s e_A -\big(\ell^{(2)}-\ell_{\sharp}^{(2)}\big) n\otimes n.
\end{equation}
In a CG, where $\ell^{(2)}=0$ and $\ell^{\sharp}=0$, this decomposition simplifies to
\begin{equation}
	\label{PAB}
	P=h^{AB} e_A\otimes e_B,
\end{equation}
so the previous identity reduces to
\begin{equation}
	\label{RJHCG}
	h^{AB}{\bm{\mc R}}_{AB}+2H=0.
\end{equation}

We conclude the section by computing two contractions of the curvature $\ol R$ with $\bm\ell$ and $n$ needed later.

\begin{lema}
	\label{lema}
	Let $\mc D=\{\mc H,\bg,\bm\ell,\ell^{(2)},\bY\}$ be CHD. Then, 
	\begin{align}
		\bm\ell_a \ol R^a{}_{bcd}&=2\ol\nabla_{[d} \bPi_{c]b}+2\bK_{b[d}\ol\nabla_{c]} \ell^{(2)}+2\ell^{(2)}\ol\nabla_{[c}\bK_{d]b}\label{lemalR}\\
			n^c \bm\ell_a \ol R^a{}_{bcd}  &= \ol\nabla_d\left(\bPi_{cb}n^c\right)-\ol\nabla_n \bPi_{db}-P^{ca}\bK_{ad}\bPi_{cb}+\bPi_{cb}\bPi_{da} n^c n^a+\bK_{db}\ol\nabla_n \ell^{(2)}\label{nlRgeneral}\\
			&\quad +\ell^{(2)}\left(\ol\nabla_n \bK_{db}+P^{ca}\bK_{cb} \bK_{ad}\right).\nonumber
	\end{align}
	\begin{proof}
		The first equation follows from the Ricci identity and the fact that $\ol\nabla_a \bm\ell_b = \bPi_{ab}-\ell^{(2)} \bK_{ab}$ (see \eqref{olnablaell}),
		\begin{align*}
			\bm\ell_a \ol R^a{}_{bcd} & = \ol\nabla_d\ol\nabla_c \bm\ell_b-\ol\nabla_c\ol\nabla_d \bm\ell_b = 2\ol\nabla_{[d} \bPi_{c]b}+2\bK_{b[d}\ol\nabla_{c]} \ell^{(2)}+2\ell^{(2)}\ol\nabla_{[c}\bK_{d]b}.
		\end{align*}
		For the second, we ``integrate'' by parts equation \eqref{lemalR} and use $\bK(n,\cdot)=0$ to find $$n^c \bm\ell_a \ol R^a{}_{bcd}  = \ol\nabla_d\left(\bPi_{cb}n^c\right) - \bPi_{cb}\ol\nabla_d n^c - \ol\nabla_n \bPi_{db}+\bK_{db}\ol\nabla_n \ell^{(2)}+\ell^{(2)}\left(\ol\nabla_n \bK_{db} + \bK_{cb}\ol\nabla_d n^c\right).$$ Inserting \eqref{olnablan} particularized to CHD, namely 
		\begin{equation}
			\label{olnablan2}
			\ol\nabla_d n^c =P^{ca}\bK_{ad}-\bPi_{da}n^c n^a,
		\end{equation}
equation \eqref{nlRgeneral} follows.
	\end{proof}
\end{lema}
Notice that in a CG equation \eqref{nlRgeneral} simplifies to
\begin{equation}
	\label{nlR}
	n^c \bm\ell_a \ol R^a{}_{bcd}=\ol\nabla_d\left(\bPi_{cb}n^c\right)-\ol\nabla_n \bPi_{db}-P^{ca}\bK_{ad}\bPi_{cb}+\bPi_{cb}\bPi_{da} n^c n^a
\end{equation}
and the identity \eqref{lemalR} becomes $\bm\ell_a \ol R^a{}_{bcd}=2\ol\nabla_{[d} \bPi_{c]b}$. Moreover in the ACG, $\bm\ell_a \ol R^a{}_{bcd}=2\ol\nabla_{[d} \bY_{c]b}$ because $\bm\ell=\dd\ul u$ (cf. Remark \ref{observación} combined with $\lambda=1$) and hence $\bF=0$.

\section{Constraint tensors on the foliation}
\label{sec_fol}
In this section we compute the constraint tensors $H$, $J(n)$, ${\bm J}_A$ and ${\bm{\mc R}}_{AB}$ in terms of the intrinsic geometry of the foliation. The first step is to define a set of tensors tangent to the foliation which capture all the information of the hypersurface data. Two such tensors (${\bm\chi}$ and ${\bm{\eta}}$) have already been introduced before. We recall their definition and introduce two additional ones.

\begin{defi}
	\label{def_tensors}
Let $\mc D$ be CHD and $\ul u$ a FF. We define the ``foliation tensors'' ${\bm\chi}$, ${\bm{\Upsilon}}$, ${\bm{\eta}}$ and $\omega$ on each leaf $\mc S_{\ul u}$ by 
\begin{equation*}
{\bm\chi} \d \phi^{\star}_{\ul u} \bK, \qquad {\bm{\Upsilon}} \d \phi^{\star}_{\ul u} \bY, \qquad {\bm{\eta}} \d\lie_n\bm\ell_{\para}- \phi_{\ul u}^{\star}\left(\bPi(n,\cdot)\right)-{\bm\chi}(\ell^{\sharp},\cdot),\qquad \omega \d \dfrac{1}{2}\bY(n,n).
\end{equation*}
\end{defi}

The motivation behind these definitions is the following. Suppose that a CHD $\mc D$ is embedded on an ambient manifold $(\mc M,g)$ with rigging $\xi$. Recall that $\bm\nu$ is the unique one-form normal to $f(\mc H)$ satisfying $\bm\nu(\xi)=1$. The vector $\nu\d g^{\sharp}(\bm\nu,\cdot)$ is the push-forward of $n$. From the definition of the tensor $\bK$ (see \eqref{defK}), it is clear that ${\bm\chi}$ is the null second fundamental form of the leaf $\mc S_{\ul u}$ w.r.t. $\nu$. For the interpretation of the remaining terms we restrict ourselves to the case when $\mc D$ is in a CG. Then the rigging vector is null and orthogonal to the leaves (see Remark \ref{rmkCG}), and hence from equation \eqref{Yembedded} we see that ${\bm{\Upsilon}}$ coincides with the null second fundamental form of $\mc S_{\ul u}$ w.r.t. $\xi$. Now let $X$ be a vector field tangent to the foliation. Taking into account decomposition \eqref{nablan},
\begin{equation}
	\label{etaambient}
	{\bm{\eta}}(X) = \bm\ell\left(\ol\nabla_n X\right) = g\left(\xi,\ol\nabla_n X\right) = g\left(\xi,\nabla_{\nu} X\right),
\end{equation} 
where in the last equality we used \eqref{nablambient} and the fact that $\bK(n,\cdot)=0$. Thus, ${\bm{\eta}}$ coincides with the torsion one-form of $\mc S_{\ul u}$ with respect to the normal null basis $\{\nu,\xi\}$ (cf. \cite{It,Book,Luk}). Finally, from equation \eqref{olnablan2} it follows
\begin{equation}
	\label{nablann=Y}
	\ol\nabla_n n = -\bY(n,n)n=-2\omega n.
\end{equation} 
So
\begin{equation}
	\label{omegambient}
	\omega =-\dfrac{1}{2} g(\ol\nabla_n n, \xi) = -\dfrac{1}{2} g(\nabla_{\nu} {\nu}, \xi)
\end{equation}  
measures the deviation of ${\nu}$ from being affinely parametrized. Thus, the tensors ${\bm\chi}$, ${\bm{\Upsilon}}$, ${\bm{\eta}}$ and $\omega$ are a generalization of the coefficient components in the double null foliation (cf. \cite{It,Chris_and_Klain,Book,Luk}) and coincide with them when the data is embedded and written in a characteristic gauge. In terms of the hypersurface data, the tensor ${\bm{\Upsilon}}$ captures the information of the tangent components of $\bY$, the tensor ${\bm{\eta}}$ encodes the normal-tangent part, whereas $\omega$ carries the information of the normal-normal component, and hence it does not depend on the specific foliation. Once the motivation of the Definition \ref{def_tensors} is clear, we compute the gauge transformation laws of the foliation tensors.
\begin{lema}
	\label{transformationtensors}
	Let $\mc D$ be CHD with FF $\ul u$, $(z,\zeta)\in\mc F^{\star}(\mc H)\times\Gamma(\mc H)$ and $\mc D' = \mc{G}_{(z,\zeta)}\mc D$. Then,
	\begin{align}
		{\bm\chi}' &= z^{-1}{\bm\chi},\label{trans_chi}\\ 
		{\bm{\Upsilon}}' &= z{\bm{\Upsilon}} + \bm\ell_{\para}\otimes_{s} \phi_{\ul u}^{\star}(\dd z)+\dfrac{1}{2}\lie_{z\zeta_{\para}}h+\zeta_nz {\bm\chi},\label{transulchi}\\
		{\bm{\eta}}' &={\bm{\eta}},\label{transeta}\\
		\omega' &= z^{-1}\omega - \dfrac{1}{2}n\left(z^{-1}\right).\label{trans_omega}
	\end{align}
Moreover, the connection $\wt\nabla$ is gauge invariant.
\begin{proof}
The first is a consequence of \eqref{Ktrans}. For the second, taking the pullback of \eqref{transY}, $${\bm{\Upsilon}}' = z{\bm{\Upsilon}} + \bm\ell_{\para}\otimes_{s} \phi_{\ul u}^{\star}(\dd z)+\dfrac{1}{2}\phi_{\ul u}^{\star}\left(\lie_{z\zeta}\bg\right).$$ Writing $\zeta=\zeta_{\para}+\zeta_n n$ and recalling that  $\lie_{f n}\bg = f \lie_n\bg$ for any scalar $f$, the transformation law \eqref{transulchi} follows because $$\dfrac{1}{2}\phi_{\ul u}^{\star}\left(\lie_{z\zeta}\bg\right) = \dfrac{1}{2}\phi_{\ul u}^{\star}\big(\lie_{z\zeta_{\para}}\bg\big) + \dfrac{1}{2}\zeta_n z \phi_{\ul u}^{\star}\left(\lie_{n}\bg\right)=\dfrac{1}{2}\lie_{z\zeta_{\para}}h +\zeta_n z {\bm\chi},$$ where we have used the following standard property of the Lie derivative $\Phi^{\star}\left(\lie_{\Phi_{\star} V} T\right) = \lie_V \left(\Phi^{\star} T \right)$ valid for any injective map $\Phi$, vector $V$ and covariant tensor $T$.  Next we prove the invariance of ${\bm{\eta}}$ under gauge transformations. Let $X$ be a vector tangent to the foliation. Firstly, from \eqref{transn}, Proposition \ref{gaugeconection} and the fact that $\bK(n,\cdot)=0$, $$\ol\nabla'_{n'} X = \ol\nabla_{n'} X = z^{-1}\ol\nabla_n X.$$ Secondly, equation \eqref{nablan} in the primed gauge reads
\begin{align*}
	\ol\nabla'_{n'} X  = \wt\nabla'_{n'} X + {\bm{\eta}}'(X) n' = z^{-1} \wt\nabla_{n} X + z^{-1} {\bm{\eta}}(X) n.
\end{align*}
Hence, we conclude that both $\wt\nabla$ and ${\bm{\eta}}$ are gauge invariant.  Finally, the transformation of $\omega$ is immediate from \eqref{transY} and \eqref{transn}.
\end{proof}
\end{lema}

When the transformation takes place between characteristic gauges, the transformation laws of ${\bm\chi}$, ${\bm{\Upsilon}}$, ${\bm{\eta}}$ and $\omega$ become particularly simple.

\begin{cor}
\label{transformationtensors_CG}
Let $\mc D$ be CHD with FF $\ul u$ written in a CG, $z\in\mc F^{\star}(\mc H)$ and $\mc D' = \mc{G}_{(z,0)}\mc D$. Then,
\begin{equation*}
{\bm\chi}' = z^{-1}{\bm\chi}, \qquad {\bm{\Upsilon}}' = z{\bm{\Upsilon}}, \qquad
{\bm{\eta}}' ={\bm{\eta}}, \qquad \omega' = z^{-1}\omega - \dfrac{1}{2}n\left(z^{-1}\right).
\end{equation*}
\end{cor}

Our next aim is to write the constraint tensors $H$, $J(n)$, $\bm J_A$ and ${\bm{\mc R}}_{AB}$ in terms of the quantities we just introduced. Since the constraints are covariant we will compute them in a CG. In order to obtain them in any gauge it suffices to employ Theorem \ref{teo_gauge} and the transformation laws in Lemma \ref{transformationtensors}. Let us begin by writing the tensors $J(n)$ and ${\bm J}_A$. The computation requires several contractions of the tensors $A$ and $B$. The computation is somewhat long and has been postponed to Appendix \ref{appendixB} in order not to interrupt the presentation. Using identities \eqref{propJ1} and \eqref{propJ2} of Lemma \ref{LemaB}, expression \eqref{momentum2} becomes
\begin{equation}
	\label{J(V)}
	\begin{aligned}
{\bm J}(V)&=(\ol\nabla_V \bPi)(n,n)-(\ol\nabla_n \bPi)(V,n)-\bPi(V,n) \tr_P \bK\\
		&\quad  +\left(\bK*\bPi\right)(V,n)- \ol\nabla_V\tr_P \bK+\div_P(\bK)(V),
	\end{aligned}
\end{equation}
for an arbitrary vector field $V$, where $\tr_P \bK\d P^{ab}\bK_{ab}$, $\left(\bK*\bPi\right)_{ca} \d P^{bd} \bK_{bc}\bPi_{da}$ and $\div_P(\bK)(V)\d P^{ab}V^c\ol\nabla_a \bK_{bc}$, so in particular 
\begin{equation*}
	J(n)=-\bY(n,n) \tr_P \bK-\ol\nabla_n \tr_P \bK+\div_P(\bK)(n).
\end{equation*}
In order to write this in terms of the geometry of the foliation, we need to express $\tr_P \bK$ and $\div_P(\bK)(n)$ in terms of the foliation tensors. For the first one, recalling equation \eqref{PAB}, $\tr_P \bK = \tr_h {\bm\chi}$. For the second we use \eqref{PAB} and \eqref{olnablan2},
\begin{equation*}
	\div_P(\bK)(n)  = -P^{bd}\bK_{ba}\ol\nabla_d n^a = -P^{bd}\bK_{ba}P^{ac}\bK_{cd}=-h^{BD}h^{AC}{\bm\chi}_{BA}{\bm\chi}_{CD} \eqqcolon -|{\bm\chi}|^2.
\end{equation*}
Then, $$-J(n)= n\left(\tr_h{\bm\chi}\right) + 2\omega \tr_h{\bm\chi} +|{\bm\chi}|^2,$$ which is the abstract data form of the Raychaudhuri equation. From Theorem \ref{teo_gauge}, \eqref{transn} and the transformations in Lemma \ref{transformationtensors} it follows that the constraint tensor $J(n)$ takes the same form in any gauge (not necessarily CG). Indeed, given gauge parameters $(z,\zeta)$,
\begin{align}
	- z^2 J'(n') & = n\left(\tr_h{\bm\chi}\right) + 2\omega \tr_h{\bm\chi} +|{\bm\chi}|^2\nonumber\\
	&= z^2 n' \left(\tr_h{\bm\chi}'\right) +n(z)\tr_h{\bm\chi}' + 2z^2\omega'\tr_h{\bm\chi}' - n(z) \tr_h{\bm\chi}' + z^2|{\bm\chi}'|^2\nonumber\\
	&= z^2\left(n' \left(\tr_h{\bm\chi}'\right) + 2\omega'\tr_h{\bm\chi}'  + |{\bm\chi}'|^2\right) .\label{Jnanygauge}
\end{align}

Next we proceed by taking $V=X$, a tangent vector to the foliation, and rewrite ${\bm J}(X)$ in terms of the foliation tensors. The term $\tr_P \bK = \tr_h{\bm\chi}$ has already been computed. To compute $\div_P(\bK)(X)$ we start by showing that $\phi^{\star}_{\ul u}(\ol\nabla_X \bK) = \nabla^{h}_X {\bm\chi}$ for every $X\in\Gamma(T\mc S_{\ul u})$. For any pair of vectors $Z_1$, $Z_2$ tangent to the foliation it holds
\begin{align*}
	(\ol\nabla_X \bK)(Z_1,Z_2) & = X(\bK(Z_1,Z_2))-\bK(\ol\nabla_X Z_1,Z_2)-\bK(Z_1,\ol\nabla_X Z_2)\\
	&=X({\bm\chi}(Z_1,Z_2))-{\bm\chi}(\nabla_X^{h} Z_1,Z_2)-{\bm\chi}(Z_1,\nabla_X^{h} Z_2)\\
	&=(\nabla^{h}_X {\bm\chi})(Z_1,Z_2),
\end{align*}
where we have inserted the decomposition \eqref{decompnabla} with $\nabla^h$ instead of $\ol\nabla^{\mc S}$ because the data is written in a CG (see Remark \ref{observación}). Consequently $$\div_P \bK (X) \d P^{ab} X^c \ol\nabla_a \bK_{bc} = h^{AB}X^C\nabla^{h}{\bm\chi}_{BC}=\div_h ({\bm\chi}) (X).$$

Substituting these two terms, the tensor ${\bm J}(X)$ becomes 
\begin{equation}
	\label{JXaux}
	\begin{aligned}
{\bm J}(X)&=(\ol\nabla_X \bPi)(n,n)-(\ol\nabla_n \bPi)(X,n)-\bPi(X,n) \tr_h {\bm\chi} \\
	&\quad+\left(\bK*\bPi\right)(X,n)- \ol\nabla_X\tr_h {\bm\chi}+\div_h({\bm\chi})(X).
	\end{aligned}
\end{equation} 
Before we elaborate this expression, we introduce the one-form ${\bm{\tau}}\in\Gamma(T^{\star}\mc S_{\ul u})$ by 
\begin{equation}
	\label{taueta}
	{\bm{{\bm{\tau}}}} \d -{\bm{\eta}} - \dd\left(\log\left|\lambda\right|\right) ,
\end{equation}
which is a combination that will appear frequently below. Taking into account the definition of ${\bm{\eta}}$ in Def. \ref{def_tensors} and equation \eqref{Pipi}, 
\begin{equation}
	\label{tau2}
	{\bm{\tau}} = \phi^{\star}_{\ul u}\left(\bPi(\cdot,n)\right)+{\bm\chi}(\ell^{\sharp},\cdot),
\end{equation} 
from which \eqref{normalndeco} gets rewritten as 
\begin{equation}
	\label{olnablatau}
	\ol\nabla_X n = \chi^{\sharp}(X)-{\bm{\tau}}(X)n.
\end{equation}
The transformation law of ${\bm{\tau}}$ follows from those of ${\bm{\eta}}$ \eqref{transeta} and $\lambda$ (see \eqref{transn} and Def. \ref{def_CHD}),
\begin{equation}
	\label{transtau}
	{\bm{\tau}}' = {\bm{\tau}} +\dd\left(\log|z|\right).
\end{equation}
As a direct consequence of the expression of ${\bm{\eta}}$ in Def. \ref{def_tensors} and \eqref{tau2}, in a CG the tensors ${\bm{\eta}}$ and ${\bm{\tau}}$ can be written as 
\begin{align}
	{\bm{\eta}} & = -\phi_{\ul u}^{\star}\left(\bPi(n,\cdot)\right),\label{etaPi}\\
	{\bm{\tau}}&= \phi^{\star}_{\ul u}\left(\bPi(\cdot,n)\right).\label{tauPi}
\end{align}
The geometric interpretation of ${\bm{\tau}}$ is similar to the one of ${\bm{\eta}}$. In the following Lemma we compute the first and second terms of \eqref{JXaux} in terms of the foliation tensors. 

\begin{lema}
	\label{lemaB2}
	Let $\mc D$ be CHD written in a characteristic gauge and $X\in\Gamma(T\mc S)$. Then,
	\begin{align}
		\left(\ol\nabla_X \bPi\right)(n,n) &= 2\nabla^{h}_X\omega -2\left({\bm\chi}\cdot{\bm{\tau}}\right) (X)+4\omega{\bm{\tau}}(X)-\left({\bm\chi}\cdot\dd\left(\log|\lambda|\right)\right)(X),\label{nablaXPinn}\\
		\left(\ol\nabla_n \bPi\right)(X,n)&=\big(\wt\nabla_n {\bm{\tau}}\big)(X)+4\omega{\bm{\tau}}(X)+2\omega\ \dd\left(\log|\lambda|\right)(X),	\label{nablanPiXn}
	\end{align}
	where $\left(T\cdot\bm\alpha\right)_A\d h^{BC}T_{BA}\bm\alpha_C$.
	\begin{proof}
		We start with the computation of $\left(\ol\nabla_X\bPi\right)(n,n)$. From equations \eqref{olnablatau} and \eqref{tauPi},
		\begin{align*}
			\left(\ol\nabla_X \bPi\right)(n,n) & = \nabla^{h}_X\left(\bY(n,n)\right)-\bPi(\ol\nabla_{X}n,n)-\bPi(n,\ol\nabla_{X}n)\\
			&=2\nabla_X^{h}\omega -\left({\bm\chi}\cdot{\bm{\tau}}\right)(X)+\bY(n,n){\bm{\tau}}(X)+\left({\bm\chi}\cdot{\bm{{\bm{\eta}}}}\right)(X)+\bY(n,n)\bm\tau(X)\\
			&=2\nabla^{h}_X\omega -2\left({\bm\chi}\cdot{\bm{\tau}}\right) (X)+4\omega{\bm{\tau}}(X)-\left({\bm\chi}\cdot\dd\left(\log|\lambda|\right)\right)(X),
		\end{align*}
		where in the third line we replaced ${\bm{\eta}}$ by ${\bm{\tau}}$ according to \eqref{taueta}. The term $\left(\ol\nabla_n \bPi\right)(X,n)$ can be computed analogously:
		\begin{align*}
			\left(\ol\nabla_n \bPi\right)(X,n) & = n\left({\bm{\tau}}(X)\right)-\bPi\big(\wt\nabla_n X+{\bm{\eta}}(X)n,n\big)-\bPi\left(X,\ol\nabla_n n\right)\\
			&= n\left({\bm{\tau}}(X)\right)-{\bm{\tau}}\big(\wt\nabla_n X\big)-\bY(n,n){\bm{\eta}}(X)+\bY(n,n)\bm\tau(X)\\
			&=\big(\wt\nabla_n {\bm{\tau}}\big)(X)+4\omega{\bm{\tau}}(X)+2\omega\ \dd\left(\log|\lambda|\right)(X),
		\end{align*}
		where we used \eqref{nablan}, \eqref{nablann=Y} and Def. \eqref{nablatildeeta} together with \eqref{taueta}. Finally, from \eqref{PAB} and \eqref{tauPi}, the term $\left(\bK*\bPi\right)(X,n)$ becomes $\left({\bm\chi}\cdot{\bm{\tau}}\right)(X)$.
	\end{proof}
\end{lema}
Using equations \eqref{PAB} and \eqref{tauPi}, the third and fourth terms in \eqref{JXaux} become $-\tr_h{\bm\chi}\ {\bm{\tau}}(X)$ and $\left({\bm\chi}\cdot{\bm{\tau}}\right)(X)$, respectively. Consequently, combining all the terms, the constraint tensor $\bm J(X)$ in \eqref{JXaux} can be written in any CG as
\begin{equation}
	\begin{split}
		\label{JACG}
{\bm J}(X)&=-\big(\wt\nabla_n{\bm{\tau}}\big)(X)+2 \nabla^{h}_X\omega -\left({\bm{\tau}}\cdot{\bm\chi}\right)(X) - \tr_h{\bm\chi} \ \bm\tau(X)\\
		&\quad\, +\div_h({\bm\chi})(X)-\nabla^{h}_X\tr_h{\bm\chi}-2\omega \nabla^h_X\log|\lambda|-\left({\bm\chi}\cdot\dd\left(\log|\lambda|\right)\right)(X).
	\end{split}
\end{equation} 

Next we compute the constraint tensor ${\bm{\mc R}}_{AB}$, which from Definition \eqref{ricci} and the symmetries in Proposition \ref{symmetries} can be written as 
\begin{equation}
	\label{RAB1}
{\bm{\mc R}}_{AB} = \left(B_{acbd} P^{cd} + (A_{bca}+A_{acb})n^c\right)e_A^ae_B^b.
\end{equation} 
These contractions of $A$ and $B$ in the right-hand side are computed in Lemma \ref{LemaB}, so introducing equations \eqref{BP} and \eqref{An} and the symmetrized version of \eqref{An} into \eqref{RAB1} we conclude
\begin{equation}
	\label{RABfinal}
{\bm{\mc R}}_{AB}  = R^{h}_{AB}-2\wt\nabla_n {\bm{\Upsilon}}_{AB}+4\omega {\bm{\Upsilon}}_{AB}-\nabla_A^{h}{\bm{\eta}}_B-\nabla^{h}_B{\bm{\eta}}_A-2{\bm{\eta}}_A{\bm{\eta}}_B-{\bm\chi}_{AB}\tr_h {\bm{\Upsilon}} -{\bm{\Upsilon}}_{AB}\tr_h {\bm\chi} .
\end{equation}

Finally recalling that $2H+h^{AB}R_{AB}=0$ \eqref{RJHCG} and taking the trace of \eqref{RABfinal} w.r.t. $h$, $$H = - \dfrac{1}{2}R^{h}+ n\left(\tr_h{\bm{\Upsilon}}\right)  +\div({\bm{\eta}}) + |{\bm{\eta}}|^2 +\left(\tr_h{\bm\chi}-2\omega\right)\tr_h{\bm{\Upsilon}},$$ where we have defined $|{\bm{\eta}}|^2\d h^{AB}{\bm{\eta}}_A{\bm{\eta}}_B$ and used that $\nabla^h h=0$ and $\wt\nabla_n h=0$. Indeed,
\begin{align*}
\big(\wt\nabla_n h\big)(X,Z) & = n\left(h(X,Z)\right) - h\big(\wt\nabla_n X,Z\big) - h\big(X,\wt\nabla_n Z\big)\\
&=\left(\ol\nabla_n\bg\right)(X,Z)+{\bm{\eta}}(X)\bg(n,Z)+{\bm{\eta}}(Z)\bg(X,n)\\
&=0,
\end{align*}
where in the second equality we used \eqref{nablan} and in the last one \eqref{olnablagamma} together with $\bK(n,\cdot)=\bg(n,\cdot)=0$.\\

We summarize the results of this section in the following Theorem.
\begin{teo}
	\label{constraints}
	Let $\mc D$ be CHD and $\ul u$ a foliation function. Then the constraint tensors take the following form in a characteristic gauge
	\begin{align}
		J(n)&= -n\left(\tr_h{\bm\chi}\right) - 2\omega \tr_h{\bm\chi} -|{\bm\chi}|^2,\label{Ray}\\
{\bm J}(X)&=-\big(\wt\nabla_n{\bm{\tau}}\big)(X)+2 \nabla^{h}_X\omega -\left({\bm\chi}\cdot{\bm{\tau}}\right)(X) - {\bm{\tau}}(X)\tr_h{\bm\chi}+\div({\bm\chi})(X)\nonumber \\
		&\quad\,-\nabla^{h}_X\tr_h{\bm\chi}-2\omega \nabla^h_X\log|\lambda|-\left({\bm\chi}\cdot\dd\left(\log|\lambda|\right)\right)(X),\label{Jdata}\\
{\bm{\mc R}}_{AB} & = R^{h}_{AB}-2\wt\nabla_n {\bm{\Upsilon}}_{AB}+4\omega {\bm{\Upsilon}}_{AB}-\nabla_A^{h}{\bm{\eta}}_B-\nabla^{h}_B{\bm{\eta}}_A-2{\bm{\eta}}_A{\bm{\eta}}_B\nonumber\\
			&\quad\, -{\bm\chi}_{AB}\tr_h {\bm{\Upsilon}} -{\bm{\Upsilon}}_{AB}\tr_h {\bm\chi} ,\label{Rdata}\\
			H &= n\left(\tr_h{\bm{\Upsilon}}\right) + |{\bm{\eta}}|^2+\div({\bm{\eta}}) - \dfrac{1}{2}R^{h}+\left(\tr_h{\bm\chi}-2\omega\right)\tr_h{\bm{\Upsilon}}.\label{Hdata}
\end{align}
\end{teo}

Under further gauge restrictions, one can show that the expressions in Theorem \ref{constraints} agree with the standard null structure equations provided the data is embedded. We do not give the explicit comparison here since it is not needed for the proof.

\section{The Harmonic Gauge}
\label{sec_HG}

When solving any initial value problem in General Relativity one has to deal with the issue of the coordinates: since GR is a geometric theory the Einstein equations cannot have a unique solution. For that reason one has to ``fix the gauge'' by choosing an appropriate coordinate system in order to solve them. The standard approach to deal with this issue, both in the classical Cauchy problem and in the characteristic one, is to write the Einstein equations in some well-chosen gauge (e.g. harmonic gauge). This yields a new system of geometric PDE, called reduced Einstein equations, which does have a well-posed initial value problem in a PDE sense, i.e., that there is a unique solution in some neighbourhood of the initial hypersurface(s) (see \cite{HawkingEllis,Wald} and \cite{Luk,Rendall}, respectively). This approach requires showing, at the very end, that the solution of the reduced system is in fact a solution of the full Einstein field equations. Thus, the first step is to translate the harmonic condition on the coordinates, namely $\square_g x^{\mu}=0$, into a condition on our data. We start with the embedded case and then promote the definition to the abstract level.
\begin{prop}
	\label{propbox}
Let $\mc D$ be embedded CHD on a spacetime $(\mc M,g)$ with rigging $\xi$ and Levi--Civita connection $\nabla$ and let $f$ be a smooth function on $\mc M$. Then, 
\begin{equation}
	\label{box}
	\square_g f\st{\mc H}{=}\square_P f + 2\ol\nabla_n\left(\xi(f)\right) + \left(\tr_P \bK - 4\omega\right) \xi(f) -\left(2P^{bc}\ol\nabla_n\bm\ell_b+n\left(\ell^{(2)}\right)n^c\right) e_c(f),
\end{equation} 
where $\square_P f \d P^{ab}\ol\nabla_a\ol\nabla_b f$.
\begin{proof}
From \eqref{inverse},
\begin{align}
	\square_g f & = g^{\mu\nu}\nabla_{\mu}\nabla_{\nu} f \nonumber\\
	& \st{\mc H}{=} \left(P^{ab}e_a^{\mu}e_b^{\nu}+n^a\xi^{\nu} e^{\mu}_a+n^a\xi^{\mu}e_a^{\nu}\right)  \nabla_{\mu}\nabla_{\nu} f \nonumber\\
	&=P^{ab}  \nabla_{e_a}\nabla_{e_b} f -P^{ab}\nabla_{\nu} f \nabla_{e_a} e_b^{\nu}+2\nabla_n\left(\xi(f)\right)-2\nabla_{\nu}f\nabla_n\xi^{\nu}.\label{box1}
\end{align}
Particularizing equation \eqref{nablatxi} to CHD, contracting it with $n^a$ and using \eqref{olnablaell} as well as the expression of $\omega$ in Def. \eqref{def_tensors},
\begin{equation}
	\label{nablanxi}
	\nabla_n\xi =2\omega \xi + \big(P^{cb}\ol\nabla_n\bm\ell_b+\dfrac{1}{2}n\big(\ell^{(2)}\big)n^c\big)e_c.
\end{equation}
Introducing \eqref{nablatt} and \eqref{nablanxi} into \eqref{box1} and using $\nabla_{e_a}\nabla_{e_b} f = e_a\left(e_b(f)\right)=\ol\nabla_{e_a}\ol\nabla_{e_b} f$, 
\begin{align*}
	\square_g f &\st{\mc H}{=} P^{ab}\left(\ol\nabla_{e_a}\ol\nabla_{e_b} f -\ol\Gamma_{ab}^c e_c(f)\right)+ P^{ab} \bK_{ab}\xi(f)+ 2\ol\nabla_n\left(\xi(f)\right) \\
	&\quad\,-4\omega\xi(f)-\left(2P^{cb}\ol\nabla_n\bm\ell_b+n^c\ol\nabla_n\ell^{(2)}\right) e_c(f),
\end{align*}
and hence equation \eqref{box} follows.
\end{proof}
\end{prop}

Let $\mc D$ be embedded CHD of dimension $m$ on a spacetime $(\mc M,g)$ with rigging $\xi$. In order to write $\square_g x^{\mu}=0$ in terms of the abstract data consider a set of independent functions $\{x^{\ul a}\}_{a=1}^m$ on $\mc H$ and extend them to $\mc M$ in such a way that $\xi(x^{\ul a})=0$. Consider also a function $u$ on $\mc M$ satisfying $u|_{\mc H}=0$ and $\xi(u)=1$. Let $\Xi_{c}{}^{\ul a} \d e_c(x^{\ul a})$ and $\Xi^{c}{}_{\ul a}$ its inverse. Since $\Xi_{c}{}^{\ul a}$ is a covector (\ul{a} is \textit{not} a tensorial index), $\Xi^{c}{}_{\ul a}$ is a vector and so it is $V^c\d \Xi^{c}{}_{\ul a} \square_P x^{\ul a}$. In terms of this vector and by virtue of Proposition \ref{propbox} the conditions $\square_g x^{\ul a}=0$ are equivalent to 
\begin{equation}
	\label{V^c}
	2P^{bc}\ol\nabla_n\bm\ell_b+n\left(\ell^{(2)}\right)n^c=V^c,
\end{equation} 
which is a covariant equation. In the following Theorem we prove abstractly that given a set of independent functions $\{x^{\ul a}\}$ on $\mc H$ there exists essentially a unique gauge in which $\tr_P \bK -4\omega=0$ and equation \eqref{V^c} holds.

\begin{teo}
	\label{teo_HG}
	Let $\mc D$ be CHD, $\{x^{\ul a}\}$ a set of $m$ independent functions and $V^c=\Xi^c{}_{\ul a}\square_P x^{\ul a}$. Select a section $\mc S\subset \mc H$ and a pair $(z_0,\zeta_0)\in\mc F^{\star}(\mc S)\times\Gamma(T\mc H)$. Then there exists a unique gauge satisfying the following conditions on $\mc H$,
	\begin{align}
		\tr_P \bK -4\omega&=0,\label{combGamma}\\
		2P\left(\ol\nabla_n\bm\ell,\cdot\right)+n\left(\ell^{(2)}\right)n&=V\label{V^c2}
	\end{align}
together with $(z,\zeta)|_{\mc S}=(z_0,\zeta_0)$. Such gauges will be called ``harmonic gauge'' (HG).
	\begin{proof}

First consider equation \eqref{combGamma} in the primed gauge. Taking into account transformation laws \eqref{Ktrans} and \eqref{trans_omega} together with $\bK(n,\cdot)=0$, $$\tr_{P'} \bK' -4\omega' = z^{-1}\tr_P \bK-4z^{-1}\omega-2z^{-2} n\left( z\right) = 0.$$ Then, there exists a unique $z$ solving the previous equation with initial condition $z|_{\mc S}=z_0$ as initial condition. Moreover, since $z_0 \neq 0$, then $z$ does not vanish in some neighbourhood of $\mc H$ containing the initial section $\mc S$. The next task is to show that there is a gauge in which \eqref{V^c2} holds. This requires determining the gauge behavior of both sides in \eqref{V^c2}. Concerning the vector $V$ it suffices to study the transformation of $\square_P F$ with $F\in\mc F(\mc H)$. From equation \eqref{gaugeP} and Proposition \ref{gaugeconection},
		\begin{align*}
\square_{P'} F & = P'{}^{ab}\ol\nabla'_a\ol\nabla'_b F\\
&=\left(P^{ab}-\zeta^an^b-\zeta^b n^a\right)\left(\ol\nabla_a\ol\nabla_b F-\zeta(F) \bK_{ab}\right)\\
&=P^{ab}\ol\nabla_a\ol\nabla_b F-\zeta(F)P^{ab} \bK_{ab}-\zeta^an^b\ol\nabla_a\ol\nabla_b F-\zeta^bn^a\ol\nabla_a\ol\nabla_b F\\
&=\square_P F -\zeta(F)\tr_P \bK  -2\zeta\left(n(F)\right) +  2\left(\ol\nabla_{\zeta} n\right)(F),
		\end{align*}
where in the third line we used $\bK(n,\cdot)=0$ and in the last equality that $\ol\nabla$ is torsion-free. Hence, applying to $F=x^{\ul a}$,
\begin{equation}
\label{Vprimac}
V'{}^c= V^c - \Psi^c{}_a \left(\zeta(x^a)\tr_P \bK  +2\zeta\left(n(x^a)\right) -  2\left(\ol\nabla_{\zeta} n\right)(x^a)\right).
\end{equation}
Concerning the LHS of \eqref{V^c2} we will use transformations \eqref{transn}, \eqref{gaugeconnection} and $\bK(n,\cdot)=0$ as well as $\ol\nabla_n \bg = 0$, which follows from \eqref{olnablagamma}. We analyze each term in \eqref{V^c2} separately. For the first one we recall $\bm\ell'_b = z^{-1}\left(\bm\ell_b+\bg_{ba}\zeta^a\right)$ (see \eqref{tranfell}) and compute
$$\ol\nabla'_{n'}\bm\ell'_b =z^{-1}\ol\nabla_n\bm\ell'_b= z^{-2} n(z)\bm\ell'_b+\ol\nabla_n\bm\ell_b+\bg_{ba}\ol\nabla_n\zeta^a.$$ Contracting with $P'{}^{bc}$ and using the primed versions of \eqref{Pell} and \eqref{Pgamma}, namely
\begin{align*}
P'{}^{bc}\bm\ell'_b =-\ell'{}^{(2)}n'{}^c,\qquad 
P'{}^{bc}\bg_{ba} =\delta^c_a-\bm\ell_a'n'{}^c,
\end{align*}
as well as the transformation law \eqref{gaugeP},
\begin{align}
	2P'{}^{bc}\ol\nabla'_{n'}\bm\ell_b' & = 2z^{-2} n(z) P'{}^{bc}\bm\ell'_b + 2P'{}^{bc}\ol\nabla_n\bm\ell_b+2P'{}^{bc}\bg_{ba}\ol\nabla_n\zeta^a\nonumber\\	
	& = -2z^{-2} n(z)\ell'{}^{(2)}n'{}^c +2P^{bc}\ol\nabla_n\bm\ell_b-2\zeta^bn^c\ol\nabla_n\bm\ell_b-2\zeta^cn^b\ol\nabla_n\bm\ell_b\nonumber\\
	&\quad\,+2\ol\nabla_n\zeta^c-2zn'{}^c\left(\bm\ell_a\ol\nabla_n\zeta^a+\bg_{ab}\zeta^b\ol\nabla_n\zeta^a\right).\label{HGaux1}
\end{align}
To compute the term $n'\left(\ell'{}^{(2)}\right) n'{}^c$ we insert $\ell'{}^{(2)} = z^2\left(\ell^{(2)}+2\bm\ell(\zeta)+\bg(\zeta,\zeta)\right)$ \eqref{transell2} and get 
\begin{equation}
	\label{HGaux2}
\hspace{-1mm}	{n'}\left(\ell'{}^{(2)}\right)n'{}^c = 2z^{-2} n(z)\ell'{}^{(2)}n'{}^c + z\left(\ol\nabla_n\ell^{(2)} + 2\zeta^a\ol\nabla_n\bm\ell_a + 2\bm\ell_a\ol\nabla_n\zeta^a + 2\bg_{ab}\zeta^a\ol\nabla_n\zeta^b\right)n'{}^c.
\end{equation}
Adding \eqref{HGaux1} and \eqref{HGaux2} and using $n^d\ol\nabla_n\bm\ell_d = 2\omega$, which follows from \eqref{olnablaell} and Def. \eqref{def_tensors},
\begin{align}
2P'{}^{bc}\ol\nabla_{n'}\bm\ell_b'+{n'}\left(\ell'{}^{(2)} \right) n'{}^c  = 2P^{cd}\ol\nabla_{n}\bm\ell_b+ n\left(\ell^{(2)}\right)n^c- 4\omega\zeta^c +2\ol\nabla_n\zeta^c.\label{LHSprima}
\end{align}
Finally taking into account \eqref{Vprimac} and \eqref{LHSprima}, equation \eqref{V^c2} constitutes a first order ODE for $\zeta$ with unique solution given $\zeta_0$.
	\end{proof}
\end{teo}

\begin{obs}
	Observe that when the data is embedded and written in the HG, Proposition \ref{propbox} ensures that the functions $\{u,x^{\ul a}\}$ satisfy $\square_g x^{\ul a}=0$ and $\square_g u=0$ on $\mc H$ provided that $\xi(x^{\ul a})\st{\mc H}{=}0$ and $\xi(u)|_{\mc H}$ is constant along each null generator of $\mc H$.
\end{obs}

Let $\mc D$ be embedded CHD on a spacetime $(\mc M,g)$ with rigging $\xi$. Consider a set of $m$ independent functions adapted to the foliation, i.e., $\{x^{\ul a}\}=\{\ul u, x^A\}$, where $\{x^A\}$ is a set of $m-1$ functions on $\mc H$ satisfying $n(x^A)=0$ and $n\left(\ul u\right)\neq 0$. We want to compute explicitly $\square_g u$, $\square_g x^A$ and $\square_g \ul u$ in terms of the foliation tensors in the embedded case and then promote these expressions into abstract definitions on $\mc H$. The first one is immediate from Proposition \ref{propbox} together with $\xi(u)=1$ and the fact that $\tr_P \bK = \tr_h{\bm\chi}$, 
\begin{equation}
	\label{boxu}
	\square_g u = \tr_h {\bm\chi}-4\omega.
\end{equation}
For the other two we first prove an intermediate result.
\begin{prop}
	\label{propbox2}
	Let $\mc D$ be CHD and $\beta\in\mc F(\mc H)$. Then,
	\begin{align}
\square_P \beta &= \square_h \beta -\ell^{\sharp}(\beta)\tr_h{\bm\chi} - 2\ell^{\sharp}\left(n(\beta)\right)+2\chi^{\sharp}(\ell^{\sharp})(\beta)-\big(\ell^{(2)}-\ell_{\sharp}^{(2)}\big)n\left(n(\beta)\right)\nonumber\\
&\quad\,+\big(\tr_h{\bm{\Upsilon}} -\left(2\omega+\tr_h{\bm\chi}\right)\big(\ell^{(2)}-\ell_{\sharp}^{(2)}\big)-2{\bm{\tau}}(\ell^{\sharp})-\div_h\bm\ell_{\para}\big)n(\beta),\label{boxpf}\\
e_c(\beta)P^{cb}\ol\nabla_n\bm\ell_b &= h^{\sharp}\left(\dd \beta, \lie_n\bm\ell+\bPi(\cdot,n)\right)-2\omega\ell^{\sharp}(\beta)\nonumber\\
&\quad\,-\big(2\omega \big(\ell^{(2)}-\ell_{\sharp}^{(2)}\big)+\left(\lie_n\bm\ell\right)(\ell^{\sharp})+\bPi(\ell^{\sharp},n)\big)n(\beta).\label{Pnablal}
\end{align}
	\begin{proof}
From decomposition \eqref{Pdecomposition} and the fact that the Hessian of a function is symmetric,
\begin{align}
	\square_P \beta & = P^{ab}\left(e_a\left(e_b(\beta)\right)  - \left(\ol\nabla_{e_a} e_b\right)(\beta) \right)\nonumber\\
	&=h^{AB} \big(e_A\left(e_B(\beta)\right)- \big(\ol\nabla_{e_A}^{\mc S} e_B\big) (\beta) + \bQ(e_A,e_B)n(\beta) \big) - 2\ell^A\left(e_A\left(n(\beta)\right)-\left(\ol\nabla_{e_A} n\right)(\beta)\right)\nonumber\\
	&\quad\,-\big(\ell^{(2)}-\ell_{\sharp}^{(2)}\big)\left(n\left(n(\beta)\right)-\left(\ol\nabla_n n\right)(\beta)\right)\nonumber\\
	&=h^{AB}\big(e_A\left(e_B(\beta)\right)  - \big(\ol\nabla_{e_A}^{h} e_B\big) (\beta) -\ell^{\sharp}(\beta){\bm\chi}_{AB} + \bQ(e_A,e_B)n(\beta) \big)\label{boxaux1}\\
	&\quad\,-2\ell^{\sharp}\left(n(\beta)\right) +2\chi^{\sharp}(\ell^{\sharp})(\beta)-2{\bm{\tau}}(\ell^{\sharp}) n(\beta) - \big(\ell^{(2)}-\ell_{\sharp}^{(2)}\big)\left(n\left(n(\beta)\right)+2\omega n(\beta)\right),\nonumber
\end{align}
where in the second equality we used \eqref{decompnabla} and in the third one Proposition \ref{nablaSandnablah}, \eqref{olnablatau} and \eqref{nablann=Y}. The term $h^{AB}\bQ(e_A,e_B)$ can be computed from \eqref{tensorQ},
\begin{align}
	h^{AB} \bQ(e_A,e_B) & = h^{AB}\big(\bY(e_A,e_B)+\bF(e_A,e_B)-\ell^{(2)}{\bm\chi}_{AB} - \big(\ol\nabla^{\mc S}_{e_A}\bm\ell_{\para}\big)(e_B)\big)\nonumber\\
	&= \tr_h{\bm{\Upsilon}} -\big(\ell^{(2)}-\ell_{\sharp}^{(2)}\big)\tr_h{\bm\chi} - \div_h\bm\ell_{\para},\label{boxaux2}
\end{align}
where in the second equality we used the definition of ${\bm{\Upsilon}}$ (Def. \ref{def_tensors}), the fact that $\bF$ is antisymmetric and Proposition \ref{nablaSandnablah}. Introducing \eqref{boxaux2} into \eqref{boxaux1}, \eqref{boxpf} follows. In order to show \eqref{Pnablal} we employ again decomposition \eqref{Pdecomposition},
\begin{align*}
	e_c(\beta) P^{cb}\ol\nabla_n\bm\ell_b & = h^{CB}e_C(\beta)\left(\ol\nabla_n\bm\ell\right)(e_B)-n(\beta)\ell^B\left(\ol\nabla_n\bm\ell\right)(e_B)\\
	&\quad -2\omega \ell^{\sharp}(\beta)-2\omega \big(\ell^{(2)}-\ell_{\sharp}^{(2)}\big) n(\beta),
\end{align*}
where we used $\left(\ol\nabla_n\bm\ell\right) (n)=2\omega$ (see \eqref{olnablaell}). Equation \eqref{Pnablal} follows after taking into account that $\left(\ol\nabla_n\bm\ell\right)(e_B)  = n\left(\bm\ell(e_B)\right) -\bm\ell\left(\ol\nabla_n e_B\right)$ and thus
\begin{align*}
\left(\ol\nabla_n\bm\ell\right)(e_B) =n\left(\bm\ell(e_B)\right)-\bm\ell\left(\lie_n e_B -\ol\nabla_{e_B} n\right)=\left(\lie_n\bm\ell\right)(e_B)+\bPi(e_B,n),
\end{align*}
where we used $\bm\ell(n)=1$ and equation \eqref{normalndeco}.
	\end{proof}
\end{prop}

From Propositions \ref{propbox} and \ref{propbox2} as well as relation \eqref{tau2} one has the following Corollary. 

\begin{cor}
	\label{corollarybox}
Let $\mc D$ be embedded CHD with rigging $\xi$ and FF $\ul u$ and let $\{x^A\}$ be a set of functions satisfying $n(x^A)=0$ and extended off $\mc H$ by means of $\xi(\ul u)=\xi(x^A)=0$. Then,
\begin{align}
\hspace{-6mm}\square_g x^A&=\square_h x^A +\left(4\omega-\tr_h{\bm\chi}\right)\ell^{\sharp}(x^A)+ 2\dd x^A\left(\chi^{\sharp}(\ell^{\sharp}) - h^{\sharp}\big(\lie_n\bm\ell +\bPi\left(\cdot,n\right),\cdot\big) \right),\label{boxA}\\
\hspace{-6mm}\square_g \ul u &= \lambda\tr_h{\bm{\Upsilon}}  + \lambda\Phi\left(\bm\ell_{\para},\omega,h,\ell^{(2)}\right),\label{boxulu}
\end{align}
where
\begin{equation}
	\label{Phi}
	\begin{aligned}
	\Phi\left(\bm\ell,\omega,h,\ell^{(2)}\right)&=\big(\ell^{(2)}-\ell_{\sharp}^{(2)}\big)\left(2\omega-\tr_h{\bm\chi}-n(\log|\lambda|)\right)-\div_h\bm\ell_{\para}\\
	&\quad\, +2\left(\lie_n\bm\ell\right)(\ell^{\sharp})-2{\bm\chi}(\ell^{\sharp},\ell^{\sharp})- n\left(\ell^{(2)}\right).
	\end{aligned}
\end{equation}
\end{cor}

Equations \eqref{boxu}, \eqref{boxA} and \eqref{boxulu} lead naturally to the definition of the following abstract functions on $\mc H$,
\begin{align}
\hspace{-3mm}	\Gamma^u_{\mc H} &\d \tr_h {\bm\chi}-4\omega,\label{Gammau0}\\
\hspace{-3mm}	\Gamma^A_{\mc H} & \d \square_h x^A +\left(4\omega-\tr_h{\bm\chi}\right)\ell^{\sharp}(x^A)+ 2\dd x^A\left(\chi^{\sharp}(\ell^{\sharp})\right) -2\left( \lie_n\bm\ell +\bPi\left(\cdot,n\right)  \right)\left(\grad_h x^A\right),\label{GammaA2}\\
\hspace{-3mm}	\Gamma^{\ul u}_{\mc H} & \d \lambda\tr{\bm{\Upsilon}}+ \lambda\Phi\left(\bm\ell,\omega,h,\ell^{(2)}\right).\label{Gammau}
\end{align}

For the proof of our main Theorem it is crucial to construct a linear combination of the constraint tensors and the functions $\Gamma^A_{\mc H}$ and $\Gamma^{\ul u}_{\mc H}$ that is hierarchically independent of the tensor $\bY$. The explicit combinations are computed in the following Lemma.

\begin{lema}
	\label{lema_indepY}
Let $\mc D$ be CHD with FF $\ul u$ and let $\{x^A\}$ be a set of functionally independent functions satisfying $n(x^A)=0$. Express all the tensors in the coordinate basis $\{\ul u,x^A\}$. Then the following combination depends neither on ${\bm{\tau}}$ nor on ${\bm{\Upsilon}}$,
\begin{equation}
	\label{combJ}
 \mc L_A\left({\bm J}_B,\Gamma^B_{\mc H},n\left(\Gamma^B_{\mc H}\right)\right)\d {\bm J}_A - \dfrac{1}{2}h_{AB}\ n\left(\Gamma^B_{\mc H}\right) +\dfrac{1}{2}\big(\Omega_A{}^B h_{CB}-{\bm\chi}_{BA}-\tr{\bm\chi}\ h_{AB}\big)\Gamma^B_{\mc H},
\end{equation}
where $\Omega_A{}^B$ are the connection coefficients of $\wt\nabla_n$ in this basis, i.e., $\wt\nabla_n \partial_{x^A} =\Omega_A{}^B \partial_{x^B}$. Moreover the combination 
\begin{equation}
	\label{combH}
\lie\left(H,\Gamma^{\ul u}_{\mc H},n(\Gamma^{\ul u}_{\mc H})\right)\d H-n\left(\lambda^{-1}\Gamma^{\ul u}_{\mc H}\right)-\lambda^{-1}\left(\tr_h{\bm\chi}-2\omega\right)\Gamma^{\ul u}_{\mc H},
\end{equation}
does not depend on ${\bm{\Upsilon}}$.
\begin{proof}
We first prove the claim assuming that the data is written in a CG and then we show that it is actually true in any gauge. Let us start by finding the linear combination $\mc L_A\left({\bm J}_B,\Gamma^B_{\mc H},n\left(\Gamma^B_{\mc H}\right)\right)$ which does not depend on the tensor ${\bm{\tau}}$. Let $X\in\Gamma(T\mc S)$. Using \eqref{tau2} the functions $\Gamma^A_{\mc H}$ defined above in the coordinates $\{x^A\}$ read 
\begin{equation}
	\label{Gammatau}
	\Gamma^A_{\mc H} = \square_h x^A +\left(4\omega-\tr_h{\bm\chi}\right)\ell^A+ 4\left(\chi^{\sharp}(\ell^{\sharp})\right)^A -2\left( \lie_n\bm\ell\right)^A-2{\bm{\tau}}^A.
\end{equation} 
Recall that the previous equation is still valid in any gauge. The expression of ${\bm J}_A$ in a CG was found in Theorem \ref{constraints} and takes the form (see \eqref{Jdata}) $${\bm J}_A = -\big(\wt\nabla_n{\bm{\tau}}\big)_A-\left({\bm\chi}\cdot{\bm{\tau}}\right)_A-\tr{\bm\chi}\ {\bm{\tau}}_A + \cdots,$$ where the dots represent terms involving metric data and $\omega$. Thus, the combination ${\bm J}_A-\frac{1}{2}h_{AB}\ n\left(\Gamma^B_{\mc H}\right)$ does not carry any derivative of $\bm\tau$ and its explicit form is $${\bm J}_A-\dfrac{1}{2}h_{AB}\ n\left(\Gamma^B_{\mc H}\right)=\Omega_A{}^B{\bm{\tau}}_B - h^{BC}{\bm\chi}_{BA}{\bm{\tau}}_C-\tr{\bm\chi}\ \tau_A + \cdots,$$ where the connection coefficients $\Omega_A{}^B$ do not depend on the tensor $\bY$ (see equation \eqref{indpendentY} and the comment below). Consequently the combination $$\mc L_A\left({\bm J}_B,\Gamma^B_{\mc H},n\left(\Gamma^B_{\mc H}\right)\right)={\bm J}_A - \dfrac{1}{2}h_{AB}\ n\left(\Gamma^B_{\mc H}\right) +\dfrac{1}{2}\Omega_A{}^B h_{CB}\Gamma^B_{\mc H} -\dfrac{1}{2}{\bm\chi}_{BA}\Gamma^B_{\mc H}-\dfrac{1}{2}\tr{\bm\chi}\ h_{AB}\Gamma^B_{\mc H}$$ does not depend on $\bm\tau$. Now suppose we change to an arbitrary gauge $\mc D'$. Since $\wt\nabla$ is gauge independent (see Lemma \ref{transformationtensors}) and $\bm\tau$ transforms as in \eqref{transtau}, the combination $\mc L_A\left({\bm J}'_B,\Gamma'{}^B_{\mc H},n'\left(\Gamma'{}^B_{\mc H}\right)\right)$, which in general will differ from $\mc L_A\left({\bm J}_B,\Gamma^B_{\mc H},n\left(\Gamma^B_{\mc H}\right)\right)$, is still independent on $\bm\tau'$, by virtue of item 2. in Theorem \ref{teo_gauge} and the transformations laws in Lemma \ref{transformationtensors}. Now we proceed analogously with the second identity. By Theorem \ref{constraints} the constraint scalar $H$ reads $$H=n\left(\tr_h{\bm{\Upsilon}}\right)+\left(\tr_h{\bm\chi}-2\omega\right)\tr_h{\bm{\Upsilon}}+\cdots,$$ where the dots are now terms which do not depend on $\bm\Upsilon$. Then the combination $$\lie\left(H,\Gamma^{\ul u}_{\mc H},n(\Gamma^{\ul u}_{\mc H})\right)=H-n\left(\lambda^{-1}\Gamma^{\ul u}_{\mc H}\right)-\lambda^{-1}\left(\tr_h{\bm\chi}-2\omega\right)\Gamma^{\ul u}_{\mc H}$$ does not depend on $\bm\Upsilon$. As before, recalling \eqref{transulchi} and item 1. in Theorem \ref{teo_gauge} it will be still true that the combination $\lie\left(H,\Gamma^{\ul u}_{\mc H},n(\Gamma^{\ul u}_{\mc H})\right)$ does not depend on $\bm\Upsilon$ in any gauge since the constraint tensors $J(n)$ and ${\bm J}_A$ do not depend on $\bm\Upsilon$ (see \eqref{Ray} and \eqref{Jdata}).
\end{proof}
\end{lema}

\section{The Characteristic Problem in General Relativity}
\label{sec_CP}
So far we have only worked with one abstract hypersurface $\mc H$. However, the characteristic problem is stated on two transverse, null hypersurfaces, so the next step is to construct the appropriate framework to formulate the characteristic problem in terms of CHD. First we define the notion of \textit{double embedding}.
\begin{defi}
	\label{def_double}
Let $\mc H$ and $\mc{\ul H}$ be two manifolds with boundary, $(\mc M,g)$ a spacetime and $\mc S_0$ an orientable manifold. Let $\mc D=\{\mc H,\bg,\bm\ell,\ell^{(2)},\bY\}$ and $\ul{\mc D}=\{\ul{\mc H},\ul\bg,\bm{\ul\ell},\ul\ell^{(2)},\ul \bY\}$ be CHD. We say that the pair $\left\{\mc D,\mc{\ul D}\right\}$ is double embedded (with intersection $\mc S_0$) provided that there exist embeddings $\varphi,\ul\varphi,i,\ul i$ making this diagram commutative
\begin{diagram}
	\mc M & \mc H \arrow[l, "\varphi", hook] \\
	\mc{\ul H}  \arrow[u, "\ul\varphi", hook] & \mc S_0 \arrow[l, "\ul i", hook] \arrow[u, "i", hook] \arrow[lu, hook]
\end{diagram}
and satisfying
\begin{enumerate}
	\item $\varphi$ and $\ul\varphi$ are embeddings for $\mc H$ and $\ul{\mc H}$ on $(\mc M,g)$ in the sense of Def. \ref{defi_embedded},
	\item $i(\mc S_0) = \partial\mc H$ and $\ul i(\mc S_0)=\partial\ul{\mc H}$,
	\item $i^{\star}\bg = \ul i^{\star}\ul\bg \eqqcolon h_0$ is a Riemannian metric on $\mc S_0$,
	\item The function $\mu\in \mc F(\mc S_0)$ defined by $\mu\d g(\nu,\ul\nu)$ is negative everywhere on $\mc S_0$, where $\nu$ and $\ul\nu$ are the push-forwards of $n$ and $\ul n$, respectively.
\end{enumerate}
In general, objects in $\mc{\ul D}$ will carry an underline.
\end{defi}

In order not to overload the notation we will identify $i(\mc S_0)=\ul i(\mc S_0)=\mc S_0$, $\varphi(\mc H)=\mc H$ and $\ul\varphi(\ul{\mc H})=\ul{\mc H}$. The precise meaning will be clear from the context. From the definition of $\mu$ it follows that under a gauge transformation with parameters $(z,\zeta)$ and $(\ul z,\ul\zeta)$ on $\mc D$ and $\mc{\ul D}$, respectively, $\mu$ transforms as 
\begin{equation}
	\label{transmu}
	\mu'=z^{-1}\ul z^{-1}\mu.
\end{equation} 

The definition of double embedded CHD imposes some additional constraints between the data of both hypersurfaces. Before writing them we prove an intermediate result. First note that setting $\bm\alpha=0$ and $f=0$ in Corollary \ref{CGcorollary} we get the following.

\begin{lema}
	\label{lemacompatible}
	Let $\mc D$ and $\ul{\mc D}$ be CHD. Then there exists a gauge in which the following conditions hold
	\begin{equation}
		\label{compatible}
		\begin{gathered}
			\bm\ell_{\para}|_{\mc S_0} = \ul{\bm\ell}_{\para}|_{\mc S_0} = 0,\\
			\ell^{(2)}|_{\mc S_0} = \ul\ell^{(2)}|_{\mc S_0} = 0.
		\end{gathered}
	\end{equation}
Moreover, the freedom of this gauge is parametrized by pairs $(z,\zeta)$ and $(\ul z,\ul\zeta)$ satisfying $\zeta|_{\mc S_0}=\ul\zeta|_{\mc S_0}=0$.
\end{lema}

\begin{prop}
	\label{prop_compatible}
	Let $\left\{\mc D,\mc{\ul D}\right\}$ be double embedded CHD written in any gauge in which conditions \eqref{compatible} hold and let $X,Z\in\Gamma(T\mc S_0)$. Then,
	\begin{equation}
		\label{nrigg}
		\ul\nu\st{\mc S_0}{=}\mu \xi, \qquad \nu \st{\mc S_0}{=} \mu\ul\xi,
	\end{equation}
and the following identities hold at $\mc S_0$
	\begin{align}
	\bPi(X,n) + \ul\bPi(X,\ul n) & =-X\left(\log|\mu|\right),\label{competa}\\
		 \bY(X,Z) & = \mu^{-1}\ul{\bK}(X,Z),\label{compchi}\\
		\ul \bY(X,Z) &= \mu^{-1} \bK(X,Z).\label{compulchi}
	\end{align}
	\begin{proof} 
By \eqref{compatible} the rigging $\xi$ is null, orthogonal to $\mc S_0$ and linearly independent to the normal $\nu$. The normal $\ul\nu$ has the same properties, so it follows that $\xi$ and $\ul\nu$ are proportional. Conditions $g(\xi,\nu)=1$ and $g(\nu,\ul\nu)=\mu$ then imply 
\begin{equation*}
	\ul\nu\st{\mc S_0}{=}\mu \xi \hspace{0.5cm} \text{and similarly} \hspace{0.5cm} \nu \st{\mc S_0}{=} \mu\ul\xi.
\end{equation*}
We can therefore interchange $\nu \longleftrightarrow \mu\ul\xi$ and $\ul\nu\longleftrightarrow \mu\xi$ in any spacetime expression at $\mc S_0$ that involves only tangential derivatives. This will be used repeatedly in the following without further notice. From equations \eqref{normalndeco} and \eqref{nablambient} it follows that $\bPi(X,n) \st{\mc S_0}{=} -g\left(\xi,\nabla_X \nu\right)$ and for the same reason $\ul\bPi(X,\ul n) \st{\mc S_0}{=} -g\left(\ul\xi,\nabla_X \ul\nu\right)$. Then, using \eqref{nrigg} it follows $\bPi(X,n)\st{\mc S_0}{=} -g\left(\ul\nu,\nabla_X\ul\xi\right) - \mu^{-1}  X(\mu)$ and thus
		\begin{align*}
			\bPi(X,n) \st{\mc S_0}{=} g\left(\ul\xi,\nabla_X\ul\nu\right)-X\left(\log|\mu|\right)\st{\mc S_0}{=} -\ul\bPi(X,\ul n)-X\left(\log|\mu|\right),
		\end{align*}
so \eqref{competa} is obtained. Concerning \eqref{compchi},
		\begin{align*}
			2{\bY}(X,Z)\st{\mc H}{=}\left(\lie_{\xi} g\right)(X,Z) \st{\mc S_0}{=}\left(\lie_{\mu^{-1}\ul\nu} g\right)(X,Z)\st{\mc S_0}{=} 2\mu^{-1}\ul{\bK}(X,Z),
		\end{align*}
where in the last equality we used $\lie_{\mu^{-1}\ul\nu} g = \mu^{-1}\lie_{\ul\nu} g$ acting on vectors tangent to $\mc S_0$. This proves \eqref{compchi} and \eqref{compulchi} is analogous.
	\end{proof}
\end{prop}

\begin{rmk}
	\label{rmk}
As a self-consistency check we prove that equations \eqref{competa}-\eqref{compulchi} stay invariant under the remaining gauge freedom. Let $(z,\zeta)$ and $(\ul z,\ul\zeta)$ be gauge parameters satisfying $\zeta|_{\mc S_0}=\ul\zeta|_{\mc S_0}=0$. From \eqref{transPiXn} the LHS of \eqref{competa} transforms as $$\text{LHS}\,'=\bPi'(X,n') +\ul\bPi'(X,\ul n') =\text{LHS}+X(\log|z|)+X(\log|\ul z|).$$ From \eqref{transmu} it is immediate that the RHS of \eqref{competa} transforms in the same way, so the gauge invariance of \eqref{competa} follows. Finally, \eqref{transY} (together with $\bm\ell_{\para}|_{\mc S_0}=0$ and $\zeta|_{\mc S_0}=0$) gives the transformation law $\bY'(X,Z)=z\bY(X,Z)$, while from $\ul{\bK}'=\ul z^{-1}\ul \bK$ we have $\mu'{}^{-1}\ul \bK' = z \mu^{-1}\ul \bK$, so the gauge invariance of  \eqref{compchi}, and by analogy of \eqref{compulchi}, follow.
\end{rmk}

Apart from the conditions \eqref{competa}-\eqref{compulchi}, there is another condition that must be fulfilled at $\mc S_0$, namely that the pullbacks to $\mc H$ and $\mc{\ul H}$ of the ambient Ricci tensor agree on $\mc S_0$. This can be translated into a completely abstract condition by recalling that the tensor $R_{ab}$ as defined in \eqref{ricci} coincides with the pullback of the ambient Ricci tensor when the data happens to be embedded. This motivates the following abstract Definition.

\begin{defi}
	\label{def_DND}
	Let $\mc H$ and $\mc{\ul H}$ be two manifolds with boundary, $\mc S_0$ an orientable manifold and $\mu\in\mc{F}(\mc S_0)$ everywhere negative. Let $i:\mc S_0\hookrightarrow\mc H$ and $\ul i:\mc S_0\hookrightarrow \mc{\ul H}$ be two embeddings. Let $\mc D=\{\mc H,\bg,\bm\ell,\ell^{(2)},\bY\}$ and $\ul{\mc D}=\{\ul{\mc H},\ul\bg,\bm{\ul\ell},\ul\ell^{(2)},\ul \bY\}$ be CHD satisfying 
	\begin{enumerate}
		\item $i(\mc S_0) = \partial\mc H$ and $\ul i(\mc S_0)=\partial\ul{\mc H}$, and
		\item $i^{\star}\bg = \ul i^{\star}\ul\bg \eqqcolon h_0$ is a Riemannian metric on $\mc S_0$.
		\item $i^{\star} {\bm{\mc R}} = \ul i^{\star}\ul{\bm{\mc R}}$.
	\end{enumerate}
	We say that the triple $\{\mc D,\mc{\ul D},\mu\}$ is double null data (DND) provided that under the gauge restrictions of Lemma \ref{lemacompatible} the following conditions hold at $\mc S_0$
	\begin{align}
		\bPi(X,n) + \ul\bPi(X,\ul n) & =-X\left(\log|\mu|\right),\label{competa2}\\
		 \bY(X,Z) & = \mu^{-1}\ul{\bK}(X,Z),\label{compchi2}\\
		 \ul \bY(X,Z) &= \mu^{-1} \bK(X,Z).\label{compulchi2}
	\end{align}
\end{defi}

\begin{rmk}
	Condition 1. implies that $\mc H$ and $\ul{\mc H}$ are of the same dimension (we refer to it as the dimension of the DND). Condition 2. implies that $\bg$ and $\ul\bg$ have signature $(0,1,...,1)$ at the boundaries $\partial\mc H$ and $\partial{\ul{\mc H}}$, respectively. Since $\bg$ and $\ul\bg$ have exactly one degeneration direction at every point it follows that they have this signature everywhere. Condition 3. will play no role in this paper because we are interested in solving the characteristic problem in vacuum and thus this condition will be automatically fulfilled.
\end{rmk}

Next we extend the notion of gauge transformation and embeddedness to the context of double null data.

\begin{defi}
	Let $\{\mc D,\mc{\ul D},\mu\}$ be DND and $z\in\mc F^{\star}(\mc H)$, $\ul z\in\mc F^{\star}(\mc{\ul H})$, $\zeta\in\Gamma(T\mc H)$ and $\ul\zeta\in\Gamma(T\mc{\ul H})$. The transformed data is given by $\{\mc D',\mc{\ul D}',\mu'\}$, where $\mc D'$ and $\mc{\ul D}'$ are the transformed CHD in the sense of Definition \ref{defi_gauge} and $\mu' \d  z^{-1}\ul z^{-1}\mu$ on $\mc S_0$.
\end{defi}

\begin{defi}
	\label{def_embDND}
	Let $\{\mc D,\mc{\ul D},\mu\}$ be DND and $(\mc M,g)$ a spacetime. We say that $\{\mc D,\mc{\ul D},\mu\}$ is embedded double null data on $(\mc M,g)$ provided that the pair $\{\mc D,\mc{\ul D}\}$ is double embedded and $\mu = g(\nu,\ul\nu)$, where $\nu$ and $\ul\nu$ are the spacetime versions of $n$ and $\ul n$, respectively.
\end{defi}

\begin{obs}
Equations \eqref{competa2}-\eqref{compulchi2} together with items 2. and 3. in Def. \ref{def_DND} can be interpreted as necessary conditions for the data in order to ``match'' in the embedded case. While conditions $\bg=\ul\bg$ and ${\bm{\mc R}}=\ul{\bm{\mc R}}$ are already gauge invariant (by virtue of \eqref{transgamma} and Theorem \ref{teo_gauge}), conditions \eqref{competa2}-\eqref{compulchi2} are not. However, Remark \ref{rmk} guarantees that \eqref{competa2}-\eqref{compulchi2} are invariant under the remaining freedom in Lemma \ref{lemacompatible} (whose existence is always granted), and thus the notion of double null data is well-defined. A definition of double null data where the compatibility conditions are written gauge-covariantly will be developed in a forthcoming paper.
\end{obs}

The next proposition shows that a gauge transformation on a double null data does not affect its embeddedness properties on a spacetime. For general hypersurface data this is a known fact (Proposition 3.5 on \cite{Marc2}). Here we show that the extra structure involved in the double null data does not spoil this property.

\begin{prop}
Let $\{\mc D,\mc{\ul D},\mu\}$ be embedded double null data in a spacetime $(\mc M,g)$ with embeddings $f,\ul f$ and riggings $\xi,\ul\xi$, respectively. For any pair of gauge parameters $(z,\zeta)$ and $(\ul z,\ul\zeta)$ belonging to the subgroup of Lemma \ref{lemacompatible}, the transformed data $\{\mc D',\mc{\ul D}',\mu'\}$ is embedded double null data in the same spacetime $(\mc M,g)$, with the same embeddings $f,\ul f$ and with riggings $\xi',\ul\xi'$ given by $$\xi' = z(\xi+f_{\star}\zeta), \qquad \ul\xi' = \ul z(\ul\xi+\ul f_{\star}\ul\zeta).$$
\begin{proof}
From the definition of embedded double null data (Def. \ref{def_embDND}) we need to check three things. Firstly, that 
\begin{equation}
	\label{gauge1}
	f^{\star}\left( g(\xi',\cdot)\right) = z\left(\bm\ell + \bg(\zeta,\cdot)\right), \qquad \ul f^{\star}\left( g(\ul\xi',\cdot)\right) = \ul z\left(\ul{\bm\ell} + \bg(\ul\zeta,\cdot)\right),
\end{equation}
\begin{equation}
	\label{gauge2}
 g(\xi',\xi') = z^2\left(\ell^{(2)} + 2\bm\ell(\zeta)+\bg(\zeta,\zeta)\right), \qquad  g(\ul\xi',\ul\xi') = \ul z^2\big(\ul\ell^{(2)} + 2\bm{\ul\ell}(\ul\zeta)+\bg(\ul\zeta,\ul\zeta)\big),
\end{equation}
and
\begin{equation}
	\label{gauge3}
	\frac{1}{2}f^{\star}\big(\lie_{\xi'} g\big) = z \bY + \bm\ell\otimes_s \dd z + \dfrac{1}{2}\lie_{z\zeta}\bg, \qquad \frac{1}{2}\ul f^{\star}\left(\lie_{\ul\xi'} g\right) = \ul z \ul \bY + \ul{\bm\ell}\otimes_s \dd \ul z + \dfrac{1}{2}\lie_{\ul z\ul\zeta}\ul\bg.
\end{equation}
Secondly, that 
\begin{equation}
	\label{gauge4}
	g(\nu',\ul\nu') = z^{-1}\ul z^{-1} g(\nu,\ul\nu),
\end{equation}
and finally, that the compatibility conditions \eqref{competa2}-\eqref{compulchi2} still hold in the new gauge. Equations \eqref{gauge1}-\eqref{gauge3} are proven in Proposition 3.5 of \cite{Marc2} in the context of general embedded hypersurface data. Concerning \eqref{gauge4}, from $\xi' = z(\xi+f_{\star}\zeta)$ and $\ul\xi' = \ul z(\ul\xi+\ul f_{\star}\ul\zeta)$ together with $g(\xi',\nu')=1$ and $g(\ul\xi',\ul\nu')=1$ it turns out that $\nu' = z^{-1}\nu$ and $\ul\nu'=\ul z^{-1}\ul\nu$, so \eqref{gauge4} also holds. Finally, from equations \eqref{gauge1}-\eqref{gauge3} and Remark \ref{rmk} one concludes that the compatibility conditions hold too.
\end{proof}
\end{prop}

\begin{lema}
	\label{lema_compHG}
	Let $\{\mc D,\mc{\ul D},\mu\}$ be DND and consider two set of independent functions $\{\ul u, x^{A}\}$ and $\{u,\ul x^{A}\}$ on $\mc H$ and $\mc{\ul H}$, respectively, satisfying $n(\ul u)\neq 0$ and $\ul n(u)\neq 0$. Then there exists a unique harmonic gauge w.r.t $\{\ul u,x^{A}\}$ and $\{u,\ul x^{A}\}$ in $\mc D$ and $\ul{\mc D}$, respectively, in which \eqref{compatible} hold together with $\mu=\lambda=\ul\lambda$ on $\mc S_0$.
	\begin{proof}
Let $\mc D$ and $\mc{\ul D}$ be CHD written in a HG. By Theorem \ref{teo_HG}, the HG is defined uniquely for each choice of $z|_{\mc S_0}$ and $\zeta|_{\mc S_0}$. From Corollary \ref{CGcorollary} we can gauge transform within the HG and the transformed data $\mc D'$ and $\mc{\ul D}'$ satisfy $\bm\ell'_{\para}=0$, $\ell'{}^{(2)}=0$, $\bm{\ul\ell}'_{\para}=0$ and $\ul\ell'{}^{(2)}=0$ on $\mc S_0$. By the same Corollary the remaining gauge freedom is parametrized by the pair $(z_0,\ul z_0)$. Recalling the transformation of $\lambda$, namely $\lambda'=z^{-1}\lambda$, and the one of $\mu$ (see \eqref{transmu}) we can choose $\ul z_0=\mu\lambda^{-1}$ and $z_0=\mu\ul\lambda^{-1}$ so that $\mu'=\lambda'=\ul\lambda'$.
	\end{proof}
\end{lema}

\begin{prop}
	\label{prop_compatible2}
Let $\{\mc D,\mc{\ul D}\}$ be double embedded CHD on a spacetime $(\mc M,g)$ with riggings $\xi$ and $\ul\xi$, respectively. Consider a set of coordinates $\{u,\ul u,x^A\}$ on $\mc M$ satisfying
	\begin{enumerate}
		\item $n\left(x^A|_{\mc H}\right)=0$ and $\ul n\left(x^A|_{\mc{\ul H}}\right)=0$,
		\item $\lambda\d n\left(\ul u|_{\mc H}\right)\neq 0$ and $\ul\lambda\d \ul n\left(u|_{\mc{\ul H}}\right)\neq 0$,
		\item $\ul u|_{\mc{\ul H}}=0$ and $u|_{\mc H}=0$,
		\item $\xi\st{\mc H}{=}\partial_u$ and $\ul\xi=\partial_{\ul u}$ on $\ul{\mc H}$.
	\end{enumerate} 
Let $X\in\Gamma(T\mc S_0)$. Then the following relations hold at $\mc S_0$,
	\begin{align}
		2\bY(n,n) &=\ul\lambda\ul n\big(\ul\ell^{(2)}\big) ,\label{compomega}\\
		2\ul \bY(\ul n,\ul n) &=\lambda n\left(\ell^{(2)}\right) ,\label{compulomega}\\
		2\bPi(X,n) &=\left(\lie_{\ul n}\bm{\ul\ell}\right)(X) -X\left(\log|\ul\lambda|\right) - \left(\lie_n\bm\ell\right)(X),\label{comptau2}\\
		2\ul\bPi(X,\ul n) &=\left( \lie_{n}\bm{\ell}\right)(X) -X\left(\log|\lambda|\right) - \left(\lie_{\ul n}\bm{\ul\ell}\right)(X).\label{compultau2}
	\end{align}
	\begin{proof}
From items (1) and (2) we have $\nu \st{\mc H}{=} \lambda \partial_{\ul u}$ and $\ul\nu = \ul\lambda \partial_{u}$ on $\mc{\ul H}$. This and (3) imply that we can replace $\nu\leftrightarrow \lambda\partial_{\ul u}$, $\xi \leftrightarrow \partial_u$ (resp. $\ul\nu\leftrightarrow \ul\lambda\partial_{u}$, $\ul\xi \leftrightarrow \partial_{\ul u}$) in any spacetime calculation at $\mc H$ (resp. $\ul{\mc H}$) that only involves tangential derivatives. We apply this without further warning. From $g(\xi,\nu)=1$ and $g(\ul\xi,\ul\nu)=1$ it follows that $\lambda\st{\mc S_0}{=}\ul\lambda$. Indeed, $1=g(\xi,\nu)= \lambda g(\partial_u,\partial_{\ul u}) \st{\mc S_0}{=} \lambda g\left(\ul\lambda^{-1}\ul\nu , \ul\xi\right) = \lambda\ul\lambda^{-1}$, so
		\begin{align*}
			2\bY(n,n)  \st{\mc H}{=} \left(\lie_{\xi} g\right)(\nu,\nu)\st{\mc H}{=}\left(\lie_{\partial_u} g\right)\left(\lambda\partial_{\ul u},\lambda\partial_{\ul u}\right)\st{\mc S_0}{=}\lambda^2\partial_u\left(g(\partial_{\ul u},\partial_{\ul u})\right)\st{\mc S_0}{=}\ul\lambda\ul n\big(\ul\ell^{(2)}\big),
		\end{align*}
		where in the third equality we used that $\lambda\partial_{\ul u}$ is null on $\mc S_0$. Hence \eqref{compomega}, and by analogy \eqref{compulomega}, follow. Next we prove \eqref{comptau2} and \eqref{compultau2}. Let $X\in\Gamma(T\mc S_0)$ and consider any extension of it outside $\mc S_0$. Equation \eqref{PiLienell} gives $2\bF(X,n)= -\left(\lie_n\bm\ell\right)(X)$ and from \eqref{Yembedded} one gets $2\bY(X,n) \st{\mc H}{=} \left(\lie_{\xi} g\right)(X,\nu)$ and thus
		\begin{align*}
 2\bY(X,n) &\st{\mc S_0}{=} \ul\lambda\left( \partial_{u}\left(g(X,\partial_{\ul u})\right) - g\left(\lie_{\partial_u}X,\partial_{\ul u}\right) - g\left(X,[\partial_u,\partial_{\ul u}]\right)\right) \\
			&\st{\mc S_0}{=} \ul n\left( g(X,\partial_{\ul u})\right) - g\left(\ul\lambda\lie_{\partial_u}X,\partial_{\ul u}\right) \\
			&\st{\mc S_0}{=}\left(\lie_{\ul n}\bm{\ul\ell}\right)(X) -X\left(\log|\lambda|\right),
		\end{align*}
		where we used $\lambda\st{\mc S_0}{=} \ul\lambda$ and in the last equality we inserted $g(\partial_{\ul u},X)= g(\xi,X)=\bm\ell(X)$ and used $\ul\nu = \ul\lambda \partial_u$ so that $\ul\lambda \lie_{\partial_u} X = \lie_{\ul n}X+X\left(\log|\ul\lambda|\right)\ul n$ on $\mc{\ul H}$. Observe that the result is independent of the extension of $X$. Hence \eqref{comptau2} (and similarly \eqref{compultau2}) follows.
	\end{proof}
\end{prop}

In the following Proposition we prove that conditions \eqref{compomega}--\eqref{compultau2} are always satisfied for any DND (i.e. at the abstract level) in the harmonic gauge of Lemma \ref{lema_compHG}.

\begin{prop}
	\label{prop_compatible3}
	Let $\{\mc D,\mc{\ul D},\mu\}$ be DND written in the harmonic gauge of Lemma \ref{lema_compHG}, $X\in\Gamma(T\mc S_0)$ and $\ul u$, $u$ functions on $\mc H$ and $\ul{\mc H}$ satisfying $\lambda\d n(\ul u)\neq 0$ and $\ul\lambda\d \ul n(u)\neq 0$. Then the following relations at $\mc S_0$ hold:
	\begin{align}
	2\bY(n,n)&=\ul\lambda\ul n\big(\ul\ell^{(2)}\big) \label{omega},\\
	2\ul \bY(\ul n,\ul n) &=\lambda n\left(\ell^{(2)}\right) \label{ulomega},\\
	2\bPi(X, n) &=\left(\lie_{\ul n}\bm{\ul\ell}\right)(X) -X\left(\log|\ul\lambda|\right) - \left(\lie_n\bm\ell\right)(X),\label{tau}\\
	2\ul\bPi(X,\ul n) &= \left(\lie_{n}\bm{\ell}\right)(X) -X\left(\log|\lambda|\right) - \left(\lie_{\ul n}\bm{\ul\ell}\right)(X).\label{ultau}
	\end{align}
	\begin{proof}
The derivative $n\left(\ell^{(2)}\right)$ on $\mc S_0$ in the HG is obtained from the condition ${\Gamma^{\ul u}=0}$. Directly from \eqref{Gammau} and the fact that $\Phi\st{\mc S_0}{=}- n\left(\ell^{(2)}\right)$ (see \eqref{Phi}) it follows that ${n\left(\ell^{(2)}\right)\st{\mc S_0}{=}\tr_h{\bm{\Upsilon}}}$. By \eqref{Gammau0} the condition $\Gamma^{\ul u}_{\ul{\mc H}}=0$ is equivalent to $4\ul\omega=\tr_h\ul{\bm\chi} \st{\mc S_0}{=}\tr_{\ul P}\ul \bK$. We can connect both by means of \eqref{compchi2} and get $$2\ul \bY(\ul n,\ul n)= 4 \ul\omega = \lambda n \left(\ell^{(2)}\right).$$ This proves \eqref{ulomega} and the analogous \eqref{omega}. Concerning the other two relations we use $\Gamma^A_{\mc H}=0$ and its underlined version. Evaluating \eqref{GammaA2} at $\mc S_0$ one gets
		\begin{equation}
			\left(\lie_n\bm\ell\right)(X) +\bPi(X,n) \st{\mc S_0}{=} \left(\lie_{\ul n}\bm{\ul\ell}\right)(X) +\ul\bPi(X,\ul n).
		\end{equation}
Taking into account \eqref{competa2} as well as $\mu\st{\mc S_0}{=}\lambda\st{\mc S_0}{=}\ul\lambda$, equations \eqref{tau} and \eqref{ultau} follow.
	\end{proof}
\end{prop}

As we already mentioned GR is a geometric theory, so the Einstein equations as a PDE problem cannot have a unique solution. The standard approach to deal with this issue is to solve the reduced Einstein equations (with cosmological constant $\Lambda$) instead, namely (cf. \cite{Luk,Rendall})
\begin{equation}
	\label{reduced}
	R^h_{\alpha\beta}\d R_{\alpha\beta} + g_{\mu(\alpha}\Gamma^{\mu}{}_{,\beta)}=\dfrac{2\Lambda}{D-2} g_{\alpha\beta},
\end{equation} 
where $D=m+1$ is the dimension of the spacetime. This system admits a well-posed initial value problem essentially because the principal symbol of \eqref{reduced} is hyperbolic. In the following Lemma we compute the tangent components of \eqref{reduced} on any embedded CHD in terms of the constraint tensors and the $\Gamma$-functions (see \eqref{Gammau0}-\eqref{Gammau}).

\begin{lema}
	\label{lema_red}
	Let $\mc D$ be embedded CHD on $(\mc M,g)$ with corresponding rigging $\xi$ and $\{u,\ul u,x^A\}$ a coordinate system on $\mc M$ satisfying $\xi(x^A)\st{\mc H}{=}\xi(\ul u)\st{\mc H}{=}0$, $\xi(u)\st{\mc H}{=}1$, $u|_{\mc H}=0$, $n(x^A|_{\mc H})=0$ and that $\ul u|_{\mc H}$ is a foliation function of $\mc D$. Suppose also that $(\mc M,g)$ is a solution of the reduced Einstein equations. Then,
	\begin{align}
		 J(n) + n\left(\Gamma^{u}_{\mc H}\right) &=0,\label{reducedRay}\\
\bm J_A+\dfrac{1}{2}\partial_{x^A}\left(\Gamma^u_{\mc H}\right) + \dfrac{1}{2} \bm\ell_A\ n\left(\Gamma^u_{\mc H}\right)+\dfrac{1}{2} h_{AB} \ n\left(\Gamma^B_{\mc H}\right)&=0,\label{reducedJA0}\\
	\hspace{-0.25cm} \bm{\mc R}_{AB}  + h_{C(B}\ \partial_{x^{A)}}\Gamma^C_{\mc H}  + \bm\ell_{(A}\ \partial_{x^{B)}}\left(\Gamma^u_{\mc H}\right) & = \dfrac{2\Lambda}{m-1} h_{AB}.\label{reducedH0}
	\end{align}
\begin{proof}
We have already shown $\nu\st{\mc H}{=}\lambda\partial_{\ul u}$ so contracting \eqref{reduced} with $\nu^{\alpha}\nu^{\beta}$ and taking into account that $\nu$ is null and satisfies $g(\nu,\partial_{x^A})\st{\mc H}{=}0$ it yields
\begin{align*}
 R^h_{\alpha\beta}\nu^{\alpha}\nu^{\beta} \st{\mc H}{=} R_{\alpha\beta}\nu^{\alpha}\nu^{\beta} + g(\partial_u,\nu) \ \nu\left(\Gamma^u_{\mc H}\right)\st{\mc H}{=} J(n)  +n\left(\Gamma^u_{\mc H}\right),
\end{align*}
where we used $R_{\alpha\beta}\nu^{\alpha}\nu^{\beta} \st{\mc H}{=} J(n)$ and that $\xi\st{\mc H}{=}\partial_u$. Since $(\mc M,g)$ is a solution of \eqref{reduced}, equation \eqref{reducedRay} follows. Similarly the ``normal-tangent'' components of \eqref{reduced} are
\begin{align*}
R^h_{\alpha\beta}\nu^{\alpha}(\partial_{x^A})^{\beta} & \st{\mc H}{=}	R_{\alpha\beta}\nu^{\alpha}(\partial_{x^A})^{\beta} + \dfrac{1}{2} g(\partial_u,\nu) \partial_{x^A} \left(\Gamma^u_{\mc H}\right) \\
&\quad + \dfrac{1}{2} g(\partial_u,\partial_{x^A}) \nu \left(\Gamma^u_{\mc H}\right) + \dfrac{1}{2} g(\partial_{x^A},\partial_{x^B})\ \nu\left(\Gamma^B_{\mc H}\right)  \\
&\st{\mc H}{=} {\bm J}_A+\dfrac{1}{2}\partial_{x^A}\left(\Gamma^u_{\mc H}\right) + \dfrac{1}{2} \bm\ell_A\ n\left(\Gamma^u_{\mc H}\right)+\dfrac{1}{2}  n\left(\Gamma^B_{\mc H}\right)h_{BA},
\end{align*}
where we used $g(\partial_u,\partial_{x^A})\st{\mc H}{=}g(\xi,\partial_{x^A})\st{\mc H}{=}\bm\ell_A$ and $g(\partial_{x^A},\partial_{x^B})\st{\mc H}{=} h_{AB}$. Finally the ``tangent-tangent'' component of \eqref{reduced} is 
\begin{align*}
	R^h_{\alpha\beta}(\partial_{x^A})^{\alpha}(\partial_{x^B})^{\beta} & = R_{\alpha\beta}(\partial_{x^A})^{\alpha}(\partial_{x^B})^{\beta} + \dfrac{1}{2} g(\partial_u,\partial_{x^A}) \partial_{x^B}\Gamma^u_{\mc H} + \dfrac{1}{2} g(\partial_u,\partial_{x^B}) \partial_{x^A}\Gamma^u_{\mc H}\\
	&\quad\, + \dfrac{1}{2} g(\partial_{x^C},\partial_{x^A}) \partial_{x^B}\Gamma^C_{\mc H} + \dfrac{1}{2} g(\partial_{x^C},\partial_{x^A}) \partial_{x^B}\Gamma^C_{\mc H} 
\end{align*}
and after using $R^h_{\alpha\beta}(\partial_{x^A})^{\alpha}(\partial_{x^B})^{\beta} =\frac{2\Lambda}{m-1} h_{AB}$ equation \eqref{reducedH0} follows.
\end{proof}
\end{lema}

We are ready to state and prove the main result of this paper.

\begin{teo}
	\label{main}
Let $\{\mc D,\mc{\ul D},\mu\}$ be double null data of dimension $m> 1$ as defined in Def. \ref{def_DND} satisfying the abstract constraint equations 
	\begin{equation}
		\label{constraintsL}
{\bm{\mc R}} = \dfrac{2\Lambda}{m-1}\bg \quad \text{and}\quad \ul{\bm{\mc R}}=\dfrac{2\Lambda}{m-1}\ul\bg,
	\end{equation}
where ${\bm{\mc R}}$ is defined in \eqref{ricci2}, $\ul{\bm{\mc R}}$ is its underlined version and $\Lambda\in\real$. Then, after restricting the data $\{\mc D,\mc{\ul D},\mu\}$ if necessary, there exists a spacetime $(\mc M,g)$ solution of the $\Lambda$-vacuum Einstein equations such that $\{\mc D,\mc{\ul D},\mu\}$ is embedded double null data on $(\mc M,g)$ in the sense of Def. \ref{def_embDND}. Moreover for any two such spacetimes $(\mc M,g)$ and $(\mc{\wh M},\wh g)$ there exist neighbourhoods of $\mc H\cup\mc{\ul H}$, $\mc U\subseteq\mc M$ and $\wh{\mc U}\subseteq\mc{\wh M}$, and a diffeomorphism $\varphi: \mc U\to \wh{\mc U}$, such that $\varphi^{\star}\wh g=g$.
\begin{proof}
We will use ${\bm{\mc R}}_{ab}=\frac{2\Lambda}{m-1}\bg_{ab}$ (and its underlined version) in the forms $J(n)=0$, ${\bm J}_A=0$ and ${\bm{\mc R}}_{AB}=\frac{2\Lambda}{m-1} h_{AB}$ (see \eqref{hamil2}, \eqref{momentum2} and \eqref{constrainttensors}). The first step of the proof is to solve the reduced Einstein equations. Let $\{\ul u, x^A\}$ and $\{u, x^A\}$ be coordinates on $\mc H$ and $\mc{\ul H}$ satisfying $\ul u\ge 0$, $\lambda\d n(\ul u)\neq 0$, $n(x^A)=0$ on $\mc H$ and $u\ge 0$, $\ul\lambda\d \ul n(u)\neq 0$, $\ul n(x^A)=0$ on $\mc{\ul H}$, as well as $\mc S_0=\{u=\ul u=0\}$. Consider a manifold $\mc N$ with coordinates $\{u,\ul u, x^A\}$ defined as $\mc N=\{u,\ul u\ge 0\}\subset \real^2\times\mc S$ and two embeddings $f:\mc H\hookrightarrow\mc N$ and $\ul f:\mc{\ul H}\hookrightarrow\mc N$ such that $f\left(\mc H\right)=\{u=0\}$, $\ul f\left(\ul{\mc H}\right)=\{\ul u=0\}$, $f(\mc H\cup\mc{\ul H})=\mc S_0$ and $\{\ul u|_{\mc H},x^A|_{\mc H}\}$ and $\{u|_{\mc H},x^A|_{\mc H}\}$ are the given coordinates on $\mc H$ and $\ul{\mc H}$, respectively. Throughout the proof we identify $\mc H$ and $\mc{\ul H}$ with their images under $f$ and $\ul f$, respectively. We want to construct a metric $g$ solution of the reduced Einstein equations on some neighbourhood $\mc M\subseteq \mc N$ of $\mc S_0$. From Theorem 1 of \cite{Rendall} we need to provide initial data for the metric $g_{\mu\nu}$ on $\mc H\cup\ul{\mc H}$ continuous at $\mc S_0$ and with smooth restrictions on $\mc H$ and $\ul{\mc H}$. In order to do so we write $\{\mc D,\mc{\ul D},\mu\}$ in the gauge of Lemma \ref{lema_compHG} w.r.t. the functions $\{\ul u, x^A\}$ and $\{u,x^A\}$ on $\mc H$ and $\mc{\ul H}$, respectively, and we provide the following initial data on $\mc H$ $$g_{u\, u}=\ell^{(2)},\qquad g_{u \, \ul u}=\lambda^{-1}, \qquad g_{u\, A} = \bm\ell_A, \qquad  g_{\ul u\, \ul u}=0, \qquad g_{\ul u\, A}=0, \qquad g_{AB}=h_{AB},$$ and on $\mc{\ul H}$, $$g_{u\, u}=0,\qquad g_{u \, \ul u}=\ul\lambda^{-1}, \qquad g_{u\, A} = 0, \qquad  g_{\ul u\, \ul u}=\ul\ell^{(2)}, \qquad g_{\ul u\, A}=\bm{\ul\ell}_A, \qquad g_{AB}={\ul h}_{AB},$$ in the coordinates $\{u,\ul u, x^A\}$. Since $\{\mc D,\mc{\ul D},\mu\}$ is written in a gauge in which \eqref{compatible} and $\lambda\st{\mc S_0}{=}\ul\lambda$ hold, the functions $g_{\mu\nu}$ are continuous on $\mc H\cup\ul{\mc H}$ and their restrictions to $\mc H$ and $\mc{\ul H}$ are smooth. Then from Rendall's Theorem 1 \cite{Rendall} there exists an open neighbourhood $U$ of $\mc S_0$ and a unique metric $g$ on $\mc M\d U\cap\mc N$ solution of the reduced Einstein equations \eqref{reduced} such that the components of $g$ in the coordinates $\{u,\ul u,x^A\}$ on $U\cap\left(\mc H\cup\ul{\mc H}\right)$ coincide with the given ones. By construction $f^{\star}\left(g(\partial_u,\cdot)\right)=\bm\ell$, $f^{\star}\left(g(\partial_u,\partial_u)\right)=\ell^{(2)}$, $\ul f^{\star}\left(g(\partial_{\ul u},\cdot)\right)=\bm{\ul\ell}$ and $\ul f^{\star}\left(g(\partial_{\ul u},\partial_{\ul u})\right)=\ul\ell^{(2)}$, so the only riggings that have a chance to make the data embedded in the sense of Definition \ref{defi_embedded} are $\xi=\partial_u$ and $\ul\xi=\partial_{\ul u}$, respectively. Let $\wt{\bY}$ and $\wt{\ul \bY}$ be defined as $\wt{\bY}\d\frac{1}{2}f^{\star}\left(\lie_{\xi} g\right)$ and $\wt{\ul \bY}\d\frac{1}{2}\ul f^{\star}\big(\lie_{\ul\xi} g\big)$. Then $\wt{\mc D}=\{\mc H,\wt\bg\d\bg,\wt{\bm\ell}\d\bm\ell,\wt\ell{}^{(2)}\d\ell^{(2)},\wt{\bY}\}$ and $\wt{\ul{\mc D}}=\{\mc{\ul H},\wt{\ul\bg}\d\ul\bg,\wt{\bm{\ul\ell}}\d\bm{\ul\ell},\wt{\ul\ell}{}^{(2)}\d\ul{\ell}^{(2)},\wt{\ul \bY}\}$ are embedded CHD on $(\mc M,g)$ with embeddings $f$, $\ul f$ and riggings $\xi$, $\ul\xi$, respectively, as in Def. \ref{defi_embedded} (in what follows we denote with a tilde the expressions depending on $\wt{\mc D}$ and $\wt{\ul{\mc D}}$). By construction the metric part of $\wt{\mc D},\wt{\ul{\mc D}}$ coincides with the one of $\{\mc D,\mc{\ul D},\mu\}$ and $g\left(\wt\nu,\wt{\ul\nu}\right)=\mu$. Hence the pair $\{\wt{\mc D},\wt{\ul{\mc D}}\}$ is double embedded in the sense of Def. \ref{def_double}. To prove that $\{\mc D,\mc{\ul D},\mu\}$ is actually embedded DND we need to show that the original tensors $\bY$ and $\ul \bY$ coincide with the embedded ones $\wt{\bY}$ and $\wt{\ul \bY}$, respectively.\\

For the existence part of the Theorem we need to prove two things: (1) that the solution of the reduced EFE is indeed a solution of the EFE and (2) that the tensors $\bY$ and $\ul \bY$ coincide with $\wt{\bY}$ and $\wt{\ul \bY}$, respectively. In order to prove (1) we will show that the coordinates are harmonic w.r.t the metric $g$. To prove (2) we will write homogeneous ODE for the tensors $\bY-\wt \bY$ and $\ul \bY-\wt{\ul \bY}$ and show that they vanish on $\mc S_0$. Both goals are achieved simultaneously.\\

From Propositions \ref{prop_compatible2} and \ref{prop_compatible3} applied to the data $\{\wt{\mc D},\wt{\ul{\mc D}}\}$ and to the original DND $\{\mc D,\mc{\ul D},\mu\}$, respectively, it follows that $\wt{\bY}(n,n)\st{\mc S_0}{=} \bY(n,n)$, $\wt{\ul \bY}(\ul n,\ul n)\st{\mc S_0}{=} \ul \bY(\ul n,\ul n)$, $\wt{\bY}(X,n) \st{\mc S_0}{=} \bY(X,n)$ and $\wt{\ul \bY}(X,n)\st{\mc S_0}{=}\ul \bY(X,\ul n)$ for every $X\in\Gamma(T\mc S_0)$, or in terms of the foliation tensors, $\wt\omega \st{\mc S_0}{=} \omega$, $\wt{\ul\omega}\st{\mc S_0}{=}\ul\omega$, $\wt{\bm{\tau}}\st{\mc S_0}{=}\bm\tau$ and $\wt{\ul{\bm{\tau}}}\st{\mc S_0}{=}\bm{\ul\tau}$. Moreover, from Proposition \ref{prop_compatible} applied to $\{\wt{\mc D},\wt{\ul{\mc D}}\}$ and Def. \ref{def_DND} applied to $\{\mc D,\mc{\ul D},\mu\}$, $\wt{{\bm{\Upsilon}}}={\bm{\Upsilon}}$ and $\wt{\ul{\bm{\Upsilon}}}=\ul{\bm{\Upsilon}}$ on $\mc S_0$. In order to prove that these equations hold everywhere and not only on $\mc S_0$ we start by considering the tilde version of equation \eqref{reducedRay} on $\mc H$, namely 
\begin{equation}
	\label{red}
	\wt J(n) +n\big(\wt\Gamma^u_{\mc H}\big)=0.
\end{equation} 
From the abstract constraint $J(n)=0$ it follows $2\omega \tr_h{\bm\chi} = -n\left(\tr_h{\bm\chi}\right)-|{\bm\chi}|^2$ (see \eqref{Ray} and recall that $J(n)$ takes the same form in any gauge, as shown in \eqref{Jnanygauge}). Since the metric data from $\mc D$ and $\mc{\wt D}$ coincides, $$-\wt{J}(n) = n\left(\tr_h{\bm\chi}\right) + 2\wt{\omega} \tr_h{\bm\chi} + |{\bm\chi}|^2 = 2(\wt\omega-\omega)\tr{\bm\chi}.$$ Recall that in the harmonic gauge $\tr{\bm\chi}-4\omega=0$ (see \eqref{combGamma}) and therefore $\wt\Gamma^u_{\mc H} = \tr_h{\bm\chi} - 4\wt\omega = 4(\omega-\wt\omega)$. Hence equation \eqref{red} can be rewritten as $- 2(\wt\omega-\omega)\tr{\bm\chi}+4n\left(\omega-\wt\omega\right)=0$, which is an homogeneous ODE for $\wt\omega-\omega$, which together with $\wt\omega \st{\mc S_0}{=} \omega$ implies $\wt\omega=\omega$ on $\mc H$ and then also $\wt\Gamma^u_{\mc H}=0$. The corresponding argument applied to $\ul{\mc H}$ gives $\wt{\Gamma}_{\mc{\ul H}}^{\ul u}=0$ and $\ul\omega=\wt{\ul\omega}$ on $\mc{\ul H}$. Taking into account $\wt\Gamma^u_{\mc H}=0$ the tilde version of equation \eqref{reducedJA0} reads
\begin{equation}
	\label{reducedJA}
\bm{\wt J}_A + \dfrac{1}{2}  n\big(\wt{\Gamma}^B_{\mc H}\big)h_{AB}=0.
\end{equation} 

Since ${\bm J}_A=0$ and the functions $\Gamma^A_{\mc H}$ \eqref{GammaA2} vanish in the harmonic gauge in which $\mc D$ is written, the combination $\lie_A\left({\bm J}_B,\Gamma^B_{\mc H}, n\left(\Gamma^B_{\mc H}\right)\right)$ defined in \eqref{combJ} also vanishes. From Lemma \ref{lema_indepY} this particular combination depends neither on ${\bm{\tau}}$ nor in ${\bm{\Upsilon}}$, so its tilde version also vanishes, $\mc L_A\big(\bm{\wt J}_B,\wt \Gamma^B_{\mc H},n\big(\wt \Gamma^B_{\mc H}\big)\big)=0$ (recall that $\omega$ is not problematic anymore). This gives $\bm{\wt{J}}_A$ in terms of $\wt{\Gamma}_{\mc H}^A$, $n(\wt{\Gamma}_{\mc H}^A)$, which inserted in \eqref{reducedJA} yields an homogeneous ODE for the functions $\wt\Gamma^A_{\mc H}$. Since $\wt{\bm{\tau}} \st{\mc S_0}{=} {\bm{\tau}}$ and $\Gamma^A_{\mc H}=0$, expression \eqref{Gammatau} gives $\wt\Gamma^A_{\mc H}\st{\mc S_0}{=}0$ and hence $\wt\Gamma^A_{\mc H}=0$ and ${\bm{\tau}}=\bm{\wt\tau}$ on $\mc H$. The same argument on $\mc{\ul H}$ proves $\wt{\Gamma}^A_{\mc{\ul H}}=0$ and $\ul{\bm{\tau}}=\bm{\ul{\wt\tau}}$ on $\mc{\ul H}$. Finally consider the trace of the tilde version of \eqref{reducedH0} w.r.t $h$, which taking into account $\wt\Gamma^u_{\mc H}=\wt\Gamma^A_{\mc H}=0$ reads 
\begin{equation}
	\label{reducedH}
\wt H =2\Lambda.
\end{equation} 
The abstract constraint equation $H=2\Lambda$ together with $\Gamma^{\ul u}_{\mc H}=0$ (since $\mc D$ is written in the HG) imply that the combination \eqref{combH} is equal to $2\Lambda$, and also is $\lie (\wt H,\wt\Gamma^{\ul u}_{\mc H}, n(\wt\Gamma^{\ul u}_{\mc H}))$ since this expression was constructed precisely so that it does not depend on ${\bm{\Upsilon}}$ (recall that $\omega$ and ${\bm{\tau}}$ are not problematic anymore). Inserting \eqref{reducedH} into the tilde version of \eqref{combH} we get an homogeneous ODE for $\wt\Gamma^{\ul u}_{\mc H}$. Since $\tr\wt{{\bm{\Upsilon}}}\st{\mc S_0}{=}\tr{\bm\Upsilon}$ and $\Gamma^{\ul u}_{\mc H}=0$, expression \eqref{Gammau} gives $\wt\Gamma^{\ul u}_{\mc H}\st{\mc S_0}{=}0$. Hence $\wt\Gamma^{\ul u}_{\mc H}=0$ and $\tr\wt{{\bm{\Upsilon}}}=\tr{\bm\Upsilon}$ everywhere on $\mc H$. The corresponding argument on $\mc{\ul H}$ gives $\tr\wt{\ul{\bm{\Upsilon}}}=\tr{\ul{\bm{\Upsilon}}}$ and $\wt\Gamma^{\ul u}_{\mc{\ul H}}=0$ on $\ul{\mc H}$. \\

The rest of the argument is standard (cf. \cite{HawkingEllis,Straumann,Wald}). As a consequence of the Bianchi identity the functions $\square_g u$, $\square_g \ul u$ and $\square_g x^A$ satisfy a homogeneous wave equation, which together with the fact that $\wt\Gamma^{\mu}_{\mc H}\st{\mc H}{=}0$ and $\wt{\Gamma}^{\mu}_{\mc{\ul H}}=0$ on $\ul{\mc H}$ yields $\square_g u=\square_g \ul u=\square_g x^A=0$ everywhere on $\mc M$. Consequently $(\mc M, g)$ is indeed a solution of the Einstein field equations with cosmological constant $\Lambda$. In order to show that $\{\mc D,\mc{\ul D},\mu\}$ is embedded DND on $(\mc M,g)$, as represented in Fig. \ref{fig1}, we still need to prove that the trace-free part of the tensors ${\bm{\Upsilon}}$ and $\ul{\bm{\Upsilon}}$ coincide with the ones of $\bm{\wt{\Upsilon}}$ and $\bm{\wt{\ul\Upsilon}}$, respectively. Since $\ric(g)=\Lambda g$ and the rest of the data coincides, the tensor $\bm{\wt{\Upsilon}}$ satisfies the same equation as the original $\bm\Upsilon$, namely ${\bm{\mc R}}_{AB}=\frac{2\Lambda}{m-1} h_{AB}$ (see \eqref{Rdata}), and thus the tensor ${\bm{\Upsilon}}-\bm{\wt{\Upsilon}}$ satisfies an homogeneous first order ODE, which together with ${\bm{\Upsilon}}\st{\mc S_0}{=}\wt{\bm{\Upsilon}}$ yields $\bm\Upsilon= \bm{\wt\Upsilon}$ on $\mc H$. The same argument on $\mc{\ul H}$ proves $\ul{\bm{\Upsilon}}= \bm{\wt{\ul\Upsilon}}$ on $\mc{\ul H}$. Then the original tensors $\bY$, $\ul \bY$ coincide with the embedded ones $\wt \bY$, $\wt{\ul \bY}$ and consequently the original abstract constraint equations are the pullback of the Einstein $\Lambda$-vacuum equations to $\mc H\cup\mc{\ul H}$.\\

\begin{figure}
	\centering
	\psfrag{a}{$\Psi$}
	\psfrag{b}{$\{\mc D,\mc{\ul D},\mu\}$}
	\psfrag{c}{$(\mc M,g)$}
	\includegraphics[width=10cm]{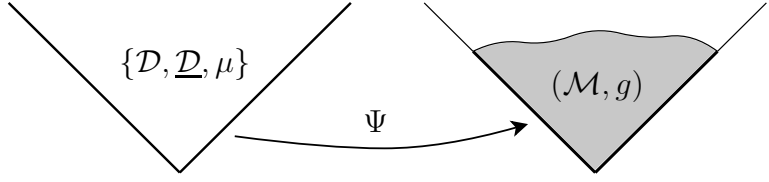} 
	\caption{Embedded double null data $\{\mc D,\mc{\ul D},\mu\}$ with embedding $\Psi$ (in the sense of Def. \ref{def_embDND}) in a spacetime $(\mc M,g)$ solution of the $\Lambda$-vacuum Einstein field equations.}
	\label{fig1}
\end{figure}

In order to prove the uniqueness part of the theorem consider $\{\mc D,\mc{\ul D},\mu\}$ as embedded DND in another spacetime $(\mc{\wh M},\wh g)$ solution of the EFE with cosmological constant $\Lambda$ with embeddings $f,\ul f$ and riggings $\xi,\ul\xi$. The aim is to show that there exist neighbourhoods of $\mc H\cup\mc{\ul H}$, $\mc U\subseteq\mc M$ and $\mc{\wh U}\subseteq\mc{\wh M}$, and a diffeomorphism ${\varphi:\mc U{\to}\ \mc{\wh U}}$, such that $\varphi^{\star}\wh g=g$. By Theorem 1 of \cite{Rendall}, for each set of independent functions $\{{\ul u},{x}^A\}$ on $\mc H$ and $\{{u},{x}^A\}$ on $\mc{\ul H}$ satisfying $\lambda\d n({\ul u})\neq 0$, $n( x^A)=0$ on $\mc H$ and $\ul\lambda\d\ul n( u)\neq 0$, $\ul n( x^A)=0$ on $\mc{\ul H}$, there exist an open neighbourhood $U$ of $\mc S_0$ and unique smooth functions $\{\wh u,\wh{\ul u},\wh x^A\}$ on $\mc{\wh U}\d U\cap\mc{\wh M}$ such that $\square_{\wh g} \wh u = \square_{\wh g} \wh{\ul u} = \square_{\wh g} \wh x^A=0$ on $\mc{\wh U}$ with the given functions on $\mc{\wh U}\cap\left(\mc H\cup\mc{\ul H}\right)$ as initial conditions and $f(\mc H)=\{\wh u=0\}$, $\ul f(\mc{\ul H})=\{\wh{\ul u}=0\}$.\\

First we prove that when the data is written in the gauge of Lemma \ref{lemacompatible}, the riggings are $\xi=\partial_{\wh u}$ and $\ul\xi=\partial_{\wh{\ul u}}$ on $f(\mc H)$ and $\ul f(\ul{\mc H})$, respectively. Proposition \ref{propbox} together with the fact that $\wh u$ is harmonic w.r.t. $\wh g$ imply $n\left(\xi(\wh u)\right)=0$ on $\mc H$. Since the data is written in a gauge in which $\bm\ell_{\para}\st{\mc S_0}{=}0$ and $\ell^{(2)}\st{\mc S_0}{=}0$, relations \eqref{nrigg} hold, so $\xi(\wh u)\st{\mc S_0}{=}\mu^{-1}\ul n(\wh u) = \mu^{-1}\ul\lambda = 1$, since $\mu=\lambda=\ul\lambda$ on $\mc S_0$ in the gauge of Lemma \ref{lema_compHG}. Consequently $\xi(\wh u)=1$ on $\mc H$, and by analogy, $\ul\xi(\wh{\ul u})=1$ on $\mc{\ul H}$. Concerning the functions $\{\wh{\ul u},\wh{x}^A\}$, Proposition \ref{propbox}, equations $\square_{\wh g} \wh{\ul u} = \square_{\wh g} \wh x^A=0$ and the fact that the gauge is harmonic imply $n\left(\xi(\wh{\ul u})\right)=0$ and $n\left(\xi(\wh{x}^A)\right)=0$ on $\mc H$. By the same argument as before, $\xi(\wh{\ul u})\st{\mc S_0}{=} \mu^{-1}\ul n(\wh{\ul u})=0$ and $\xi(\wh{x}^A)\st{\mc S_0}{=} \mu^{-1}\ul n(\wh{x}^A)=0$, from where one concludes that $\xi(\wh{\ul u})=\xi(\wh{x}^A)=0$ everywhere on $\mc H$ (and similarly $\ul\xi(\wh{u})=\ul\xi(\wh{x}^A)=0$ on $\mc{\ul H}$). This proves $\xi=\partial_{\wh u}$ and $\ul\xi=\partial_{\wh{\ul u}}$ on $f(\mc H)$ and $\ul f(\ul{\mc H})$, respectively. Then, from the definition of embedded data, $$g_{\wh u\, \wh u}=\ell^{(2)},\qquad g_{\wh u \, \wh{\ul u}}=\lambda^{-1}, \qquad g_{\wh u\, \wh A} = \bm\ell_{A}, \qquad  g_{\wh{\ul u}\, \wh{\ul u}}=0, \qquad g_{\wh{\ul u}\, \wh A}=0, \qquad g_{\wh A\wh B}=h_{ A B},$$ on $\mc H$ and $$g_{\wh u\, \wh u}=0,\qquad g_{\wh u \, \wh{\ul u}}=\ul\lambda^{-1}, \qquad  g_{\wh u\, \wh A} = 0, \qquad  g_{\wh{\ul u}\, \wh{\ul u}}=\ul\ell^{(2)}, \qquad g_{\wh{\ul u}\, \wh A}=\bm{\ul\ell}_A, \qquad g_{\wh A\wh B}={\ul h}_{AB}$$ on $\mc{\ul H}$. After restricting $\mc{\wh U}$ further, if necessary, there exists a neighbourhood $\mc U\subseteq\mc M$ and a diffeomorphism $\varphi:\mc U\to\mc{\wh U}$ defined by $x^{\mu} = \wh{x}^{\mu}\circ\varphi$. By construction $\mc U\cap \left(\mc H\cup\mc{\ul H}\right)\neq\emptyset$. Since $0=\varphi^{\star}\left(\square_{\wh g}\wh x^{\mu}\right)=\square_{\varphi^{\star}\wh g} x^{\mu}$, the coordinates $x^{\mu}$ are harmonic w.r.t. $\varphi^{\star}\wh g$. Moreover, from the fact that $\wh g$ is a solution of the $\Lambda$-vacuum EFE, $\ric\left[\wh g\right] = \frac{2\Lambda}{m-1}\wh g$, so $$\varphi^{\star}\left(\ric\left[\wh g\right]\right) =\ric\left[\varphi^{\star}\wh g\right] =\dfrac{2\Lambda}{m-1}\varphi^{\star}\wh g,$$ and therefore $\varphi^{\star}\wh g$ is a solution of the reduced equations in the coordinates $\{x^{\mu}\}$, just like $g$. In order to prove that $\varphi^{\star}\wh g$ and $g$ are actually the same, by Theorem 1 of \cite{Rendall} we only need to show that their restrictions on $\mc U\cap\left(\mc H\cup\mc{\ul H}\right)$ agree. This follows directly from the fact that the push-forward $\varphi_{\star}$ is the identity, because $\varphi_{\star}\partial_u = \partial_{\wh u}$, $\varphi_{\star}\partial_{\ul u} = \partial_{\wh{\ul u}}$ and $\varphi_{\star}\partial_{x^A} = \partial_{\wh x^A}$, and therefore $(\varphi^{\star}\wh g)_{\mu\nu} = g_{\mu\nu}$ on $\mc U\cap\left(\mc H\cup\mc{\ul H}\right)$.

\end{proof}
\end{teo}

\begin{rmk}
The argument in Theorem \ref{main} can be immediately generalized to matter fields admitting a well-posed characteristic initial value problem in which their energy-momentum tensor on the hypersurfaces have the following dependence on the initial data: (i) $T_{n\, n}$ depends on the matter field, the metric data, and it is algebraic on $\bY(n,n)$; (ii) $T_{n\, A}$ depends on the matter field, the metric data, $\bY(n,n)$ and is algebraic on $\bY_{n\, A}$; (iii) The combination $h^{AB}T_{AB}-\frac{m-1}{2}\left(P^{ab}T_{ab}+2T(n,\xi)\right)$ depends on the matter field, the metric data, $\bY(n,n)$, $\bY_{n\, A}$ and it is algebraic on $h^{AB}\bY_{AB}$; (iv) $T_{AB}$ depends on the matter field, the metric data, $\bY(n,n)$, $\bY_{n\, A}$, $h^{AB}\bY_{AB}$ and it is algebraic on the trace-free part of $\bY_{AB}$. The third requirement follows from the fact that, from \eqref{inverse},
\begin{align*}
	g^{\mu\nu} T_{\mu\nu}  \st{\mc H}{=} \left(P^{ab} e^{\mu}_a e_b^{\nu} + 2n^a e_a^{\mu}\xi^{\nu}\right) T_{\mu\nu}
	\st{\mc H}{=} P^{ab} T_{ab} + 2T(n,\xi).
\end{align*}
\end{rmk}

\begin{appendices}
	\section{Gauge-covariance of the tensors $A$ and $B$}
	\label{appendix}
In this appendix we prove that the tensors $A$ and $B$ as defined in \eqref{A} and \eqref{B} transform as
\begin{equation}
	\label{transAyB}
	A'_{abc} = z(A_{abc}+\zeta^d B_{dabc}),\hspace{1cm} B'_{abcd}=B_{abcd}
\end{equation} 
under a gauge transformation $(z,\zeta)$. These are the expected transformations if one thinks the data as embedded, but we prove this statement in full generality.
\begin{prop}
	\label{covariance1}
	Let $\mc D=\{\mc H,\bg,\bm\ell,\ell^{(2)},\bY\}$ be null hypersurface data and $(z,\zeta)$ gauge parameters. Let $A$ and $B$ be the tensors defined in \eqref{A} and \eqref{B}, respectively. Then
	\begin{enumerate}
		\item $\mc G_{(z,\zeta)}(A) = z(A+i_{\zeta}B),$
		\item $\mc G_{(z,\zeta)}(B) = B$.
	\end{enumerate}
	\begin{proof}
By the composition law $(z,\zeta)=(z,0)\circ (1,\zeta)$ (see \eqref{group}) it suffices to prove the result with $(z,0)$ and $(1,\zeta)$ independently. We start by assuming that $\mc D$ is null hypersurface data written in a CG (we shall deal later with the general case). Then,
		\begin{align}
			A_{bcd}&=\bm\ell_a\ol{R}^a{}_{bcd},\label{Achar}\\
			B_{abcd}&=\bg_{af}\ol{R}^f{}_{bcd}-2\ol\nabla_{[c}\left(\bK_{d]b}\bm\ell_a\right).\label{Bchar}
		\end{align}
By the transformation laws \eqref{tranfell}, \eqref{Ktrans} and \eqref{gaugeconnection} the expression of $B$ above is insensitive to the transformations $(z,0)$, and thus we only need to show the invariance of $B$ under transformations of the form $(1,\zeta)$. Denote with a prime the transformed data,
		\begin{align}
			\label{Bprima}
			B_{abcd}'  = \bg_{af}\ol R'^f{}_{bcd}-2\ol\nabla'_{[c}\left(\bK_{d]b}'\bm\ell_a'\right)+2\ell'{}^{(2)}\bK_{a[d}'\bK'_{c]b}.
		\end{align}
		Since no products of $\bK$ appear in equation \eqref{Bchar}, a good strategy is to identify the elements with $\bK\cdot \bK$ in the two first terms and see that they cancel out with those in the third one. The first term is given by Proposition \ref{curvatura},
		\begin{equation}
			\label{aux2}
			\bg_{af}\ol R'^f{}_{bcd}  = \bg_{af}\left(\ol R^f{}_{bcd}+2\ol\nabla_{[c}\left(\zeta^f \bK_{d]b}\right)+2\zeta^f\zeta^g \bK_{g[c}\bK_{d]b}\right).
		\end{equation}
		For the second one we apply \eqref{tranfell} and \eqref{Ktrans}, as well as Proposition \ref{gaugeconection},
		\begin{equation*}
			\begin{aligned}
				2\ol\nabla'_{[c}\left(\bK_{d]b}'\bm\ell_a'\right) & = 2\ol\nabla_{[c}\left(\bK_{d]b}\bm\ell_a\right)+2\ol\nabla_{[c}\left(\bK_{d]b}\bg_{af}\zeta^f\right)\\
				&\quad+2\left(\bm\ell_a+\bg_{af}\zeta^f\right)\zeta^g \bK_{b[d}\bK_{c]g}+2\left(\bm\ell(\zeta)+\bg(\zeta,\zeta)\right) \bK_{a[d}\bK_{c]b}.
			\end{aligned}
		\end{equation*}
		Expanding the second term and inserting \eqref{olnablagamma},		
		\begin{equation}
			\label{aux3}
			\begin{aligned}
				2\ol\nabla'_{[c}\left(\bK_{d]b}'\bm\ell_a'\right) & = 2\ol\nabla_{[c}\left(\bK_{d]b}\bm\ell_a\right)+ 2\bg_{af}\ol\nabla_{[c}\left(\zeta^f\bK_{d]b}\right)\\
				&\quad +2\left(2\bm\ell(\zeta)+\bg(\zeta,\zeta)\right)\bK_{a[d}\bK_{c]b}+2\bg_{af}\zeta^f\zeta^g \bK_{g[c}\bK_{d]b} .
			\end{aligned}
		\end{equation}
		Introducing \eqref{aux2} and \eqref{aux3} into \eqref{Bprima}, and taking into account $\ell'{}^{(2)} = z^2\left(2\bm\ell(\zeta)+\bg(\zeta,\zeta)\right)$,
		\begin{align}
			\label{aux4}
			B'_{abcd} = \bg_{af} \ol R^f{}_{bcd} - 2\ol\nabla_{[c}\left(\bK_{d]b}\bm\ell_a\right) ,
		\end{align}
so the gauge invariance of $B$ follows. Now we study the transformation of $A$ in the same way. In the primed gauge, 
		\begin{equation}
			\label{Aprima}
			A_{bcd}'=\bm\ell_a'\ol{R}'{}^a{}_{bcd}+2\ell'{}^{(2)}\ol{\nabla}'_{[d} \bK'_{c]b} +\bK'_{b[c}\ol\nabla_{d]} \ell'{}^{(2)},
		\end{equation} 
		while in a characteristic gauge $A$ takes the form \eqref{Achar}. From \eqref{tranfell}, \eqref{transell2}, \eqref{gaugeconnection} and \eqref{Ktrans} it follows $\mc G_{(z,0)}(A) = zA$. Concerning transformations of the form $(1,\zeta)$, our strategy is to get rid of the terms with products of $\bK$ and products of $\zeta$. Firstly from \eqref{olnablagamma} it follows 
		\begin{equation}
			\label{nablagammagamma}
			\ol\nabla_a \left(\bg(\zeta,\zeta)\right) = 2\bg_{cb}\zeta^c\ol\nabla_a\zeta^b-2\bK_{ab}\bm\ell(\zeta)\zeta^b.
		\end{equation} 
		Since $\ell'{}^{(2)}=2\bm\ell(\zeta)+\bg(\zeta,\zeta)$, the second term of \eqref{Aprima} can be written as
		\begin{align*}
			2\ell'{}^{(2)}\ol\nabla'_{[d}\bK'_{c]b} = 2\left(2\bm\ell(\zeta)+\bg(\zeta,\zeta)\right)\left(\ol\nabla_{[d}\bK_{c]b}-\zeta^a\bK_{b[d}\bK_{c]a}\right),
		\end{align*}
		while using \eqref{nablagammagamma} the third one is	
		\begin{align*}
			\bK'_{b[c}\ol\nabla'_{d]}\ell'{}^{(2)} &=2z\bK_{b[c}\left(\bg_{fg}\zeta^f\ol\nabla_{d]}\zeta^g-\bK_{d]f}\bm\ell(\zeta)\zeta^f\right)+2z\bK_{b[c}\left(\zeta^f\ol\nabla_{d]}\bm\ell_f+\bm\ell_f\ol\nabla_{d]}\zeta^f\right).
		\end{align*}
For the first term in \eqref{Aprima} we use Proposition \ref{curvatura} and equation \eqref{tranfell}. Putting everything together and simplifying,
		\begin{align*}
			z^{-1}	A'_{bcd} & = \bm\ell_a \ol R^a{}_{bcd}+\bg_{af}\zeta^f\ol R^a{}_{bcd}+2\bm\ell_a \ol\nabla_{[c}\left(\zeta^a \bK_{d]b}\right)\\
			&\quad\, +4\bm\ell(\zeta)\ol\nabla_{[d} \bK_{c]b}+2\zeta^f \bK_{b[c}\ol\nabla_{d]}\bm\ell_f + 2\bm\ell_f \bK_{b[c}\ol\nabla_{d]}\zeta^f\\
			&=\bm\ell_a \ol R^a{}_{bcd}+\bg_{af}\zeta^f\ol R^a{}_{bcd}+2\bm\ell(\zeta)\ol\nabla_{[d}\bK_{c]b}+2\zeta^f \bK_{b[c}\ol\nabla_{d]}\bm\ell_f,
		\end{align*}
		and the claim $z^{-1} A'_{bcd} = A_{bcd}+\zeta^a B_{abcd}$ follows because $\zeta^a\ol\nabla_{[d}\left(\bK_{c]b}\bm\ell_a\right) = \bm\ell(\zeta)\ol\nabla_{[d}\bK_{c]b}+\zeta^f \bK_{b[c}\ol\nabla_{d]}\bm\ell_f$. To finish the proof we still need to show that the assumption that the initial gauge is characteristic does not spoil the generality of the argument. Let $\mc D$ be CHD and $\mc D''=\mc{G}_{(z'',\zeta'')}\mc D$. From Proposition \ref{propcaracteristico} there always exist gauge parameters $(z,\zeta)$ such that $\mc D' \d \mc G_{(z,\zeta)}\mc D$ is in a CG. Let $(z',\zeta')$ be the parameters making the following diagram commutative
		
		\begin{diagram}
			\mc D \arrow[rr, "\mc G_{(z'',\zeta'')}"]  \arrow[rd, bend right, "\mc G_{(z,\zeta)}"] & & \mc D''\\
			& \mc D' \arrow[ur, bend right, "\mc G_{(z',\zeta')}"] & 
		\end{diagram}
		
		\vspace{0.3cm}
		
		In other words, $(z',\zeta') =(z'',\zeta'')\circ (z,\zeta)^{-1} = (z'' z^{-1}, z(\zeta''-\zeta))$, where we have made use of \eqref{group} and \eqref{gaugelaw}. We already know that $\mc G_{(z',\zeta')} (A') = A'' = z'(A'+i_{\zeta'}B')$ and $\mc G_{(z',\zeta')}(B')=B'' = B'$, as well as $\mc G_{(z,\zeta)^{-1}} (A') = A = z^{-1}(A'+i_{-z\zeta}B)$ and $\mc G_{(z,\zeta)^{-1}} (B')=B=B'$, after using \eqref{gaugelaw} again. Then, since $A'' = z'(A'+i_{\zeta'}B) $ it follows
		\begin{align*}
			A'' =z'' \left(z^{-1} A' + i_{\zeta''}B' -i_{\zeta} B'\right)
			 = z''\left(A+i_{\zeta} B' +i_{\zeta''} B'-i_{\zeta} B' \right)
			= z''\left(A+i_{\zeta''} B\right),
		\end{align*}
		and hence $\mc G_{(z'',\zeta'')}(A) = z''\left(A+i_{\zeta''} B\right)$. The fact that $\mc G_{(z'',\zeta'')}(B) = B$ is obvious.
	\end{proof}
\end{prop}

The next natural step is to study the symmetries of $A$ and $B$. When the data is embedded, these symmetries are the ones inherited from the curvature tensor of the ambient space. Here we establish them without assuming embeddedness of the data.

\begin{prop}
	\label{symmetries}
	The tensors $A$ and $B$ possess the following symmetries:
	\begin{enumerate}
		\item $A_{bcd}=-A_{bdc}$,
		\item $A_{bcd}+A_{cdb}+A_{dbc}=0$,
		\item $B_{abcd} + B_{acdb} + B_{adbc}=0$,
		\item $B_{abcd}=-B_{abdc}=-B_{bacd}$,
		\item $B_{abcd}=B_{cdab}.$
	\end{enumerate}
	\begin{proof}
		The symmetries $A_{bcd}=-A_{bdc}$ and $B_{abcd}=-B_{abdc}$ are obvious. The second one is a consequence of the first Bianchi identity for $\ol R$ and the fact that $\bK$ is symmetric. The third one is analogous. In order to prove $B_{abcd}+B_{bacd}=0$ we first compute the symmetrization of the first term in \eqref{B} which, taking into account \eqref{olnablagamma}, is
		\begin{align*}
			\bg_{af}\ol R^f{}_{bcd}+\bg_{bf}\ol R^f{}_{acd} & = \ol\nabla_d\ol\nabla_c\bg_{ab}-\ol\nabla_c\ol\nabla_d\bg_{ab}\\
			&=-\ol\nabla_d\left(\bK_{ca}\bm\ell_b+\bK_{cb}\bm\ell_a\right)+\ol\nabla_c\left(\bK_{da}\bm\ell_b+\bK_{db}\bm\ell_a\right)\\
			&= 2\ol\nabla_{[c}\left(\bK_{d]a}\bm\ell_b\right)+2\ol\nabla_{[c}\left(\bK_{d]b}\bm\ell_a\right).
		\end{align*}
		For any symmetric tensor it holds $\bK_{b[d}\bK_{c]a}+\bK_{a[d}\bK_{c]b}=0$, so the symmetrization relative to the indices $a,b$ in the third term in \eqref{B} vanishes. Thus,
		\begin{align*}
			B_{abcd}+B_{bacd}  = 2\ol\nabla_{[c}\left(\bK_{d]a}\bm\ell_b\right)+2\ol\nabla_{[c}\left(\bK_{d]b}\bm\ell_a\right)-2\ol\nabla_{[c}\left(\bK_{d]b}\bm\ell_a\right)-2\ol\nabla_{[c}\left(\bK_{d]a}\bm\ell_b\right)=0.
		\end{align*}
		The fifth symmetry is consequence of items (3) and (4). Indeed, let $\accentset{\circ}{B}_{abcd}$ be the cyclic permutation $\accentset{\circ}{B}_{abcd} \d B_{abcd} + B_{acdb} + B_{adbc}$, which by the third item vanishes. Then, taking into account item (4), $$\hspace{-5mm}0  = \accentset{\circ}{B}_{abcd} - \accentset{\circ}{B}_{bcda} -\accentset{\circ}{B}_{cdab}+\accentset{\circ}{B}_{dcab}= B_{abcd} - B_{bacd} - B_{cdab} + B_{dcab}=2B_{abcd} - 2B_{cdab}.$$ \vskip -9mm
	\end{proof}
\end{prop}

\section{Some contractions of the tensors $A$ and $B$}
\label{appendixB}

In this appendix we compute the contractions of the tensors $A$ and $B$ that are needed in Section \ref{sec_fol} to write down the constraint tensors in terms of the foliation tensors.

\begin{lema}
	\label{LemaB}
Let $\mc D$ be CHD written in a characteristic gauge, $\{e_A\}$ a basis of $\Gamma(T\mc S)$, $V\in\Gamma(T\mc H)$ and let $A$ and $B$ the tensors defined in \eqref{A} and \eqref{B}. Then,
\begin{align}
A_{bcd} n^b n^d V^c & = (\ol\nabla_n \bPi)(V,n)-(\ol\nabla_V \bPi)(n,n),\label{propJ1}\\
B_{abcd} n^a P^{bd} V^c & =-\bPi(V,n) \tr_P \bK +\left(\bK*\bPi\right)(V,n) - \ol\nabla_V \tr_P \bK+\div_P(\bK)(V),\label{propJ2}\\
B_{cadb} P^{cd}e_A^ae_B^b &= {}^h R_{AB} - \tr_h{\bm\chi}\ {\bm{\Upsilon}}_{AB}-\tr_h{\bm{\Upsilon}}\ {\bm\chi}_{AB}+\left({\bm\chi}\cdot{\bm{\Upsilon}}\right)_{AB}+\left({\bm{\Upsilon}}\cdot{\bm\chi}\right)_{AB},\label{BP}\\
A_{bca}n^ce_A^ae_B^b &= -\nabla^h_A{\bm{\eta}}_B + 2\omega {\bm{\Upsilon}}_{AB}-\wt\nabla_n {\bm{\Upsilon}}_{AB}-{\bm{\eta}}_A{\bm{\eta}}_B-\left({\bm\chi}\cdot{\bm{\Upsilon}}\right)_{AB},\label{An}
\end{align}
where we define $\tr_P \bK\d P^{ab}\bK_{ab}$, $\left(\bK*\bPi\right)_{ca} \d P^{bd} \bK_{bc}\bPi_{da}$, $\div_P(\bK)(V)\d P^{ab}V^c\ol\nabla_a \bK_{bc}$ and $(S\cdot T)_{AB}\d h^{CD} S_{AC}T_{BD}$ for any pair of two-covariant tensors $S$ and $T$.
\begin{proof}
The first one follows from the expression of $A$ in \eqref{Achar} and Lemma \ref{lema} in a CG. Now, from \eqref{Bchar} and taking into account $\bg_{af}\ol R^f{}_{bcd}n^a=0$,
\begin{align*}
	B_{abcd} n^a  = 2\bm\ell_a \bK_{b[d}\ol\nabla_{c]}n^a-2\ol\nabla_{[c} \bK_{d]b}=    2\bK_{b[c}\bPi_{d]a}n^a - 2\ol\nabla_{[c} \bK_{d]b},
\end{align*} 
where in both equalities we used $\bm\ell(n)=1$ and in the second we inserted \eqref{olnablan2} and used $P(\bm\ell,\cdot)=0$ (see \eqref{Pell}). Contracting with $P^{bd} V^c$ and using $\bK_{bc}\ol\nabla_a P^{bc}=0$, which is a direct consequence of \eqref{olnablaP} and $\bK(n,\cdot)=0$, $$B_{abcd} n^a P^{bd} V^c  = -P^{bd} \bK_{bd} \bPi(V,n) + P^{bd} \bK_{bc}V^c\bPi_{da}n^a - \ol\nabla_V\left(P^{bd} K_{bd}\right)+P^{bd}V^c\ol\nabla_d \bK_{bc},$$ and hence \eqref{propJ2} follows. Next we proceed with the third one. Inserting \eqref{olnablagamma} into \eqref{Bchar}, the tensor $B$ in a CG is $B_{cadb} = \bg_{cf}\ol R^f{}_{adb}-2\bK_{a[b} \bPi_{d]c} - 2\bm\ell_c\ol\nabla_{[d}\bK_{b]a}$, so using \eqref{PAB} and that $\bPi_{DC}=\bY_{DC}$ in a CG (see Remark \ref{observación}) the contraction with $P^{cd}e^a_Ae^b_B$ becomes $$B_{cadb} P^{cd}e_A^ae_B^b  = \bg\left(e_C,\ol R(e_D,e_B)e_A\right) h^{CD} - 2{\bm\chi}_{A[B}\bY_{D]C}h^{CD},$$ which is \eqref{BP} after using the Gauss identity \eqref{gaussfacil}. Finally, in order to compute the term $A_{bca}n^ce_A^ae_B^b$ we first recall the expression \eqref{Achar} for $A$ in a CG and equation \eqref{nlR}, namely 
\begin{equation}
	\label{auxRAB}
	A_{bca}n^c=n^c\bm\ell_d\ol R^d{}_{bca} = \ol\nabla_a\left(\bPi_{cb}n^c\right)-\ol\nabla_n \bPi_{ab} - P^{cd}\bK_{da}\bPi_{cb}+\bPi_{cb}\bPi_{ad}n^cn^d.
\end{equation}
From decomposition \eqref{nablan}, Remark \ref{observación} and the definitions of ${\bm{\eta}}$, ${\bm{\Upsilon}}$ and $\omega$, the term $e^a_Ae^b_B\ol\nabla_n\bPi_{ab}$ is given by 
\begin{align*}
	e^a_Ae^b_B\ol\nabla_n\bPi_{ab} &=n\big(\bPi(e_A,e_B)\big) - \bPi\big(\wt\nabla_n e_A+{\bm{\eta}}(e_A)n, e_B\big)- \bPi\left(e_A, \wt\nabla_n e_B+{\bm{\eta}}(e_B)n,\right)\\
	&=\wt\nabla_n {\bm{\Upsilon}}_{AB} +{\bm{\eta}}_A{\bm{\eta}}_B -{\bm{\tau}}_A{\bm{\eta}}_B\\
	&= \wt\nabla_n {\bm{\Upsilon}}_{AB}+2{\bm{\eta}}_A{\bm{\eta}}_B+{\bm{\eta}}_B\nabla^{h}_A\log|\lambda|,
\end{align*}
where in the second line we employed equations \eqref{etaPi} and \eqref{tauPi} and in the last equality we replaced ${\bm{\tau}}$ by ${\bm{\eta}}$ according to \eqref{taueta}. The term $e_A^ae_B^b  \ol\nabla_a\left(\bPi_{cb}n^c\right)$ can be computed using decomposition \eqref{decompnabla}, Remark \ref{observación} and the definitions of $\omega$ and ${\bm{\Upsilon}}$, 
\begin{align*}
	e_A^ae_B^b  \ol\nabla_a\left(\bPi_{cb}n^c\right) = e_A\left(-{\bm{\eta}}_B\right)-\bPi\left(n,\nabla^h_{e_A} e_B-{\bm{\Upsilon}}_{AB}n\right)=-\nabla^h_A{\bm{\eta}}_B + 2\omega {\bm{\Upsilon}}_{AB},
\end{align*}
where again \eqref{etaPi} has been taken into account. From \eqref{PAB}, \eqref{etaPi} and \eqref{tauPi}, $$\bPi_{cb}\bPi_{ad}n^cn^d e^a_A e^b_B=-{\bm{\eta}}_B{\bm{\tau}}_A,$$ which becomes ${\bm{\eta}}_A{\bm{\eta}}_B+{\bm{\eta}}_B\nabla_A^h\log|\lambda|$ after using \eqref{taueta}. Contracting \eqref{auxRAB} with $e_A^ae_B^b$ and inserting the expressions above, as well as \eqref{PAB}, equation \eqref{An} follows. 
\end{proof}
\end{lema}

\end{appendices}

\section*{Acknowledgements}

We thank P.T. Chruściel for very useful comments on the uniqueness part of the main result which helped us improving the presentation. This work has been supported by Projects PGC2018-096038-B-I00, PID2021-122938NB-I00 (Spanish Ministerio de Ciencia e Innovación and FEDER ``A way of making Europe'') and SA096P20 (JCyL). G. Sánchez-Pérez also acknowledges support of the PhD. grant FPU20/03751 from Spanish Ministerio de Universidades.

\begingroup
\let\itshape\upshape

\renewcommand{\bibname}{References}
\bibliographystyle{acm}

\bibliography{biblio}

\end{document}